\documentclass[a4paper,11pt]{article}
\usepackage{fontawesome}
\pdfoutput=1 

\usepackage{jcappub}
\usepackage{changepage}

\usepackage{amsmath}
\usepackage{mathrsfs}
\usepackage{appendix}
\usepackage{graphicx}
\usepackage{xcolor}
\usepackage{listings}
\usepackage{comment}
\usepackage{urwchancal}
\usepackage{multirow}
\usepackage{mathrsfs}
\usepackage{float}

\usepackage{cleveref}

\newcommand{\github}[1]{%
   \href{#1}{\faGithubSquare}%
}
\makeatother

\usepackage[T1]{fontenc} 
\def\ilaria#1{\textcolor{red}{[Ilaria: #1]}}

\title{\texttt{GWBird}: a toolkit for the characterization of the Stochastic Gravitational Wave Background for Ground, Space, and Pulsar Timing Array detectors}

\author{Ilaria Caporali}
\author{, Angelo Ricciardone}
\affiliation{Dipartimento di Fisica ``Enrico Fermi'',  Università di Pisa, Largo Bruno Pontecorvo 3, Pisa
I-56127, Italy}
\affiliation{INFN, Sezione di Pisa, Largo Bruno Pontecorvo 3, Pisa I-56127, Italy}

\emailAdd{ilaria.caporali@phd.unipi.it}
\emailAdd{angelo.ricciardone@unipi.it}

\abstract{The detection of the Stochastic Gravitational Wave Background (SGWB) is one of the most challenging tasks for both current and next-generation detectors. Successfully distinguishing the SGWB from instrumental noise and environmental effects requires accurate and flexible analysis tools capable of detecting the signal and determining its origin. In this paper, we introduce a unified framework and a user-friendly tool for SGWB characterization: \texttt{GWBird} (Gravitational Wave Background Inventory of Response functions for Detectors). This code enables the computation of overlap reduction functions (ORFs), power-law integrated sensitivity curves (PLS), angular response functions, and angular PLS (APLS).  It supports the full range of gravitational wave polarization modes (tensor, scalar, and vector), allowing for the characterization of both isotropic and anisotropic SGWB components for all the polarizations. Additionally, the code includes functions for circular polarization characterization, which is particularly relevant for probing parity-violating signals. The framework integrates analyses for ground-based, space-based, and Pulsar Timing Array (PTA) detectors, offering a versatile framework for SGWB analysis.   The \texttt{GWBird} code is publicly available at:~\github{https://github.com/ilariacaporali/GWBird}}

\begin{document}
\maketitle
\flushbottom

\section{Introduction}

The groundbreaking direct detection of a gravitational wave (GW) signal in 2015 by the LIGO-Virgo-KAGRA (LVK) collaboration revolutionize our understanding of the universe~\cite{LIGOScientific:2016aoc, LIGOScientific:2016lio}. From then, many additional GW resolved signals, linked to astrophysical phenomena, have been detected~\cite{LIGOScientific:2016sjg, LIGOScientific:2017bnn, LIGOScientific:2017ycc, LIGOScientific:2017vwq, LIGOScientific:2020iuh} and have provided crucial information on the astrophysical properties of compact objects in the solar mass range. Recently, another relevant step in the GW field, has been done by the Pulsar Timing Array (PTA) collaboration (NANOGrav, EPTA/InPTA, PPTA, and CPTA), which has claimed evidence supporting the detection of a Stochastic Gravitational Wave Background~\cite{NANOGrav:2023gor, EPTA:2023fyk, Reardon:2023gzh, Xu:2023wog}, whose origin --cosmological or astrophysical-- is still unclear.  These findings mark a significant landmark in the field of astrophysics and cosmology, presenting great opportunities for delving into high-energy physics, early universe cosmology, and fundamental physics. All these detections have relied on a perfect knowledge of the layout and noise of the detectors which allowed for the extraction of the signal from the data stream.

Now, we are in the transition between second and third generation (3G) detectors, which will have better sensitivity and will also cover larger frequency bands. The path for such third generation detectors, such as the Laser Interferometer Space Antenna (LISA)~\cite{LISA:2017pwj, LISA:2024hlh}, Einstein Telescope (ET)~\cite{Maggiore:2019uih} and Cosmic Explorer (CE)~\cite{Reitze:2019iox, Evans:2021gyd}, is still long, especially for ET, since the final configuration is still under discussion (see~\cite{Branchesi:2023mws, Abac:2025saz, Caporali:2025mum, Begnoni:2025oyd} for recent paper where different configurations have been compared). However, it is already known that, when such detectors will be operative a change of paradigm will be required, both in the GW searches, and in the parameter estimation, since we will face with data stream where several (overlapping) sources will be present at the same time. Ongoing studies on mission science goals and development of advanced tools for the parameter estimation, such as the {\it{Global fit}}~\cite{Littenberg:2020bxy,Littenberg:2023xpl, Katz:2024oqg, Deng:2025wgk, Strub:2024kbe} pipelines, have to rely on flexible and fast detector characterization tools.  This will be of extreme importance both to detect hundred of thousands of resolved GW events, and also the SGWB from astrophysical and cosmological sources, which will be a guaranteed target for 3G detectors. For this reason, having a unique tool for both ground and space-based detector would be extremely useful. 

Such forthcoming 3G interferometers will also enable more stringent tests of General Relativity (GR)~\cite{LISACosmologyWorkingGroup:2019mwx, Abac:2025saz}. A key property lies in their ability to disentangle different polarizations with higher precision~\cite{Nishizawa:2009bf, Callister:2017ocg, Amalberti:2021kzh, Hu:2022byd}. In standard GR, GWs are restricted to two transverse-traceless tensor polarizations; however, several beyond-GR theories predict additional degrees of freedom (scalar or vector polarizations)~\cite{LIGOScientific:2018czr}. Detecting, or placing stringent upper limits on these extra polarization modes would significantly constrains such beyond-GR models, thereby providing deeper insight into the fundamental nature of gravity. In addition to scalar and vector modes, circularly polarized GWs, which are typically associated to parity-violating processes in the early universe, or also astrophysical background~\cite{ValbusaDallArmi:2023ydl}, constitute another potential signature of new physics.
Moreover, 3G interferometer networks, featuring extended detector arms and advanced baseline configurations, will offer enhanced angular resolution~\cite{Alonso:2020rar, Mentasti:2020yyd,LISACosmologyWorkingGroup:2022kbp, Schulze:2023ich, Mentasti:2023gmg}. This improvement is pivotal for detecting and accurately mapping the angular distribution of the SGWB, a key observable for its characterization. In particular, measuring the SGWB anisotropies provides a powerful means for distinguishing and disentangling astrophysical contributions from cosmological ones. Indeed, various studies have estimated the levels of anisotropy expected for both astrophysical~\cite{Cusin:2017fwz,Jenkins:2018nty, Bertacca:2019fnt,Bellomo:2021mer, Capurri:2022lze, Capurri:2021zli, Allen:2024mtn} and cosmological SGWBs~\cite{ValbusaDallArmi:2023nqn,ValbusaDallArmi:2024hwm,Mierna:2024pkh,Schulze:2023ich}, underscoring the central role that angular resolution and multi-detector cross-correlation methods will play in next-generation detectors~\cite{Ricciardone:2021kel, ValbusaDallArmi:2024mxz, Capurri:2021prz}.

We have developed a unified framework for characterizing the SGWB that accommodates a broad range of detector classes, from ground-based detectors and space-based missions to PTAs. At the core of this framework is \texttt{GWBird}, a fast, flexible, and accurate Python-based code composed of multiple modules dedicated to evaluating the essential components that characterize the SGWB detectability.

In particular, \texttt{GWBird} derives the {\it detector response} functions for detector network, starting from their specified geographic and geometric configurations. The code is explicitly designed to incorporate tensor, scalar, and vector polarizations, providing comprehensive coverage of GR as well as extended theories of gravity. In addition, it supports both isotropic and anisotropic SGWB signals, enabling users to generate and analyze the angular response of each detector. This modular approach not only facilitates detailed comparisons across various detector types but also streamlines the process of constructing cross-correlation analyses for multi-detector networks, thereby offering a robust platform for future SGWB searches and parameter estimation efforts.

A key tool in assessing the detectability of a SGWB is the {\it Power Law Sensitivity} (PLS), first introduced in 2013~\cite{Thrane:2013oya}. This graphical methodology estimates detector or network performance under the assumptions of stationarity, Gaussianity, isotropy, and a SGWB power-law spectral shape. In \texttt{GWBird}, we have implemented a dedicated PLS module that goes one step further, by accommodating multiple polarization states—tensor, scalar, and vector—thus expanding the usual GR-focused approach. Additionally, \texttt{GWBird} features an {\it Angular Power Law Sensitivity} (APLS) module, which extends the PLS to account for angular anisotropies in the SGWB. By evaluating the detector or network response to different multipoles, the APLS module provides crucial insights into the angular structure of the background.
Since the code is tailored also for 3G ground-based detectors, which relative orientation in the case of 2L detectors has not been defined yet, it gives the possibility to set the orientation of the detectors in a network and compute the detector response, the PLS and the APLS for different orientations.

Although the detection strategy for PTAs differs from that of interferometric detectors, \texttt{GWBird} also includes a dedicated module for PTA-based SGWB searches. In particular, it provides both the Response and Angular Response functions for PTA, extending the concept also to tensor, scalar, and vector polarizations. The code introduces the PLS definition derived from the NANOGrav pulsar catalog~\cite{nanograv15yr}, enabling a realistic estimation of PTA sensitivity. Analogously to the interferometer module, \texttt{GWBird} supports an angular sensitivity assessment for PTAs, evaluating the detector response to various multipoles. Such angular characterization can be crucial for distinguishing between an astrophysical SGWB from supermassive black hole binaries and a cosmological background~\cite{Depta:2024ykq}. 

Since parity violation is a key observable for distinguishing the origin of the SGWB, we have implemented in \texttt{GWBird} several functions to probe the circular polarization. We have provided the detector response, the PLS and the APLS for the circular polarization both for interferometric and PTA detectors.

\texttt{GWBird} is user-friendly, easily customizable to other detectors, and freely accessible to the scientific community.
The paper is structured as it follows. 
In Section~\ref{Formalism} we present the formalism of cross-correlation searches for the SGWB.
In Section~\ref{code} we present the code \texttt{GWBird} and its functionalities.
In Section~\ref{Results} we display our results obtained for all the classes of detectors and we discuss them. Lastly, we give our conclusions in Section~\ref{Conclusion}.

\section{Formalism}
\label{Formalism}

In this section, we give the fundamental definitions and tools for performing cross correlation searches for the SGWB detection.
\subsection{Detector response}

GWs are time-varying perturbations of the spacetime metric that travel at the speed of light. In the transverse-traceless (TT) gauge, these metric perturbations can be written as the superposition of plane waves with frequency $f$~\cite{Allen:1997ad, Thrane:2013oya, Romano:2016dpx, Maggiore:2007ulw, Christensen:1992wi, Finn:2008vh} 
\begin{equation}
    h_{ij} (t, \vec{x}) = \int_{-\infty}^{+ \infty} df \int_{4 \pi} d \Omega_{\hat{k}}^2 \sum_{\lambda = +, \times} h_{\lambda}(f, \hat{k}) \, e^{\lambda}_{ij}(\hat{k}) \, e^{i2\pi f \left( t - \frac{\hat{k} \cdot \vec{x}}{c} \right)}\,,
    \label{h_ij}
\end{equation}
where $\lambda = +, \times$ are the GW polarizations,  $h_{\lambda}(f, \hat{k})$ are the Fourier modes of the SGWB for each polarization $\lambda$ at a given frequency $f$ and direction of propagation specified by the unit vector $\hat{k}$, which are random fields \footnote{Their probability distributions define the statistical properties of the SGWB. }, $e_{ij}^{\lambda} (\hat{k})$ are the GW polarization tensors, $\hat{k}$ is the GW propagation direction and $c$ is the speed of light.

For an unpolarized, isotropic, and Gaussian SGWB, the two-point correlation function in Fourier space is given by~\cite{Thrane:2013oya} 
\begin{equation}
    \left\langle h_{\lambda}(f, \hat{k}) \,  h_{\lambda'}^*(f', \hat{k'})  \right\rangle = \frac{1}{16 \pi} \delta(f - f') \delta_{\lambda \lambda'} \delta^2 (\hat{k}, \hat{k}') S_{h}(f)\,,
\end{equation}
where $\delta(f - f')$ enforces the stationarity, $\delta_{\lambda \lambda'}$ the unpolarized nature and $\delta^2 (\hat{k}, \hat{k}')$ the isotropy.
$S_{h}(f)$ is the one-sided GW strain power spectral density, defined as
\begin{equation}
    S_{h}(f) = \frac{3 H_0^2}{2 \pi^2} \frac{\Omega_{\rm GW}(f)}{f^3}\,,
\end{equation}
where $H_0$ is the Hubble parameter today, and $\Omega_{\rm GW}$ is the fractional energy density spectrum,  which is defined as
\begin{equation}
    \Omega_{\rm GW}(f) = \frac{1}{\rho_c} \frac{d\rho_{\rm GW}}{d \, \log f}\,,
\end{equation}
with $\rho_c = 3 c^2 H_0^2 / 8\pi G$ the critical energy density and $\rho_{\rm GW}$ the energy density of GWs.

The detectors response to a passing GW signal $h(t)$ is the convolution of the metric perturbations $h_{ij}$ with the detector tensor $D^{ij}$ \cite{Romano:2016dpx}
\begin{equation}
    h(t, \vec{x}) =  \int_{-\infty}^{+ \infty} d\tau \int d^3 y \, D^{ij}(\tau, \vec{y}) h_{ij}(t-\tau, \vec{x}- \vec{y})\,.
    \label{h}
\end{equation}
By plugging eq.~\eqref{h_ij} in eq.~\eqref{h}, we get
\begin{align}
    h(t) &=  \int_{-\infty}^{+ \infty} d\tau \int d^3 y   \int_{-\infty}^{+ \infty} df \int_{4 \pi} d \Omega_{\hat{k}}^2 \sum_{\lambda = +, \times} h_{\lambda}(f, \hat{k}) \;D^{ij}(\tau, \vec{y}) \; e^{\lambda}_{ij}(\vec{k}) \; e^{-i2\pi f \left( \tau - \frac{\hat{k} \cdot \vec{y}}{c} \right)} \;  e^{i2\pi f \left( t- \frac{\hat{k} \cdot \vec{x}}{c} \right)} \,.
\end{align}
Now, we define 
\begin{equation}
    F^{\lambda} (f, \vec{k}) = \int_{-\infty}^{+ \infty} d\tau \int d^3 y \; D^{ij}(\tau, \vec{y}) \; e^{\lambda}_{ij}(\vec{k}) \; e^{-i2\pi f \left( \tau - \frac{\hat{k} \cdot \vec{y}}{c} \right)}\,,
\end{equation}
and we get \footnote{Note that here we make the assumption that the signal is observed for an infinite amount of time. If this assumption does not hold, look at \cite{Romano:2016dpx}.}
\begin{equation}
    h(t, \vec{x}) =  \int_{-\infty}^{+ \infty} df \; e^{i2\pi f t} \int d\Omega_{\hat{k}}^2 \sum_{\lambda = +, \times} h_{\lambda}(f, \hat{k}) \;F^{\lambda} (f, \vec{k}) \;  e^{-i2\pi f \left( \frac{\hat{k} \cdot \vec{x}}{c} \right)} \,.
\end{equation}
So the projected GW in Fourier space reads
\begin{equation}
    \tilde{h}(f) = \int d\Omega_{\hat{k}}^2 \sum_{\lambda = +, \times} h_{\lambda}(f, \hat{k}) \;F^{\lambda} (f, \vec{k}) \;  e^{-i2\pi f \left( \frac{\hat{k} \cdot \vec{x}}{c} \right)} \,.
    \label{response_strain}
\end{equation}
\subsection{Detector Tensor and Angular Pattern Function}

Here, we derive the detector tensor $D_{ij}$ and the angular pattern function $F^{\lambda}(\theta, \phi)$. First, let us assume the following orthonormal coordinate system for the detector~\cite{Nishizawa:2009bf, Romano:2016dpx}
\begin{equation}
    \begin{cases}
        \hat{x} = (1,0,0) \\
        \hat{y} = (0,1,0) \\
        \hat{z} = (0,0,1)\,. \\
    \end{cases}
\end{equation}
Then, we consider the GW coordinate system rotated by angles $(\phi, \theta)$ 
\begin{equation}
    \begin{cases}
        \hat{u} = (\cos{\theta} \cos{\phi}, \, \cos{\theta}\sin{\phi}, \, - \sin{\theta}) \\
        \hat{v} = (-\sin{\phi}, \, \cos{\phi},0) \\
        \hat{k} = (\sin{\theta}\cos{\phi}, \, \sin{\theta} \sin{\phi}, \, \cos{\theta}) \,.\\
    \end{cases}
\end{equation}
In order to consider the most general choice of coordinates, we consider the polarization angle $\psi$ (this consists in a rotation with respect to the GW propagating axis). Thus we get
\begin{equation}
    \begin{cases}
    \hat{m} = \hat{u} \cos{\psi} \, + \, \hat{v}\sin{\psi}\\
    \hat{n} = -\hat{u} \sin{\psi} \, + \, \hat{v}\cos{\psi}\\
    \hat{k} = \hat{k}\,.
    \end{cases}
\end{equation}
A generic metric of gravity in four dimensions allows, at most, six polarization modes of a GW \footnote{ If a spacetime includes extra dimensions, the number of polarization modes can be more than six. However, once the polarization are projected onto our 3-space, the polarization we observe are degenerate and only 6 of them survives}. For a GW propagating in the $\hat{z}$ direction, the basis of the six polarizations is defined by
\begin{align}
    & e^{+}_{ij} = \begin{pmatrix}
                    1 & 0 & 0 \\
                    0 & -1 & 0 \\
                    0 & 0 & 0 \\
                 \end{pmatrix} , \; \; \; 
    e^{\times}_{ij} = \begin{pmatrix}
                    0 & 1 & 0 \\
                    1 & 0 & 0 \\
                    0 & 0 & 0 \\
                 \end{pmatrix} , \nonumber  \\
    & e^{x}_{ij} = \begin{pmatrix}
                    0 & 0 & 1 \\
                    0 & 0 & 0 \\
                    1 & 0 & 0 \\
                 \end{pmatrix} , \; \; \; 
   \;\; e^{y}_{ij} = \begin{pmatrix}
                    0 & 0 & 0 \\
                    0 & 0 & 1 \\
                    0 & 1 & 0 \\
                 \end{pmatrix} , \nonumber  \\
    & e^{b}_{ij} = \begin{pmatrix}
                    1 & 0 & 0 \\
                    0 & 1 & 0 \\
                    0 & 0 & 0 \\
                 \end{pmatrix} , \; \; \; 
    e^{l}_{ij} = \sqrt{2}\begin{pmatrix}
                    0 & 0 & 0 \\
                    0 & 0 & 0 \\
                    0 & 0 & 1 \\
                 \end{pmatrix} ,    
\end{align}
where $+, \times$ denotes the tensor, $x, y$ the vector and $b, \ell$ (breathing and longitudinal) the scalar polarizations, respectively. 
These polarization representations are defined in a 3-dimensional space. Each polarization mode is orthogonal to one another and it is normalized such as $e^{\lambda}_{ij} e^{ij}_{\lambda'} = 2 \delta_{\lambda \lambda'}$ with $\lambda= +, \times, x, y, b, l$. \footnote{Note here that $e^{l}_{ij}$ has a $\sqrt{2}$ factor, following the convention used in~\cite{Nishizawa:2009bf, Amalberti:2021kzh, Isi:2015cva}. }

In order to give an expression for the polarization modes, we choose the most general orthonormal basis $\{ \hat{m}, \hat{n}, \hat{k}\}$~\cite{Nishizawa:2009bf}
\begin{align}
    e^{+}_{ij} &= m_i m_j - n_i n_j \,, \nonumber\\
    e^{\times}_{ij} &= m_i n_j + n_i m_j \,,\nonumber\\
    e^{x}_{ij} &= m_i k_j + k_i m_j \,,\nonumber\\
    e^{y}_{ij} &= n_i k_j + k_i n_j \,,\nonumber\\
    e^{b}_{ij} &= m_i m_j + n_i n_j\,,\nonumber \\
    e^{l}_{ij} &= \sqrt{2} k_i k_j \,. 
    \label{eq:pol_tensors}
\end{align}
The detector response to an incoming GW signal is represented by the angular pattern function (APF), a scalar quantity defined as 
\begin{equation}
    F^{\lambda} (\theta, \phi) = D^{ij}(\theta, \phi) e^{\lambda}_{ij} (\theta, \phi)\,,
    \label{apf}
\end{equation}
where $D^{ij}$ is the detector tensor, which contains all the information about the detector geometry, and it is defined as
\begin{equation}
    D^{ij}= \frac{1}{2} \{ u^i u^j -  v^i v^j \}\,, 
\end{equation}
where $u , \, v$ are the unit vectors along each interferometer arm. This definition is valid in the so-called  ``low-frequency approximation''. In general, one should also take into account a transfer function of the following type~\cite{Romano:2016dpx}
\begin{align}
    \mathcal{T}(\hat{u}  \cdot \hat{k}, f) =  e^{-2 \pi i f \frac{L}{c}} \left[  e^{- \pi i f \frac{L}{c} (1 - \hat{u}\cdot \hat{k})} \text{sinc} \left( \frac{\pi f L}{c} (1 + \hat{u}\cdot \hat{k} ) \right) \right. \nonumber \\
     + \left. e^{- \pi i f \frac{L}{c} (1 + \hat{u}\cdot \hat{k})} \text{sinc} \left( \frac{\pi f L}{c} (1 - \hat{u}\cdot \hat{k} ) \right)   \right] \,.
     \label{transfer_eq}
\end{align}
In this way, the detector tensor becomes
\begin{align}
    D^{ij} = \frac{1}{2} \left[ u^i u^j   \mathcal{T}(\hat{u}  \cdot \hat{k}, f) -  v^i v^j   \mathcal{T}(\hat{v}  \cdot \hat{k}, f) \right] \, . 
\end{align}
The transfer function will mainly affect interferometers where the low frequency approximation cannot be applied, i.e., the frequency of the measured GW signal is comparable or higher than the characteristic frequency of the detector which is defined as $f_{*} = c / 2 \pi L$, where $L$ is the length of the detector arm.
 While the transfer function is usually neglected in LIGO, Virgo and ET, since it starts to be dominant at frequencies higher than the working ones, this is not the case for the LISA detector, for instance. However, \texttt{GWBird} naturally implements the transfer function term for all types of interferometers, both ground and space-based. \\

\subsection{Overlap Reduction Function}
\label{ORF}

The overlap reduction function characterizes the two-point correlation of the gravitational-wave strain between detectors $i$ and $j$. From eq.~\eqref{response_strain} one obtains
\begin{align}
    \left< \tilde{h}_i(f) \tilde{h}_j^{*}(f') \right> =& \frac{1}{2}   \delta(f - f')  \frac{3H_0^2}{10 \pi^2 f^3} \, \Omega_{\rm GW}(f) \gamma_{ij}(f)\,,
    \label{2ptstrain}
\end{align}
where $\gamma_{ij}(f)$ is the overlap reduction function that is defined as \footnote{Note that, in our convention, $\gamma_{ij}$  includes a normalization factor $5/8\pi$. This ensures that for two co-located and perfectly aligned detectors, the overlap reduction function evaluates to one.}
\begin{align}
    \gamma_{ij}(f) = \frac{5}{8 \pi} \int_{4\pi} d^2\Omega_{\hat{k}}  \;  \sum_{\lambda} F_i^{\lambda}(f, \hat{k})   F_j^{\lambda *}(f, \hat{k}) \,   e^{-i2\pi f \frac{(\vec{x}_i - \vec{x}_j)\cdot \hat{k}}{c}}\,.
\end{align}
We can extend the definition of the overlap reduction function (ORF) to the all the polarization modes~\cite{Nishizawa:2009bf, Amalberti:2021kzh, Callister:2017ocg} \footnote{We adopt a notation in which the polarization modes are indicated by their lower case initials: $t$ for tensors, $v$ for vectors, and $s$ for scalars.}
\begin{align}
    \gamma_{ij}^{t}(f) &= \frac{5}{8\pi}  \int d^2\Omega_{\hat{k}}  \;  \sum_{\lambda= +, \times} F_i^{\lambda}(f, \hat{k})   F_j^{\lambda*}(f, \hat{k}) \,   e^{-i2\pi f \frac{(\vec{x}_i - \vec{x}_j)\cdot \hat{k}}{c}}\,, \label{orf_t}  \\
    \gamma_{ij}^{v}(f) &= \frac{5}{8\pi}  \int d^2\Omega_{\hat{k}}  \;  \sum_{\lambda= x, y} F_i^{\lambda}(f, \hat{k})   F_j^{\lambda*}(f, \hat{k}) \,   e^{-i2\pi f \frac{(\vec{x}_i - \vec{x}_j)\cdot \hat{k}}{c}}\,, \label{orf_v} \\
    \gamma_{ij}^{s}(f) &=  \frac{5}{4\pi(1 + \mathcal{K})} \int  d^2\Omega_{\hat{k}}  \;  \left(F_i^{b}(f, \hat{k})   F_j^{b*}(f, \hat{k}) + \mathcal{K}F_i^{l}(f, \hat{k})   F_j^{l*}(f, \hat{k}) \right) \,   e^{-i2\pi f \frac{(\vec{x}_i - \vec{x}_j)\cdot \hat{k}}{c}}\,, \label{orf_s}
\end{align}
where $\mathcal{K}$ is the ratio of the energy density between the longitudinal and breathing mode, due to their degeneracy~\cite{Nishizawa:2009bf, Amalberti:2021kzh, Chatziioannou:2021mij} \footnote{We follow an hybrid notation between~\cite{Nishizawa:2009bf} and~\cite{Callister:2017ocg}. Namely, we include the prefactor $\xi$ in the normalization coefficient of $\gamma_{ij}^{Scalar}(f)$ as in~\cite{Callister:2017ocg} but  we drop the assumption made in~\cite{Callister:2017ocg} of having an unpolarized SGWB of scalar modes, by introducing this $\mathcal{K}$ factor in the scalar longitudinal sector, as in~\cite{Nishizawa:2009bf}. Throughout all the results of the paper, we set $\mathcal{K}=0$. Since interferometric detectors cannot distinguish between breathing and longitudinal scalar modes, they are considered degenerate. Therefore, following standard convention~\cite{Nishizawa:2009bf, Chatziioannou:2021mij}, we consider only one and will refer to it as the scalar mode from this point onward, in the context of interferometric detectors.}.

For detectors with a triangular configuration, such as LISA, given that they are composed by three detectors (or channels), more combinations can be realized. Typically, there are two choices for detector combinations: one is the XYZ basis, where X, Y, Z are the three TDI (Time Delay Interferometry) channels, and one is the AET basis, where signal or noise can be diagonalized, when equal arm lengths and identical noise levels are considered (see~\cite{Hartwig:2023pft, Kume:2024sbu} for the case where unequal arm and noise levels are considered).  As shown in~\cite{Flauger:2020qyi}, when the three interferometers have the same noise properties, the transformation matrix from the XYZ basis to the AET basis is given by
\begin{equation}
R= \begin{pmatrix} 
-\frac{1}{\sqrt{2}} & 0 & \frac{1}{\sqrt{2}} \\ 
\frac{1}{\sqrt{6}} & -\frac{2}{\sqrt{6}} & \frac{1}{\sqrt{6}} \\ 
\frac{1}{\sqrt{3}} & \frac{1}{\sqrt{3}} & \frac{1}{\sqrt{3}}  
\end{pmatrix} \, .
\end{equation}
Under this rotation, the overlap reduction function matrix in the AET basis becomes
\begin{equation} 
\gamma_{\alpha \beta}^{AET} \equiv \begin{pmatrix} 
\gamma_{ii}-\gamma_{ij} & 0 & 0 \\ 
0 & \gamma_{ii}-\gamma_{ij} & 0 \\ 
0 & 0 & \gamma_{ii}+2\gamma_{ij}  
\end{pmatrix} \, ,
\label{orf_AET}
\end{equation}
where the terms $\gamma_{ii}$ and $\gamma_{ij}$ are respectively the auto and cross overlap reduction function and are defined in  eqs.~\eqref{orf_t},~\eqref{orf_v},~\eqref{orf_s},  $i, j = X, Y, Z$, with $i\neq j$ and $\alpha, \beta= A, E, T$ with $\alpha = \beta$.\footnote{This holds for every polarization mode.}

\subsubsection{Chiral Overlap Reduction Function}

In eq. \eqref{h_ij} we consider the decomposition in the $\{ +, \times\}$ basis. However, we can also give an alternative decomposition to the circular polarization basis, defined by the right handed and the left handed modes, as in~\cite{Seto:2007tn, Crowder:2012ik}
\begin{align}
    e^{R}_{ij} &= \frac{e^{+}_{ij} + i e^{\times}_{ij}}{\sqrt{2}},  \quad \quad \quad  e^{L}_{ij} = \frac{e^{+}_{ij} - i e^{\times}_{ij}}{\sqrt{2}}\,.
    \label{pol}
\end{align}
The corresponding GW amplitudes will be given by
\begin{equation}
    h_{R} = \frac{h_{+} - ih_{\times}}{\sqrt{2}}\,,
    \quad \quad\quad
    h_{L} = \frac{h_{+} + ih_{\times}}{\sqrt{2}}\,,
    \label{leftampl}
\end{equation}
and the two-point correlation functions is \cite{Seto:2007tn}
\begin{align}
    \langle h_{R}(f, \hat{k}) \,  h_{R}^*(f', \hat{k'}) \rangle &= \frac{\delta(f-f')\, \delta^2(\hat{k}, \hat{k'})}{4 \pi} \left( I(f, \hat{k}) + V(f, \hat{k})\right)\,,
    \label{2pt_right}\\
    \langle h_{L}(f, \hat{k}) \,  h_{L}^*(f', \hat{k'}) \rangle &= \frac{\delta(f-f')\, \delta^2(\hat{k}, \hat{k'})}{4 \pi} \left( I(f, \hat{k}) - V(f, \hat{k})\right)\,,
    \label{2pt_left}
\end{align}
where $I(f, \hat{k})$ represents the total GW intensity, defined in terms of the energy density spectrum as $\Omega_{\rm GW}$ as $I(f) = \Omega_{\rm GW}(f)\rho_c/4 \pi^2 $ (with $\rho_c$ the critical energy density of the universe) and $V(f, \hat{k})$ is a real function that characterize the asymmetry between the amplitude of the right and left handed GW components.

The corresponding overlap reduction functions for the intensity $\gamma_{ij}^I(f)$ and the circular polarization $\gamma_{ij}^V(f)$ are given by~\cite{Seto:2007tn}
\begin{align}
    \gamma_{ij}^I(f) &= \frac{5}{8\pi} \int d^2 \Omega_{\hat{k}} \left(  F_i^{+}(f, \hat{k})   F_j^{+*}(f, \hat{k}) + F_i^{\times}(f, \hat{k})   F_j^{\times *}(f, \hat{k}) \right)e^{-i2\pi f \frac{(\vec{x}_i - \vec{x}_j)\cdot \hat{k}}{c}}\,, \label{orf_I}\\
    \gamma_{ij}^V(f) &= \frac{5i}{8\pi} \int d^2 \Omega_{\hat{k}} \left(  F_i^{+}(f, \hat{k})   F_j^{\times*}(f, \hat{k}) - F_i^{\times}(f, \hat{k})   F_j^{+ *}(f, \hat{k}) \right)e^{-i2\pi f \frac{(\vec{x}_i - \vec{x}_j)\cdot \hat{k}}{c}} \label{orf_V}\,.
\end{align}
\subsection{Angular Response}
\label{angularresponse}

In analogy with Section~\ref{ORF} and following~\cite{LISACosmologyWorkingGroup:2022kbp}, we introduce the angular overlap reduction function for different polarizations as 
\begin{align}
    \gamma_{ij}^{\ell m, \, t} &= \frac{5}{8\pi} \int d \Omega_{\hat{k}} \sum_{\lambda=+, \times} F_i^{\lambda}(f, \hat{k})   F_j^{\lambda}(f, \hat{k}) \,  e^{-2\pi i f \frac{\hat{k} \dot (\vec{x_i}- \vec{x_j})}{c}} Y^{\ell m} (\theta, \phi) \sqrt{4\pi} \label{eq:aorf_t}\,, \\
    \gamma_{ij}^{\ell m, \, v} &=\frac{5}{8\pi} \int d \Omega_{\hat{k}} \sum_{\lambda=x, y} F_i^{\lambda}(f, \hat{k})   F_j^{\lambda}(f, \hat{k}) \,  e^{-2\pi i f \frac{\hat{k} \dot (\vec{x_i}- \vec{x_j})}{c}} Y^{\ell m} (\theta, \phi) \sqrt{4\pi} \label{eq:aorf_v}\,, \\
    \gamma_{ij}^{\ell m, \, s} &=   \frac{5}{4\pi(1 + \mathcal{K})}\int d \Omega_{\hat{k}} \left(F_i^{b}(f, \hat{k})   F_j^{b}(f, \hat{k}) + \mathcal{K} F_i^{l}(f, \hat{k})   F_j^{l}(f, \hat{k}) \right) \, \nonumber \\
    &e^{-2\pi i f \frac{\hat{k} \dot (\vec{x_i}- \vec{x_j})}{c}} Y^{\ell m} (\theta, \phi)  \sqrt{4\pi} \label{eq:aorf_s}\,,\\
    \gamma_{ij}^{\ell m,\,  V}(f) &= \frac{5i}{8\pi} \int d^2 \Omega_{\hat{k}} \left(  F_i^{+}(f, \hat{k})   F_j^{\times*}(f, \hat{k}) - F_i^{\times}(f, \hat{k})   F_j^{+ *}(f, \hat{k}) \right) \nonumber \\
    &e^{-i2\pi f \frac{(\vec{x}_i - \vec{x}_j)\cdot \hat{k}}{c}} \, Y^{\ell m} (\theta, \phi)  \sqrt{4\pi} \label{eq:aorf_V}\,, 
\end{align}
where $Y_{\ell m}(\hat{k}) \equiv \sqrt{4\pi}\,\tilde{Y}^*_{\ell m}(\hat{k})$ with $\tilde{Y}^*_{\ell m}(\hat{k})$ the standard spherical harmonics \footnote{This different normalization for the spherical harmonics is justified by the fact that $\gamma_{ij}^{00, \, Tensor}=1$ for co-aligned detectors.}.

At this point, it is possible to define the angular response functions for different channels/detectors combination~\cite{LISACosmologyWorkingGroup:2022kbp}
\begin{align}
    R_{ij}^{\ell} (f) = \left( \sum_{m = -\ell}^{\ell} \left| \gamma_{ij}^{\ell m}(f) \right|^2 \right)^{1/2}\,, 
    \label{R_ell}
\end{align}
that is invariant under rotations.

It can be shown~\cite{LISACosmologyWorkingGroup:2022kbp} that, under a rigid rotation of the instrument, the angular overlap reduction functions in eqs. \eqref{eq:aorf_t}, \eqref{eq:aorf_v}, \eqref{eq:aorf_s}, \eqref{eq:aorf_V} transform as spherical harmonics. Specifically, considering LISA, if $\mathcal{R}$ is a rotation under which the position of the three satellites changes according to $\vec{x}_i \rightarrow \mathcal{R}\vec{x}_i $, we have that the angular decomposition of the response function will change as
\begin{equation}
    \gamma^{\ell m}_{\mathcal{R}_i \mathcal{R}_j } (f) = \sum_{m' = -\ell}^{\ell} \left[ D^{(\ell)}_{m m'}(\mathcal{R})\right]^* \, \gamma_{ij}^{\ell m'} (f)\,,
\end{equation}
where $D^{(\ell)}_{m m'}(\mathcal{R})$ are the elements of the Wigner D-matrix.

As shown in~\cite{LISACosmologyWorkingGroup:2022kbp}, it is useful to actually consider the angular response function in the AET basis, rather than in the XYZ basis. In such a basis, for even $\ell$ they read
\begin{equation*}
R_{AA}^{\ell } (f) =  R_{EE}^{\ell} (f) = \left[ \frac{1}{4} \sum_{m= -\ell}^{+ \ell} \left| (1 + e^{-4/3 i m \pi}) \gamma_{XX}^{\ell m} - 2 \gamma_{XY}^{\ell m } \right|^2 \right]^{1/2}\,, 
\end{equation*}
\begin{equation*}
R_{TT}^{\ell } (f) = \left[ \frac{1}{9} \sum_{m= -\ell}^{+ \ell} \left[  1+ 2\cos \left( \frac{2 m \pi}{3} \right)^2\right] \left|  \gamma_{XX}^{\ell m} + 2 \gamma_{XY}^{\ell m } \right|^2 \right]^{1/2}\,, 
\end{equation*}
\begin{equation*}
R_{AE}^{\ell } (f) = \left[ \frac{1}{3} \sum_{m= -\ell}^{+ \ell} \sin^2 \left( \frac{m \pi}{3} \right) \left| (1 + e^{2/3 i m \pi}) \gamma_{XX}^{\ell m} - 2 \gamma_{XY}^{\ell m } \right|^2 \right]^{1/2}\,, 
\end{equation*}
\begin{equation*}
R_{AT}^{\ell } (f) = R_{ET}^{\ell}(f) = \left[ \frac{2}{3} \sum_{m= -\ell}^{+ \ell} \sin^2 \left( \frac{m \pi}{3} \right) \left| (1 + e^{2/3 i m \pi}) \gamma_{XX}^{\ell m} + 2 \gamma_{XY}^{\ell m } \right|^2 \right]^{1/2}\,, 
\end{equation*}
while for odd $\ell$ 
\begin{equation*}
R_{AA}^{\ell } (f) =  R_{EE}^{\ell} (f) = R_{TT}^{\ell} (f) = 0,
\end{equation*}
\begin{equation*}
    R_{AE}^{\ell} (f) = \left[ \frac{1}{3} \sum_{m= -\ell}^{\ell} \left[ 
    1 + 2\cos \left( \frac{2 m \pi}{3} \right)\right]^2 \left|\gamma_{XY}^{\ell m}  \right|^2 \right]^{1/2},
\end{equation*}
\begin{equation*}
    R_{AT}^{\ell} (f) = R_{ET}^{\ell} (f) =  \left[ \frac{1}{3} \sum_{m= -\ell}^{\ell} \sin^2 \left( \frac{m\pi}{3} \right) \left|\gamma_{XY}^{\ell m}  \right|^2 \right]^{1/2}.
\end{equation*}
It is important to notice that once we go to higher multipoles than the monopole it is no longer true that the response is diagonal. 

\subsection{Power Law Integrated Sensitivity Curves}
\label{PLS}

The signal-to-noise ratio (SNR) for a cross correlation search for an unpolarized and isotropic SGWB is given by~\cite{Maggiore:2007ulw, Thrane:2013oya} 
\begin{equation}
    {\rm SNR} = \frac{3(H_0/h)^2}{10\pi^2} \sqrt{T} \left[ 2 \int_0^{\infty}df \,  \sum_{i}^{N_{\rm det}}\sum_{j>i}^{N_{\rm det}}\frac{\left( \gamma_{ij}(f) \, h^2\Omega_{\rm GW}(f)\right)^2}{f^6 N_i(f) N_j(f)} \right]^{1/2},
    \label{SNR}
\end{equation}
where $i, j$ are the indices running over the detector network \footnote{Note that when LISA is considered in the AET basis, the sum happens on the auto correlation channels.}, $N_{\rm det}$ is the number of detectors in the network, $T$ is the observation time (assumed to be taken the same for all the detectors in the network), $N_i(f), \, N_j(f)$ are the auto power spectral densities (PSD) for the noise of the detectors, $\gamma_{ij}$ is the overlap reduction function defined in eq.~\eqref{orf_t}, $\Omega_{\rm GW}$ is the energy density spectrum, $H_0$ is the Hubble parameter today and $h=0.7$.
So the SNR depends on multiple factors, including the number of frequency bins, the noise PSD, the detector's geometry (such as its separation and relative orientation), and most importantly, the spectral shape of the signal. Thus, the PSD alone does not provide a complete description. This means that an additional tool is needed to represent the detector's sensitivity to a SGWB signal.

Assuming that the SGWB signal can be approximated by a power law, namely 
\begin{equation}
h^2\Omega_{\rm GW} (f) = h^2\Omega_{\beta} \left( \frac{f}{f_{*}} \right)^{\beta},
\label{powerlaw}
\end{equation}
as shown in~\cite{Thrane:2013oya}, graphical methods can be used to build sensitivity curves for power-law backgrounds based on the expected SNR. From eq.~\eqref{SNR}, one can define the effective energy density spectrum for the a detector network
            \begin{equation}
                h^2\Omega_{{\rm eff}}(f) = \frac{10 \pi^2}{3(H_0/h)^2} \left[\sum_{i=1}^{N_{\rm det}} \sum_{j>i}^{N_{\rm det} }\frac{\gamma_{ij}^2(f)}{f^6\,N_i(f)N_j(f)} \right]^{-1/2}.
            \end{equation}
For a set of power law indices (i.e., $\beta = \{-40, ..., 40\}$) and some choice of pivot frequency $f_{*}$, by plugging eq.~\eqref{powerlaw} into eq.~\eqref{SNR}, one can evaluate the value of the amplitude $\Omega_{\beta}$ such that the integrated SNR has some fixed value (i.e., $\rm SNR_{th} = 1$ for ground based detectors such as LIGO Hanford, LIGO Livingston or ET\footnote{Section \ref{Results} explicitly specifies the value of $\rm SNR_{th}$ for each detector network considered.} ); namely
        \begin{equation}
            h^2\Omega_{\beta} = \frac{\rm SNR_{th}}{\sqrt{2T}} \left[ \int_{f_{min}}^{f_{max}} df \, \frac{(f/f_{*})^{2\beta}}{h^4\Omega_{\rm eff}^2(f)} \right]^{-1/2}\,.
        \end{equation}
For each pair of values for $\beta$ and $\Omega_{\beta}$, then one can plot
        \begin{equation}
            h^2\Omega_{\rm GW}(f) = h^2\Omega_{\beta} \left( \frac{f}{f_{*}} \right)^{\beta}\,.
        \end{equation}
Finally, the power law integrated sensitivity curve is evaluated, for each frequency $f$, as the maximum of $\Omega_{\rm GW}(f)$ among all the $\beta$s.

In this work, we have implemented a numerical class for computing PLSs for all types of detectors.  Complementary to this numerical methodology, an analytical study has been done in~\cite{Belgacem:2025oom}, where general features of PLSs are derived. 

\subsubsection{SGWB made of Tensor, Vector and Scalar modes}
\label{Mpol}

The discussion up to this point holds if we consider a SGWB made of tensor, vector or scalar polarization modes separately. However, in the most general scenario, a SGWB might be made of a combination of these three at the same time. In order to distinguish the different polarization modes, we need to consider a detectors network involving at least three interferometers~\cite{Nishizawa:2009bf}. The SNR expression for tensor, vector and scalar polarization modes separately can be derived \cite{Amalberti:2021kzh} and we have
\begin{equation}
\resizebox{\textwidth}{!}{$
    {\rm SNR_M} = \frac{3(H_0/h)^2}{10 \pi^2} \sqrt{2T} \left\{  \int_{0}^{\infty} df \, \frac{ \left(\Pi (f)\, h^2\Omega_{\rm GW}^M (f)\right)^2}{f^6 \left[ (\alpha_i^M(f))^2 N_i(f)N_j(f) +  (\alpha_j^M(f))^2 N_j(f)N_k(f) + (\alpha_k^M(f))^2 N_k(f)N_i(f)\right]}\right\}^{1/2},
$}
\label{SNR_M}
\end{equation}
where $M = t, v, s$ and
\begin{equation}
    \Pi(f) = \gamma_{ij}^t\left(  \gamma_{jk}^s \gamma_{ki}^v - \gamma_{ki}^s \gamma_{jk}^v\right) + \gamma_{jk}^t\left(  \gamma_{ki}^s \gamma_{ij}^v - \gamma_{ij}^s \gamma_{ki}^v\right) + \gamma_{ki}^t\left(  \gamma_{ij}^s \gamma_{jk}^v - \gamma_{jk}^s \gamma_{ij}^v\right)\,, 
\end{equation}
are the frequency dependent coefficients needed to isolate one specific polarization mode. The coefficients $\alpha_i$, for different polarizations, are given by
\begin{itemize}
    \item tensor modes 
    \begin{align}
        \alpha_i^t &= \gamma_{jk}^s \gamma_{ki}^v - \gamma_{ki}^s \gamma_{jk}^v\,,\nonumber \\
        \alpha_j^t &= \gamma_{ki}^s \gamma_{ij}^v - \gamma_{ij}^s \gamma_{ki}^v \,,\nonumber \\
        \alpha_k^t &= \gamma_{ij}^s \gamma_{jk}^v - \gamma_{jk}^s \gamma_{ij}^v\,,
    \end{align}

    \item vector modes 
    \begin{align}
        \alpha_i^v &= \gamma_{jk}^s \gamma_{ki}^t - \gamma_{ki}^s \gamma_{jk}^t\,,\nonumber \\
        \alpha_j^v &= \gamma_{ki}^s \gamma_{ij}^t - \gamma_{ij}^s \gamma_{ki}^t\,,\nonumber \\
        \alpha_k^v &= \gamma_{ij}^s \gamma_{jk}^t - \gamma_{jk}^s \gamma_{ij}^t\,,
    \end{align}

    \item scalar modes 
    \begin{align}
        \alpha_i^s &= \gamma_{jk}^t \gamma_{ki}^v - \gamma_{ki}^t \gamma_{jk}^v \,,\nonumber\\
        \alpha_j^s &= \gamma_{ki}^t \gamma_{ij}^v - \gamma_{ij}^t \gamma_{ki}^v\,,\nonumber \\
        \alpha_k^s &= \gamma_{ij}^t \gamma_{jk}^v - \gamma_{jk}^t \gamma_{ij}^v\,.
    \end{align}
\end{itemize}
From eq.~\eqref{SNR_M}, tensor, vector, and scalar polarization modes can be separated algebraically, provided that the condition $\Pi(f) \neq 0$ holds.
In a similar fashion to Section~\ref{PLS}, one can find the PLS curve, which will set, at a given $\rm SNR_{th}$ and a given observation time, what is the minimum amplitude observable for a SGWB made of tensor, vector or scalar polarization modes among a SGWB made of every possible polarization. 

\subsubsection{SGWB made of Tensor and X-polarization modes}
\label{Xpol}

We now consider a different case from the previous one, specifically we take into account a SGWB made of tensor and X-polarization modes where $X = v, s$. Namely we consider vector-tensor theories and scalar-tensor theories of gravity respectively. Even in this case we need at least three interferometers to work with to separate the polarizations. One can then obtain~\cite{Amalberti:2021kzh, Omiya:2020fvw} a SNR expression for a SGWB made of X polarization modes among a SGWB made of tensor and X-polarization mode
\begin{equation}
    {\rm SNR_X} = \frac{3(H_0/h)^2}{10 \pi^{2}} \sqrt{T} \left\{ \int_0^{\infty} df \frac{ \left[ \left( \gamma_{ij}^t (f) \gamma_{ik}^X(f) - \gamma_{ik}^t(f) \gamma_{ij}^X(f)\right) \, h^2\Omega_{\rm GW}^{X}(f) \right]^2}{f^6 \left[ (\gamma_{ij}^t(f))^2 N_i(f) N_k(f) + (\gamma_{ik}^t(f))^2 N_i (f) N_j(f)\right]} \right\}^{1/2}\,,
    \label{SNR_X}
\end{equation}
As a consistency check, if only tensor modes are present, then $X=t$, which implies that the numerator of the integrand  vanishes, which results in a null SNR across all considered GW frequency ranges. This cancellation method effectively erases the contribution of tensor modes to the SNR defined for the SGWB, allowing us to isolate the additional X-polarization modes. To analyze these modes, we also require the numerator of the integrand to be nonzero, implying that the following condition must be satisfied
\begin{equation}
    \gamma_{ij}^t (f) \gamma_{ik}^X(f) - \gamma_{ik}^t(f) \gamma_{ij}^X(f) \neq 0\,.
\end{equation}
Note that interferometer $i$ acts as the ``dominant'' detector, as it influences all ORFs in eq.~\eqref{SNR_X}.

In a similar way to Section~\ref{PLS}, one can find the PLS curve, which will set, at a given SNR threshold and a given observation time, what is the minimum amplitude observable of a SGWB made of vector or scalar polarization modes among a SGWB made of tensor and the polarization modes.

\subsection{Sensitivity to \texorpdfstring{$\ell$}{l} multipoles: Angular Power Law Integrated Sensitivity Curves}
\label{sensitivity to multipoles}

Similarly to~\cite{LISACosmologyWorkingGroup:2022kbp}, we start by considering the relative sensitivity of a pair of detectors to different $\ell$ multipoles by assuming that a single multipole primarily dominates the SGWB, and that multipoles with the same $\ell$ but different $m$ follow the same Gaussian statistics. The expected SNR can be written as
\begin{equation}
    \left< {\rm SNR} \right> = \sqrt{ \sum_{\ell} \left< {\rm SNR} \right>_{\ell}^2 }\,,
\end{equation}
where 
\begin{equation}
     \left< {\rm SNR} \right>_{\ell} = \frac{3 (H_0/h)^2}{10 \pi^2 } \sqrt{2 T \int_{0}^{\infty} df  \, \frac{h^4\Omega_{\rm GW}^2 (f)}{f^6 \, N_i(f) \, N_j(f)} C_{\ell}^{\rm GW} \left( R_{ij}^{\ell}(f) \right)^2 }\,, 
     \label{mean_SNR}
\end{equation}
with $R_{ij}^{\ell}(f)$ defined in eq.~\eqref{R_ell} and $C_{\ell}^{\rm GW}$ is the angular power spectrum value at each multipole $\ell$ \footnote{Note that the $C_{\ell}^{\rm GW}$ are normalized to the monopole, namely $C_{\ell=0}^{GW}=1$.}. From eq.~\eqref{mean_SNR} we get the effective energy density spectrum of the noise 
\begin{equation}
    h^2\Omega_{{\rm GW}, \, ij, \, n}^{\ell} (f) = \frac{10 \pi^2 f^3 }{3 (H_0/h)^2}  \frac{\sqrt{N_i (f) N_j(f)}}{R_{ij}^{\ell}(f)}\,.
\end{equation}
In the case where multiple channels are present, such as LISA and ET in its triangular configuration or networks with more than two interferometers, we can write the optimally weighted sum over the channels as
\begin{equation}
    h^2\Omega_{{\rm GW}, n}^{\ell} (f) = \left\{ \sum_{i,j} \left[ \frac{1}{h^2\Omega_{{\rm GW}, \, ij, \,  n}^{\ell}(f)} \right]^2 \right\}^{-1/2}\,.
\end{equation}
Then, the SNR can be rewritten as
\begin{equation}
    \langle {\rm SNR}\rangle_\ell = \sqrt{2T \int_0^{\infty} df \left[ \frac{\sqrt{C_\ell^{\rm GW}} h^2\Omega_{\rm GW}(f)}{h^2\Omega_{\rm GW, \, n}(f)}\right]^2}\,.
\end{equation}
At this point, it is easy to obtain the angular power law integrated sensitivity (APLS) curve in a similar fashion to Section~\ref{PLS}. The obtained sensitivity curve will assess, at a fixed observation time and SNR threshold, what is the minimum level of the SGWB monopole amplitude in order to detect anisotropies at a given multipole $\ell$.

\subsection{Pulsar Timing Array Response and Angular Response}
\label{PTA}

PTAs measure the timing residual $\Delta t_i$ induced by GWs along the line of sight between a pulsar $i$ (located in the direction $\hat{p}_i$ at distance $D_i$) and the Earth.

Following~\cite{Depta:2024ykq}, the time shift induced on a photon reaching the Earth at time $t$ reads
\begin{equation}
    \Delta t _i = \frac{1}{2} \frac{ \hat{p}^a_i \hat{p}^b_i}{1 + \hat{p}_i \cdot \hat{k}} \int_0^{D_i} h_{ab}(t(s), \vec{x}(s))\,, 
    \label{time_residual}
\end{equation}
where $s$ is a affine parameter parametrizing the geodesic connecting the pulsar to the Earth, $t(s) = t-(D_i -s)$ and $\vec{x}= (D_i -s)\hat{p}_i$. 

It can be shown that the timing residual for the pulsar $i$ is~\cite{Hazboun:2019vhv, Depta:2024ykq} 
\begin{equation}
    \Delta t _i = \int_{-\infty}^{+ \infty} df \int d\Omega_{\hat{k}}^2 \sum_{\lambda} \frac{1- e^{-2\pi i f D_i\frac{(1+ \hat{k}\cdot \vec{p_i})}{c}}}{2 \pi i f} \frac{1}{2} \frac{ \hat{p}^a_i \hat{p}^b_i}{1 + \hat{p}_i \cdot \hat{k}} e_{ab}^{\lambda}(\hat{k}) h_{\lambda}(f, \hat{k}) e^{2\pi i ft}\,.
\end{equation}
At this point, we can evaluate the correlation function of timing residuals from pulsar $i$ and pulsar $j$ observed at times $t_i$ and $t_j$ which reads 
\begin{equation}
   \langle \Delta t_i (t_i) \, \Delta t_j(t_j) \rangle = \int_{-\infty}^{+\infty} df \, \gamma_{ij} (f)\frac{S_h(f)}{24 \pi^2 f^2} e^{2\pi i f (t_i - t_j)}\,, 
\end{equation}
where $\gamma_{ij}$ is the overlap reduction function \footnote{Properly normalized such that, for tensor polarization modes, $\gamma_{ii}=1$, i.e., the overlap reduction function is equal to unity if the two pulsars are the same.}, defined as
\begin{equation}
    \gamma_{ij}(f) = \frac{3}{8 \pi} \int d \Omega_{\hat{k}}^2 \left( 1 - e^{-2\pi i fD_i \frac{(1+ \hat{p}_i\cdot \hat{k})}{c}} \right) \, \left( 1 - e^{2\pi i fD_j \frac{(1+ \hat{p}_j\cdot \hat{k})}{c}} \right)\sum_{\lambda} F_i^{\lambda} (\hat{p}_i, \hat{k}) F_j^{\lambda} (\hat{p}_j, \hat{k})\,. 
    \label{orf_pulsar}
\end{equation}
Here we have introduced
\begin{equation}
    F_i^{\lambda} (\hat{p}_i, \hat{k}) = \frac{1}{2} \frac{ \hat{p}^a_i \hat{p}^b_i}{1 + \hat{p}_i \cdot \hat{k}} e_{ab}^{\lambda}(\hat{k})\,, 
\end{equation}
which can be seen as the angular pattern function (similar to \eqref{apf} for interferometers) for pulsar timing arrays.

In analogy with Section~\ref{ORF} for ground and space-based detectors, we can define the overlap reduction function for different polarizations 
\begin{align}
    \gamma_{ij}^{t}(f) &= \frac{3}{8 \pi} \int d \Omega_{\hat{k}}^2 \left( 1 - e^{-2\pi i fD_i \frac{(1+ \hat{p}_i\cdot \hat{k})}{c}} \right) \, \left( 1 - e^{2\pi i fD_j \frac{(1+ \hat{p}_j\cdot \hat{k})}{c}} \right) \nonumber \\
    &\sum_{\lambda= +, \times} F_i^{\lambda} (\hat{p}_i, \hat{k}) F_j^{\lambda} (\hat{p}_j, \hat{k})\,,\\
    \gamma_{ij}^{v}(f) &= \frac{3}{8 \pi} \int d \Omega_{\hat{k}}^2 \left( 1 - e^{-2\pi i fD_i \frac{(1+ \hat{p}_i\cdot \hat{k})}{c}} \right) \, \left( 1 - e^{2\pi i fD_j \frac{(1+ \hat{p}_j\cdot \hat{k})}{c}} \right) \nonumber \\
    &\sum_{\lambda= x, y} F_i^{\lambda} (\hat{p}_i, \hat{k}) F_j^{\lambda} (\hat{p}_j, \hat{k}), \\
    \gamma_{ij}^{s-b}(f) &= \frac{3}{8 \pi} \int d \Omega_{\hat{k}}^2 \left( 1 - e^{-2\pi i fD_i \frac{(1+ \hat{p}_i\cdot \hat{k})}{c}} \right) \, \left( 1 - e^{2\pi i fD_j \frac{(1+ \hat{p}_j\cdot \hat{k})}{c}} \right) \nonumber \\
    & F_i^{b} (\hat{p}_i, \hat{k}) F_j^{b} (\hat{p}_j, \hat{k})\,, \\
    \gamma_{ij}^{s-l}(f) &= \frac{3}{8 \pi} \int d \Omega_{\hat{k}}^2 \left( 1 - e^{-2\pi i fD_i \frac{(1+ \hat{p}_i\cdot \hat{k})}{c}} \right) \, \left( 1 - e^{2\pi i fD_j \frac{(1+ \hat{p}_j\cdot \hat{k})}{c}} \right) \nonumber \\
    & F_i^{l} (\hat{p}_i, \hat{k}) F_j^{l} (\hat{p}_j, \hat{k})\,, \\
    \gamma_{ij}^{V}(f) &= \frac{3i}{8 \pi} \int d \Omega_{\hat{k}}^2 \left( 1 - e^{-2\pi i fD_i \frac{(1+ \hat{p}_i\cdot \hat{k})}{c}} \right) \, \left( 1 - e^{2\pi i fD_j \frac{(1+ \hat{p}_j\cdot \hat{k})}{c}} \right) \nonumber  \\
    &\left(  F_i^{+}(f, \hat{k})   F_j^{\times*}(f, \hat{k}) - F_i^{\times}(f, \hat{k})   F_j^{+ *}(f, \hat{k}) \right)\,, 
    \label{orf_pulsar_polarization}
\end{align}
where in the case of PTA, we keep the two scalar modes separate, as often done in literature as in~\cite{Chamberlin:2011ev, daSilvaAlves:2011fp, Bernardo:2022xzl}.

We can notice that, the pulsar terms $\left( 1 - e^{\pm 2\pi i fD_i \frac{(1+ \hat{p}_i\cdot \hat{k})}{c}}\right)$, $\left( 1 - e^{2\pi i fD_j \frac{(1+ \hat{p}_j\cdot \hat{k})}{c}} \right)$ could potentially be neglected, as in~\cite{Depta:2024ykq}, in the case of transverse polarization modes of GW (namely $+, \times \,$ and $b$). However, for longitudinal polarization modes this is not the case. As noticed in~\cite{Chamberlin:2011ev, daSilvaAlves:2011fp}, the frequency dependency actually has an impact in close pulsar pairs. 

For tensor modes, the $\gamma_{ij}^{Tensor}$ reduces to the Hellings-Downs correlation~\cite{Hellings:1983fr}
\begin{equation}
    \gamma_{ij}(\alpha_{ij}) = (1+ \delta_{ij}) \left[ \frac{1}{2} + \frac{3(1+ \cos(\alpha_{ij}))}{4} \left( \ln \frac{1- \cos(\alpha_{ij})}{2}  -\frac{1}{6}\right)\right],
    \label{eq:hellingsdowns}
\end{equation}
which is only a function of the angular separation between the pulsars $\alpha_{ij}$ 
Similarly, a closed form also for scalar-breathing mode has been provided in~\cite{Lee_2008, Chamberlin:2011ev} \footnote{We used a different normalization constant, in order to be consistent with our notation.}
\begin{equation}
    \gamma_{ij}^{Scalar-b}(\alpha_{ij}) = \frac{1 + \delta_{ij}}{8} \left( 3 + \cos(\alpha_{ij})\right)\,.
    \label{eq:hellingsdowns_scalars}
\end{equation}
However, despite being true that for vector modes a closed form cannot be found, we provide an expression of an approximated fitting function that interpolates the behavior of the overlap reduction function for vector modes as a function of the angular separation $\alpha_{ij}$ between the pulsars
\begin{equation}
    \gamma_{ij}^{v}(\alpha_{ij}) = A \log\left(B \alpha_{ij} \right) + C \cos\left(D\alpha_{ij} + E\right)\,,
    \label{eq:approximated_HD_vectors}
\end{equation}
where $A = -\pi, \, B=0.674, \, C= -3, \,  D=0.813, \,  E=0.2$. \footnote{As pointed out by~\cite{Romano:2016dpx}, this expression should have in principle a dependency over the frequency as well. However, the impact of frequency dependency happens only at very small $\alpha_{ij}$, as pointed out by~\cite{Chamberlin:2011ev}. This means that this function can be trusted for $\alpha_{ij}>\pi/10$.}

We define also the angular overlap reduction functions for PTA, namely
\begin{align}
    \gamma_{ij}^{\ell m, \, t}(f) &= \frac{3}{8 \pi} \int d \Omega_{\hat{k}}^2 \left( 1 - e^{-2\pi i fD_i \frac{(1+ \hat{p}_i\cdot \hat{k})}{c}} \right) \, \left( 1 - e^{2\pi i fD_j \frac{(1+ \hat{p}_j\cdot \hat{k})}{c}} \right)\,,  \nonumber\\
    & \sum_{\lambda= +, \times} F_i^{\lambda} (\hat{p}_i, \hat{k}) F_j^{\lambda} (\hat{p}_j, \hat{k})\, Y^{\ell, m}(\theta, \phi) \sqrt{4\pi} \,, \\
    \gamma_{ij}^{\ell m, \, v}(f) &= \frac{3}{8 \pi} \int d \Omega_{\hat{k}}^2 \left( 1 - e^{-2\pi i fD_i \frac{(1+ \hat{p}_i\cdot \hat{k})}{c}} \right) \, \left( 1 - e^{2\pi i fD_j \frac{(1+ \hat{p}_j\cdot \hat{k})}{c}} \right) \nonumber \\
    & \sum_{\lambda= x, y} F_i^{\lambda} (\hat{p}_i, \hat{k}) F_j^{\lambda} (\hat{p}_j, \hat{k})\, Y^{\ell, m}(\theta, \phi) \sqrt{4\pi}\,, \\
    \gamma_{ij}^{\ell m, \, s-b}(f) &= \frac{3}{8 \pi} \int d \Omega_{\hat{k}}^2 \left( 1 - e^{-2\pi i fD_i \frac{(1+ \hat{p}_i\cdot \hat{k})}{c}} \right) \, \left( 1 - e^{2\pi i fD_j \frac{(1+ \hat{p}_j\cdot \hat{k})}{c}} \right) \nonumber \\ 
    & F_i^{b} (\hat{p}_i, \hat{k}) F_j^{b} (\hat{p}_j, \hat{k}) \, Y^{\ell, m}(\theta, \phi) \sqrt{4\pi}\,,  \\
    \gamma_{ij}^{\ell m, \, s-l}(f) &= \frac{3}{8 \pi} \int d \Omega_{\hat{k}}^2 \left( 1 - e^{-2\pi i fD_i \frac{(1+ \hat{p}_i\cdot \hat{k})}{c}} \right) \, \left( 1 - e^{2\pi i fD_j \frac{(1+ \hat{p}_j\cdot \hat{k})}{c}} \right) F_i^{l} (\hat{p}_i, \hat{k}) \nonumber \\
    & F_j^{l} (\hat{p}_j, \hat{k}) \, Y^{\ell, m}(\theta, \phi) \sqrt{4\pi}\,, \\
    \gamma_{ij}^{\ell m, \, V}(f) &= \frac{3i}{8 \pi} \int d \Omega_{\hat{k}}^2 \left( 1 - e^{-2\pi i fD_i \frac{(1+ \hat{p}_i\cdot \hat{k})}{c}} \right) \, \left( 1 - e^{2\pi i fD_j \frac{(1+ \hat{p}_j\cdot \hat{k})}{c}} \right)  \nonumber \\
    &\left(  F_i^{+}(\hat{p}_i, \hat{k})   F_j^{\times*}(\hat{p}_j, \hat{k}) - F_i^{\times}(\hat{p}_i, \hat{k})   F_j^{+ *}(\hat{p}_j, \hat{k}) \right) Y^{\ell, m}(\theta, \phi) \sqrt{4\pi}\,, 
    \label{orf_pulsar_polarization_angular}
\end{align}
and, as done in eq.~\eqref{R_ell}, we can define the angular response functions by summing in quadrature over all the $m$ at a fixed $\ell$.

\subsubsection{Pulsar Timing Array Noise treatment}

In order to derive the PLS and the APLS for PTA, we need to specify the noise treatment adopted in the code. We assume that all the pulsars have identical white timing noise power spectral densities. This noise effect is intrinsic to all pulsars because of the finite SNR of match filtering process adopted in order to extract the times of arrival. This white noise contribution for a single pulsar can be written as~\cite{Thrane:2013oya, Babak:2024yhu}
\begin{equation}
    N_{\rm WN}(f) = 2 \Delta t \, \sigma^2\,,
\end{equation}
where $1/\Delta t$ is the cadence of the measurements (taken to be $20 \,\rm yr^{-1}$, assuming that all the pulsars in the catalog have the same cadence) and $\sigma$ is the uncertainty on the pulse time of arrival, i.e., the root mean square timing noise, taken to be $100 \, \rm ns$. 

We do not include effects that take into account the pulsar period and the spindown rate, which would introduce non-stationarity in the timing residuals and a general loss of sensitivity. 

Other colored noise contributions could be taken into account as in~\cite{Babak:2024yhu} and~\cite{Hazboun:2019vhv}, in which they apply Bayesian techniques in order to estimate the PTA sensitivity.

\begin{figure}[t!]
    \centering
    \includegraphics[width=0.45\linewidth]{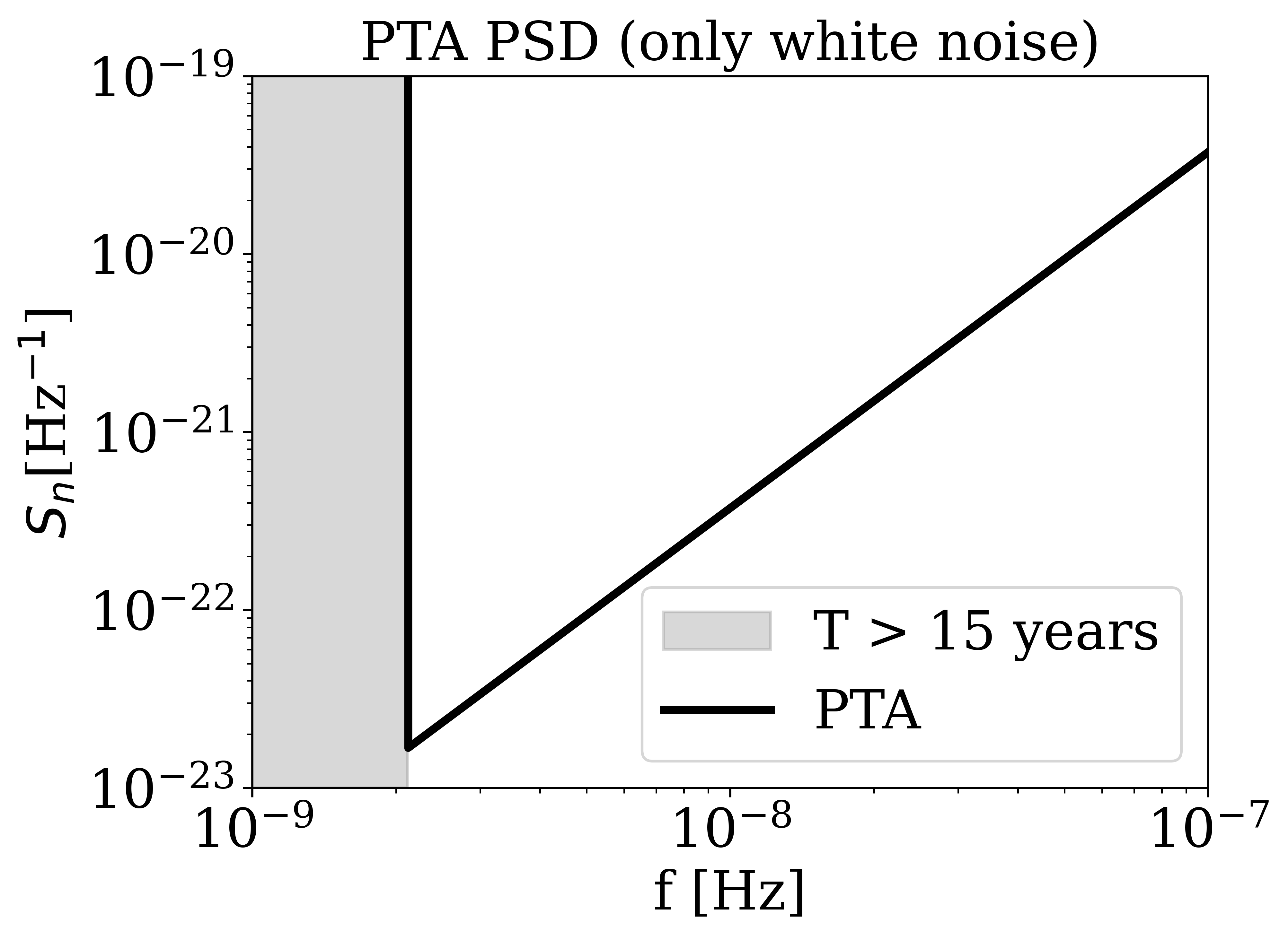}
    \caption{PSD of PTA considering $T= 15 \, \rm yrs$, $\sigma= 100 \, \rm ns$,  $\Delta t = 20 \rm \, yr^{-1}$.}
    \label{fig:Sn_PTA}
\end{figure}

Given this, the timing noise power spectral density will become
\begin{equation}
    S_{n}(f) = 12 \pi^2 f^2 \, N_{\rm WN}\,,
\end{equation}
where the minimum frequency to take into account is given by the observation time considered, namely, for $T=15 \, \rm yrs$. The corresponding PSD is shown in Figure~\ref{fig:Sn_PTA}

\section{Description of the code}
\label{code}

\texttt{GWBird} is a Python package that allows to compute the response and sensitivity of a network of detectors or pulsars to a stochastic gravitational wave background. The package is based on the formalism described in Section~\ref{Formalism}. The code is structured in six main modules: \texttt{detectors.py}, \texttt{skymap.py}, \texttt{overlap.py}, \texttt{pls.py}, \texttt{anisotropy.py} and \texttt{snr.py}.
\texttt{GWBird} is publicly available at this URL:~\github{https://github.com/ilariacaporali/GWBird} 

Hereafter we describe briefly the content of each module and the corresponding functions.
\begin{figure}[t!]
    \centering
    \includegraphics[width=1\linewidth]{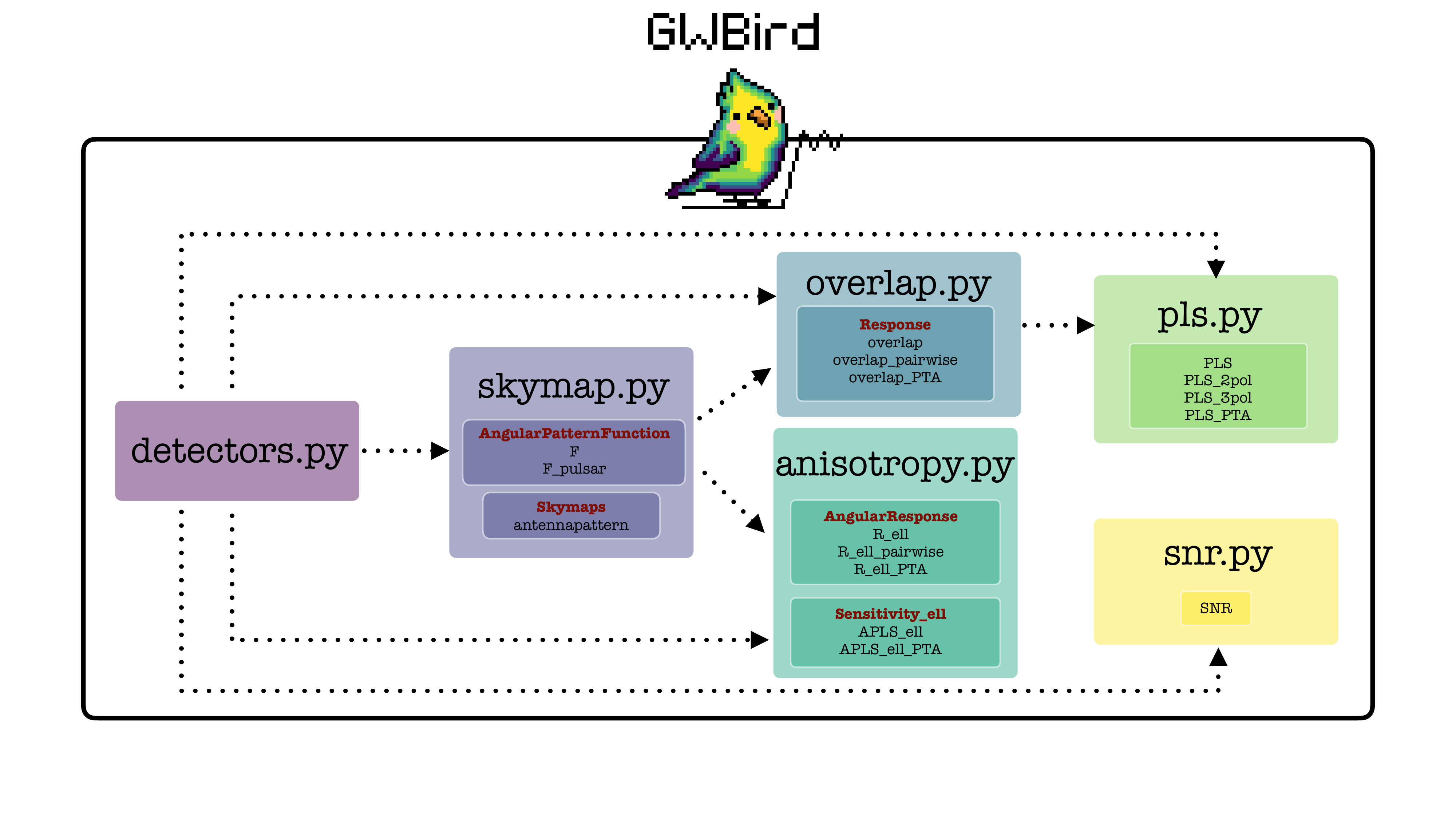}
    \caption{Concept map of \texttt{GWBird} code with the main functions the user can easily call.}
    \label{fig:GWBirdmap}
\end{figure}
\subsection{\texttt{detectors.py}}
The \texttt{detectors.py} module provides functionalities related to gravitational wave detectors and pulsar timing arrays. This module is designed to provide information about GW detectors (ground-based and space-based), perform 3D rotations for coordinate transformations, retrieve noise curves and extract pulsar data for the use in pulsar timing arrays. 

\subsection{\texttt{skymap.py}}

This module is designed for computing numerically the GW angular pattern function, both for interferometers and PTA, and provide a visualization of the Antenna Pattern. It includes several classes to model the propagation of GW signals, their polarization states, the response of detectors, and the resulting skymaps.

The module defines several classes that encapsulate different aspects of GW detection

\begin{itemize}

\item The \texttt{Basis} class defines orthonormal basis in the direction of an incoming GW signal.

\item The \texttt{PolarizationTensors} class computes GW polarization tensors in a general orthonormal basis.

\item The \texttt{TransferFunction} class models the frequency-dependent response of a GW detector based on its arm length.

\item The \texttt{AngularPatternFunction} class computes the detector and pulsar APF for different GW polarizations. It accounts for the relative orientation of the detector or pulsar and the incoming GW signal.

\item The \texttt{Skymaps} class generates skymaps of the detectors antenna pattern function.

The following code snippet represents the function call to \texttt{antennapattern}.

\begin{lstlisting}
from gwbird.skymap import Skymaps

selected_map = Skymaps.antennapattern(det1="LIGO H", 
                                      det2="LIGO L", 
                                      f=25, 
                                      psi=0, 
                                      pol="t", 
                                      nside=32
                                      )
\end{lstlisting}


\end{itemize}

\subsection{\texttt{overlap.py}}

This module provides functions to compute numerically the overlap reduction function for interferometers and pulsar timing arrays.

The \texttt{Response} class provides methods to compute the overlap reduction function for different detector pairs and pulsar timing arrays.

\begin{itemize}

\item \texttt{overlap} computes the overlap reduction function between two GW detectors.

\begin{lstlisting}
import numpy as np
from gwbird.overlap import Response  

frequencies = np.logspace(0, 3, 1000)
gamma_HL = Response.overlap(det1="LIGO H",
                            det2="LIGO L", 
                            f=frequencies,
                            pol='t', 
                            psi=0 
                            )
\end{lstlisting}

Overlap reduction function can be implemented also in the AET basis in the case of triangular detectors such as LISA

\begin{lstlisting}
import numpy as np
from gwbird.overlap import Response  

frequencies = np.logspace(-5, 0, 1000)
gamma_AA = Response.overlap(det1="LISA A",
                            det2="LISA A", 
                            f=frequencies, 
                            pol='t',
                            psi=0 
                            )
\end{lstlisting}

Custom detectors can also be provided. For details on how to implement custom detectors, please refer to the documentation accompanying this paper.
Additionally, the Einstein Telescope in the 2L configuration—with arbitrary orientation between the two detectors—can be implemented in all functions of the code. For further information, consult the documentation released with this paper.

\item \texttt{overlap\_pairwise} computes the overlap reduction function between two pulsars.

\item \texttt{overlap\_PTA}  computes the effective overlap reduction function for PTA using the NANOGrav catalog~\cite{nanograv15yr}.

\begin{lstlisting}
import numpy as np
from gwbird.overlap import Response  

frequencies = np.logspace(-9, -7, 100)
gamma_PTA = Response.overlap_PTA(f=frequencies, 
                                  pol='t',
                                  psi=0
                                  )
\end{lstlisting}

\end{itemize}

\subsection{\texttt{pls.py}}
The \texttt{pls.py} module evaluates the power law integrated sensitivity curves under different assumptions and configurations. 
This module contains the following functions
\begin{itemize}
    \item \texttt{PLS}: Evaluate the sensitivity (PLS) of a pair of detectors or a network of detectors to a SGWB signal.

    \begin{lstlisting}
    import numpy as np
    from gwbird import pls 

    frequency = np.logspace(0, 3, 1000)
    fref = 25 #Hz
    snr = 1
    Tobs = 1 #yr
    
    pls_LIGO = pls.PLS(det1='LIGO H', 
                       det2='LIGO L', 
                       f=frequency, 
                       fref=fref, 
                       pol='t', 
                       snr=snr, 
                       Tobs=Tobs, 
                       psi=0
                       )
    \end{lstlisting}

    It is possible to consider custom PSDs. For more information, check the documentation released along with this paper.
    If custom detectors want to be provided, custom PSDs need to be provided as well.

    The PLS can be also evaluated for detector networks such as LISA in the following way

    \begin{lstlisting}
    import numpy as np
    from gwbird import pls 

    frequency = np.logspace(-5, 0, 1000)
    fref = 1e-2 #Hz
    snr = 10
    Tobs = 3 #yr
    
    pls_LISA = pls.PLS(det1='LISA', 
                       det2='Network', 
                       f=frequency, 
                       fref=fref, 
                       pol='t', 
                       snr=snr, 
                       Tobs=Tobs, 
                       psi=0
                       )
    \end{lstlisting}

    \item \texttt{PLS\_2pol}: Evaluate the sensitivity (PLS) of a network of three detectors to a SGWB signal to vector or scalar contribution when one of those is present along with the tensor one.
    \begin{lstlisting}
    import numpy as np
    from gwbird import pls 

    frequency = np.logspace(0, 4, 1000)
    fref = 10 #Hz
    snr = 1
    Tobs = 1 #yr
    
    pls_v_ETCE =  pls.PLS_2pol(det1="ET X", 
                               det2="ET Y",
                               det3="CE",
                               f=frequency, 
                               fref=fref, 
                               pol='v', 
                               snr=snr, 
                               Tobs=Tobs, 
                               psi=0
                               )
    \end{lstlisting}

    \item \texttt{PLS\_3pol}: Evaluate the sensitivity (PLS) of a network of three detectors to a SGWB signal to tensor, vector or scalar contribution, when all of them are present.
    \begin{lstlisting}
    import numpy as np
    from gwbird import pls 

    frequency = np.logspace(0, 4, 1000)
    fref = 10 #Hz
    snr = 1
    Tobs = 1 #yr
    
    pls_v_ETCE =  pls.PLS_3pol(det1="ET X", 
                               det2="ET Y",
                               det3="CE",
                               f=frequency, 
                               fref=fref, 
                               pol='v', 
                               snr=snr, 
                               Tobs=Tobs, 
                               psi=0
                               )
    \end{lstlisting}
    
    \item \texttt{PLS\_PTA}: Evaluate the sensitivity (PLS) for PTA using the NANOGrav catalog~\cite{nanograv15yr}.
    \begin{lstlisting}
    import numpy as np
    from gwbird import pls 

    frequency = np.logspace(-9, -7, 100)
    snr = 1
    Tobs = 15 #yr
    
    pls_NANOGrav = pls.PLS_PTA(f=frequencies,
                                snr=snr, 
                                Tobs=Tobs,
                                pol='t', 
                                psi=0
                                )
    \end{lstlisting}
    
\end{itemize}

\subsection{\texttt{anisotropy.py}}

The \texttt{anisotropy.py} module provides functions to evaluate angular response functions and angular PLS (APLS), which are useful for assessing the feasibility of a detector network in forecasting SGWB anisotropies.  The module contains the following classes
\begin{itemize}
    \item \texttt{AngularResponse}: Class for the angular response function. The functions that can be called are
    \begin{itemize}
        \item \texttt{R\_ell} computes the angular response function between two GW detectors
        \begin{lstlisting}
        import numpy as np
        from gwbird.anisotropy import AngularResponse
        
        frequencies = np.logspace(0, 3, 1000)
        ell = 2

        R2_LIGO = AngularResponse.R_ell(ell=ell, 
                                        det1="LIGO H", 
                                        det2="LIGO L", 
                                        f=frequencies, 
                                        pol="t",
                                        psi=0
                                        )
        \end{lstlisting}

        Angular responses can be implemented also in the AET basis in the case of triangular detectors, such as LISA 

        \begin{lstlisting}
        import numpy as np
        from gwbird.anisotropy import AngularResponse
        
        frequencies = np.logspace(-5, 0, 1000)
        ell = 2

        R2_LISA_AE = AngularResponse.R_ell(ell=ell, 
                                           det1="LISA A", 
                                           det2="LISA E", 
                                           f=frequencies, 
                                           pol="t",
                                           psi=0
                                           )
        \end{lstlisting}

        Custom detector can be provided as well. For more information about the way custom detectors can be implemented, check the documentation released along with this paper.

        \item \texttt{R\_ell\_pairwise} computes the angular response function between two pulsars.
        
        \item \texttt{R\_ell\_PTA} computes the angular response function for a PTA using the NANOGrav catalog~\cite{nanograv15yr}.

        \begin{lstlisting}
        import numpy as np
        from gwbird.anisotropy import AngularResponse
        
        frequencies = np.logspace(-9, -7, 100)
        ell = 2

        R2_PTA = AngularResponse.R_ell_PTA(ell=ell, 
                                           f=frequencies, 
                                           pol="t", 
                                           psi=0
                                           )
        \end{lstlisting}

    \end{itemize}
    \item \texttt{Sensitivity\_ell}: Class to evaluate the angular sensitivity (APLS) under different assumptions and configurations. The functions that can be called are
    \begin{itemize}
        \item \texttt{APLS\_ell}: Evaluate the angular sensitivity (APLS) of a pair of detectors or a detectors network such as LISA. Namely, it gives the minimum level of the SGWB monopole in order to see a specific $C^{\ell}_{\rm GW}$ at a given multipole $\ell$ with $\langle\rm SNR_{th} \rangle_{\ell}$, considering an observation time $T$. 
        \begin{lstlisting}
        import numpy as np
        from gwbird.anisotropy import Sensitivity_ell
        
        frequencies = np.logspace(0, 3, 1000)
        ell = 2
        Cl = 1e-3
        Tobs = 1 #yr
        fref = 25 #Hz
        snr = 1

        apls2_LIGO = Sensitivity_ell.APLS_ell(det1="LIGO H", 
                                             det2="LIGO L", 
                                             ell=ell, 
                                             f=frequencies, 
                                             pol='t', 
                                             psi=0, 
                                             fref=fref, 
                                             snr=snr, 
                                             Tobs=Tobs, 
                                             Cl=Cl
                                             )
        
        \end{lstlisting}
        If custom detectors want to be provided, custom PSDs need to be provided as well.
        The  APLS can be also evaluated for detectors network, such as LISA, in the following way
        \begin{lstlisting}
        import numpy as np
        from gwbird.anisotropy import Sensitivity_ell
        
        frequencies = np.logspace(-5, 0, 1000)
        ell = 2
        Cl = 1e-3
        Tobs = 3 #yr
        fref = 1e-2 #Hz
        snr = 10

        apls2_LISA = Sensitivity_ell.APLS_ell(det1="LISA", 
                                             det2="Network", 
                                             ell=ell, 
                                             f=frequencies, 
                                             pol='t', 
                                             psi=0, 
                                             fref=fref, 
                                             snr=snr, 
                                             Tobs=Tobs, 
                                             Cl=Cl
                                             )
        
        \end{lstlisting}

        \item \texttt{APLS\_ell\_PTA} Evaluate the angular sensitivity (APLS) for a PTA using the NANOGrav catalog~\cite{nanograv15yr}
        \begin{lstlisting}
        import numpy as np
        from gwbird.anisotropy import Sensitivity_ell
        
        frequencies = np.logspace(-9, -7, 100)
        ell = 2
        Cl = 1e-3
        Tobs = 15 #yr
        snr = 1

        apls2_PTA = Sensitivity_ell.APLS_ell_PTA(ell=ell,
                                                f=frequencies,
                                                snr=snr, 
                                                Tobs=Tobs, 
                                                Cl=Cl, 
                                                pol="t", 
                                                psi=0
                                                )
        
        \end{lstlisting}

    \end{itemize}
    
\end{itemize}

\subsection{\texttt{snr.py}}

The \texttt{snr.py} module provides a function to evaluate the SNR of a SGWB signal for a network of interferometers.

\begin{itemize}
    \item \texttt{snr} function allows the computation of the SNR for a SGWB signal observed by one or more pairs of detectors, as well as a PTA (considering the NANOGrav catalog~\cite{nanograv15yr}). The calculation takes into account the observation time $T$, the frequency range and a SGWB spectral model that can be defined either through parameters (if one wants to consider a single power-law spectrum) or a custom function. The function supports different types of detectors, including space-based, pulsar timing arrays, and ground-based interferometers.

    \begin{lstlisting}
    import numpy as np
    from gwbird import snr
    
    frequencies = np.logspace(0, 3, 1000)
    Tobs = 1 #yr
    det_list = ['LIGO H', 'LIGO L']
    logA_gw = -10
    fref=25
    n_gw = 0
    gw_params=[logA_gw, n_gw, fref]
    
    def Omega_GW(f):
        logA_gw, n_gw, fref = gw_params
        return 10**logA_gw * (f / fref) ** n_gw
    
    
    snr_LIGO = snr.SNR(Tobs=Tobs, 
                       f=frequencies, 
                       gw_params = None,
                       detectors_list=det_list, 
                       pol="t", 
                       psi=0, 
                       shift_angle=None, 
                       gw_spectrum_func=Omega_GW
                       )
    \end{lstlisting}

\end{itemize}

\section{Results}\label{Results}

In this section, we present all the results obtained with \texttt{GWBird}. In the code repository we provide also an example notebook to reproduce all the results that are shown in this section.

\subsection{Overlap Reduction Function}

\subsubsection{LIGO}

\begin{figure}[!ht]
    \centering
    \includegraphics[scale=0.45]{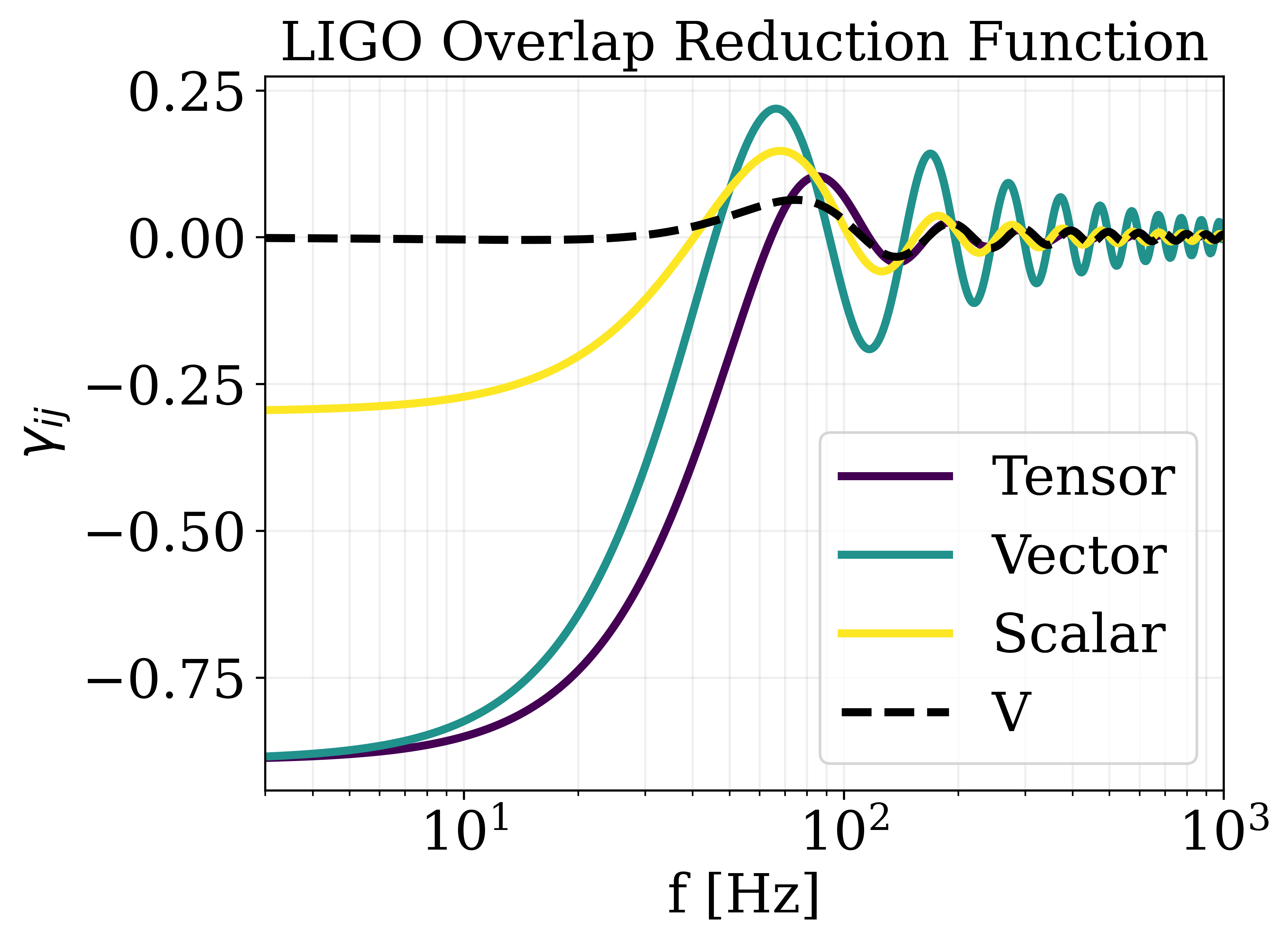}
    \caption{Overlap reduction functions for LIGO Hanford - LIGO Livingston network. Purple solid line: tensor modes. Turquoise solid line: vector modes. Yellow solid line: scalar modes. Black dashed line: circular polarization.}
    \label{fig:LIGOorf}
\end{figure}

In Figure~\ref{fig:LIGOorf} we plot the overlap reduction function for different polarization modes in the case of LIGO Hanford - LIGO Livingston. The overlap reduction functions start to oscillate and decay rapidly above $f \sim 64 \, \rm Hz$.

These curves are in agreement with~\cite{Allen:1997ad,Thrane:2013oya,Romano:2016dpx} for the tensor modes, with~\cite{Nishizawa:2009bf} for vector modes, with~\cite{Callister:2017ocg} for scalar modes \footnote{Note that here, with respect to~\cite{Nishizawa:2009bf}, the amplitude of the overlap reduction function for scalar modes is $1/3$ of their result. This comes from the multiplicative factor $\xi$ as commented in Section \ref{ORF}.} and with~\cite{Crowder:2012ik} for the circular polarization. 

\subsubsection{Einstein Telescope}

ET is a third-generation detector that is expected to be built in the next decade~\cite{Maggiore:2019uih}. Its location and design are still under discussion~\cite{Branchesi:2023mws, Abac:2025saz}. Given this, we propose two different case studies that take into account different shapes: a triangular configuration, which we assumed to be placed in Sardinia in the Sos Enattos site, and a two L-shaped detectors configuration with one detector located in Sardinia (Sos Enattos site) and another one in the Netherlands (Euregio Meuse-Rhine site). In this latter case, we consider both the so called ``aligned'' configuration, where the two detectors are parallel, i.e., $\alpha=0^{\circ}$ \footnote{$\alpha$ is the shift angle between the two detectors, namely the separation angle between the first arm of the two detectors if they where in the same location.} and the ``misaligned'' one, where the two detectors are rotated by $\alpha=45^{\circ}$.

\begin{figure}[t!]
    \centering
    \includegraphics[scale=0.45]{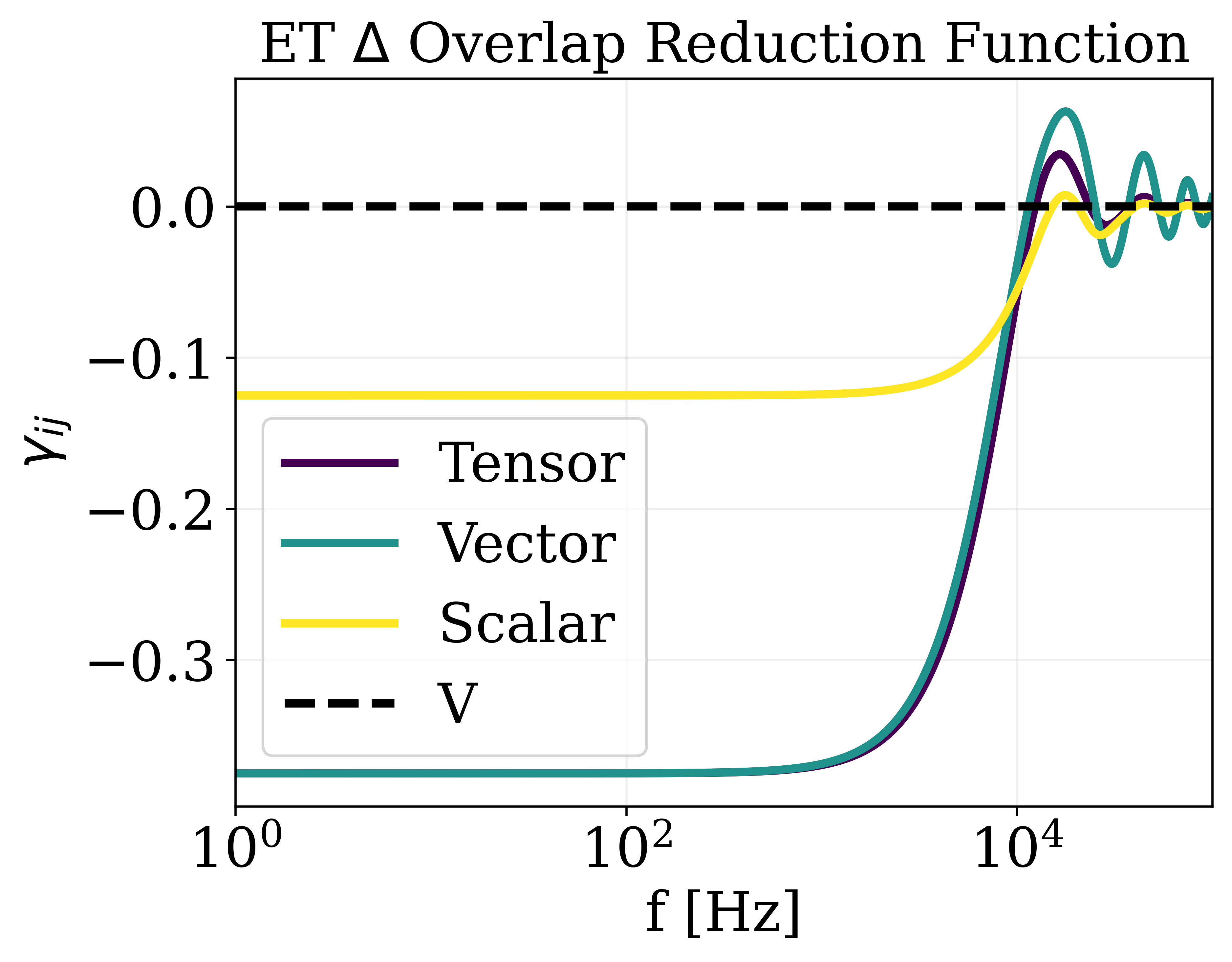}
    \includegraphics[scale=0.45]{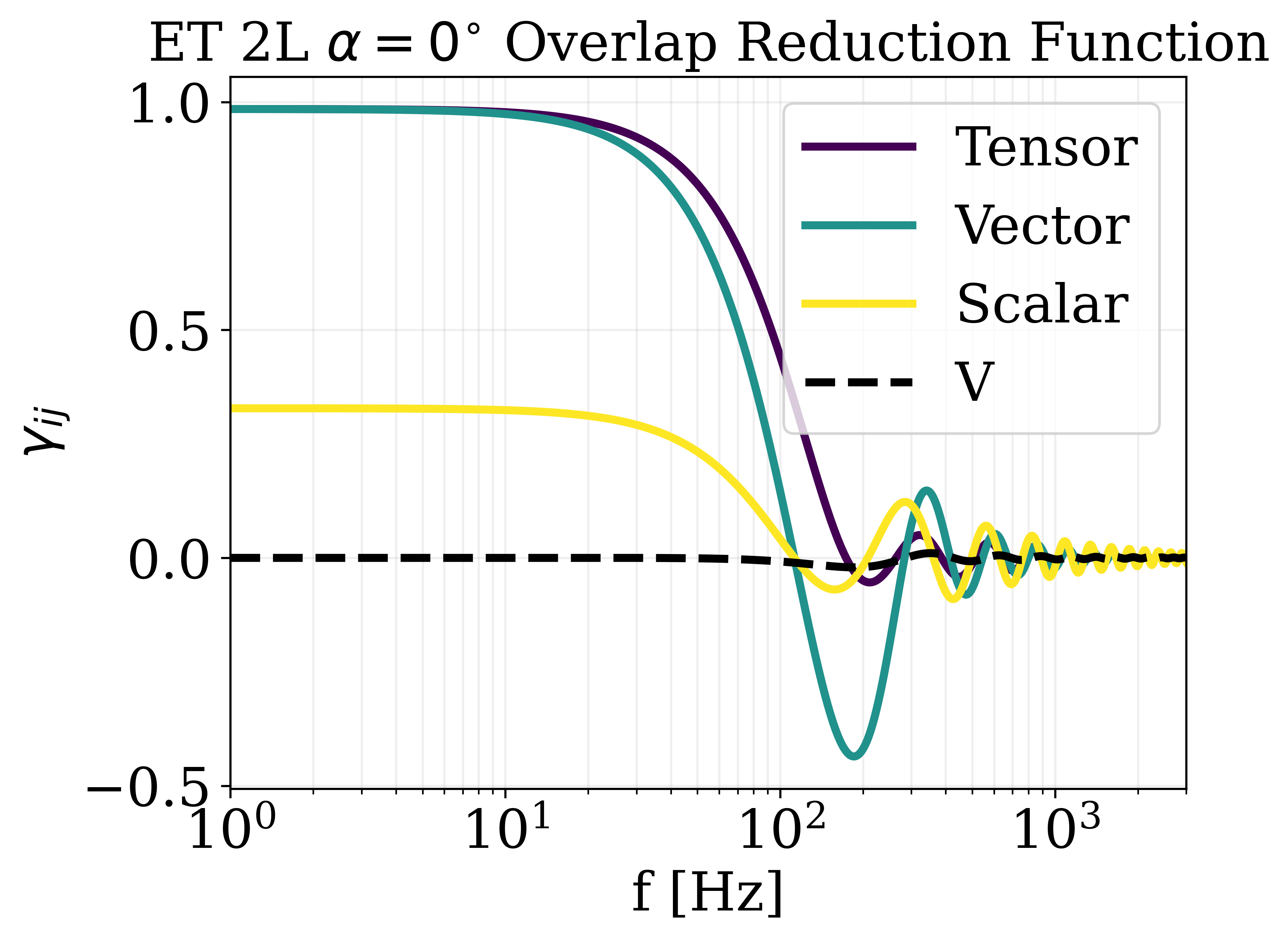}
    \includegraphics[scale=0.45]{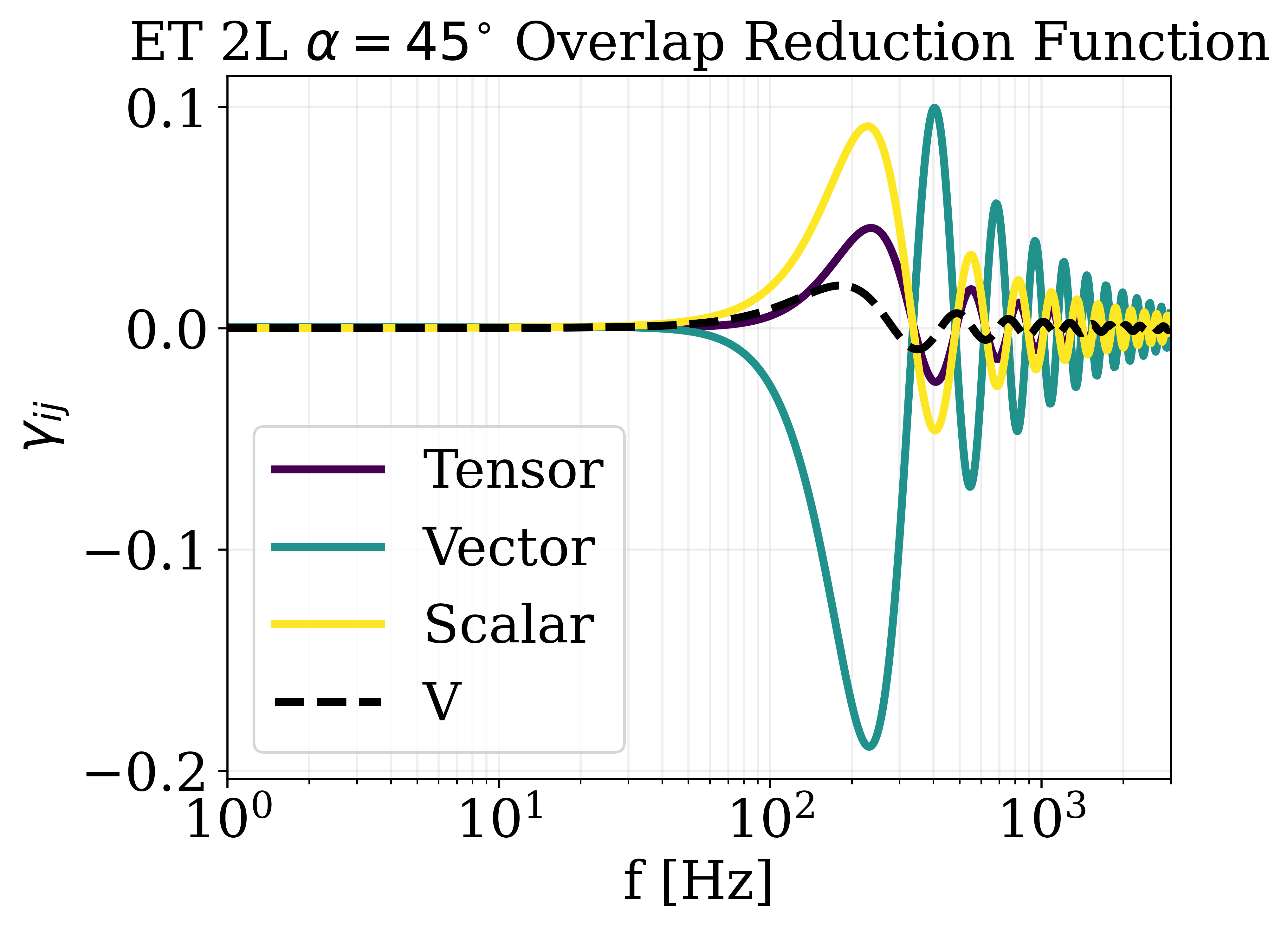}
    \caption{Overlap reduction function for ET in a triangular configuration (upper left), in the 2L shaped one aligned (bottom center), in the 2L shaped one misaligned (upper right). Purple solid lines: tensor modes. Turquoise solid lines: vector modes. Yellow solid lines: scalar modes. Black dashed lines: circular polarization. }
    \label{fig:ETorf}
\end{figure}

We show the overlap reduction function for the triangular shape configuration in Figure~\ref{fig:ETorf} (left panel) considering the X and Y channels. As it can be seen, vector modes are enhanced at higher frequencies. However, over most of the frequency band to which ET is sensitive, they nearly coincide with the tensor ones. The scalar modes, instead, are in the usual $1/3$ ratio with the other two. The stability in the high frequency part of the overlap reduction function is due to the fact that the two detectors are quite close to each other ($|\Delta \vec{X}| \sim 10 \rm{km}$). This is a common feature for all the polarization modes. The results are in agreement with~\cite{Amalberti:2021kzh} for tensor and vector modes, while for the scalar ones there is the $1/3$ factor of difference that comes from including the $\xi$ factor inside the definition of overlap reduction function. For the circular polarization mode, they are in agreement with~\cite{Abac:2025saz} \footnote{Although in~\cite{Abac:2025saz} the ET triangular configuration is considered at a different location than ours, we still find agreement. This is because, when evaluating the overlap reduction function, only the relative distance and orientation between the detectors matter.}. In this latter case, the overlap reduction function is of the order $10^{-5}$.

 We show our result for the 2L configuration in Figure~\ref{fig:ETorf} (central and right panel) regarding the overlap reduction function. We address how the relative orientation of the two arms affects the sensitivity to the background. If we consider them to be parallel ($\alpha = 0^{\circ}$), we see that the overlap is maximized in the lower frequency range ($\gamma_{ij}\sim 1$ up to $30 \, \rm{Hz}$) for tensor, vector and scalar modes. When the detectors are shifted by $\alpha = 45 ^{\circ}$, the overlap at low frequencies is effectively null ($\gamma_{ij} \sim 10^{-3}$ below $ 100 \, \rm{Hz}$) for all polarization modes.  Consequently, in this rotated configuration the two detectors will be less sensitive to a SGWB in every polarization channel. The results for tensor modes are in agreement with~\cite{Branchesi:2023mws} for the tensor modes.


\subsubsection{LISA}
\label{sec:overlapLISA}

The Laser Interferometer Space Antenna (LISA) is a space-based gravitational wave observatory developed by ESA (European Space Agency) in collaboration with NASA (National Aeronautics and Space Administration)~\cite{LISA:2024hlh}. It will consist of three spacecraft forming an equilateral triangular interferometer with 2.5 million km-long arms, orbiting the Sun in a heliocentric formation. LISA is designed to detect low-frequency GWs, which are emitted by sources such as merging massive black holes~\cite{LISA:2022yao}, compact binaries~\cite{Buscicchio:2024asl, Pozzoli:2024wfe}, and possible signals from the early universe~\cite{LISACosmologyWorkingGroup:2022jok, Caprini:2019egz, Caprini:2024hue, LISACosmologyWorkingGroup:2024hsc}. 

\begin{figure}[t!] 
    \centering
    \includegraphics[scale=0.4]{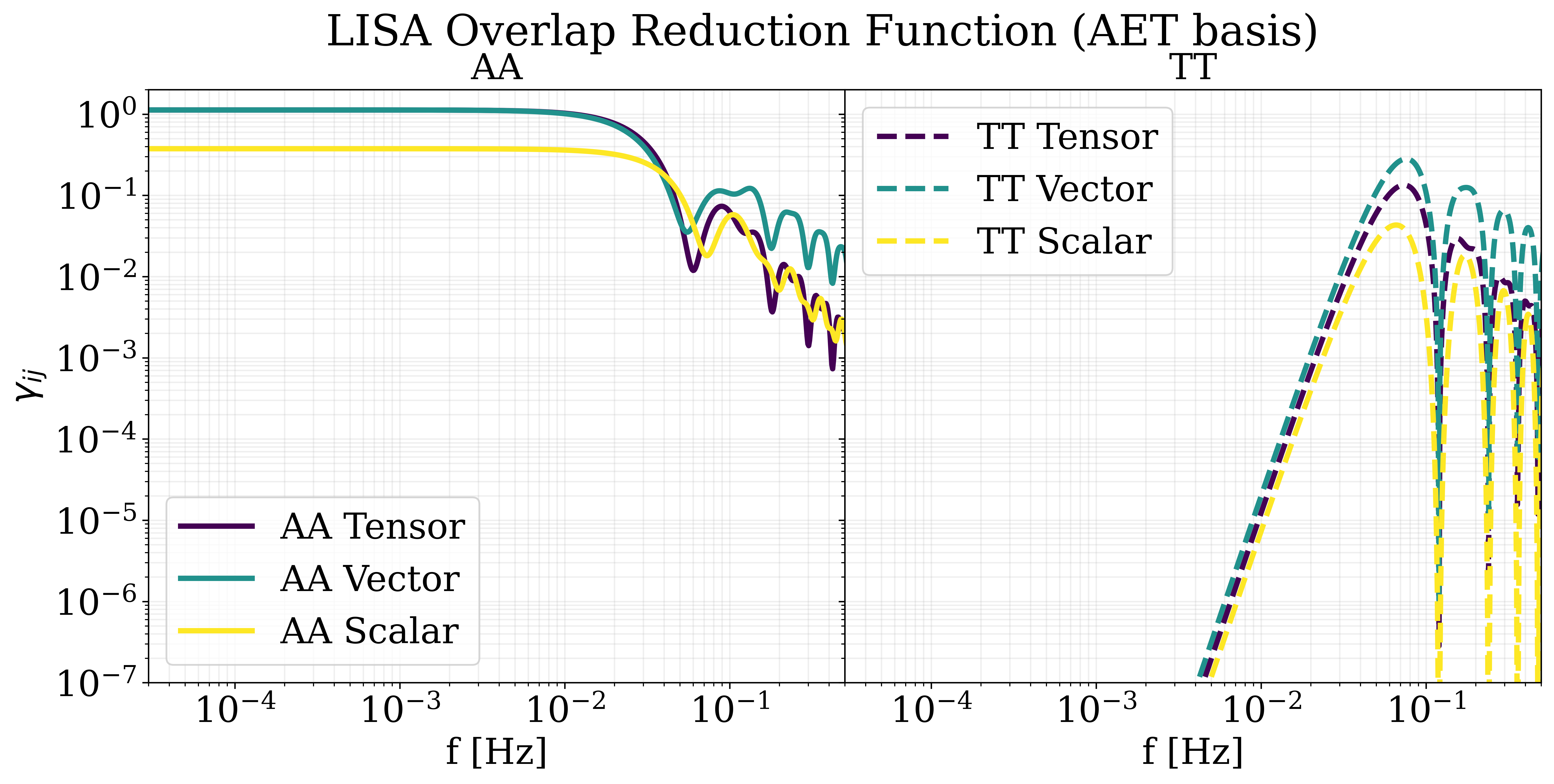}
    \caption{LISA response function (in the AET basis) for channels AA (left, solid lines) and TT (right, dashed lines). Purple lines: tensor modes. Turquoise lines: vector modes. Yellow lines: scalar modes.}
    \label{fig:LISAorf_AET}
\end{figure}

We show the overlap reduction function for the different polarization modes in the AET basis (eq.~\eqref{orf_AET}) in Figure~\ref{fig:LISAorf_AET}. As it can be seen, at lower frequencies all the polarization modes show the same behavior, despite the 1/3 reduction factor between scalar and tensor (or vector) modes (see Section \ref{ORF} for the normalization adopted), while at higher frequencies the vector modes have a slightly higher response with respect to the other two polarization modes. This happens both for the AA (and EE) channel and for the TT channel. The oscillations which happen at higher frequencies are given by the transfer function in eq.~\eqref{transfer_eq}.
Circular polarization mode here is not shown for representative purposes, as it has an amplitude below $10^{-16}$, making LISA insensitive to a chiral isotropic SGWB. The situation changes when  the motion of the solar system with respect to the cosmic reference frame is taken into account. In this case planar, triangular detectors become sensitive to a circular polarization~\cite{Domcke:2019zls}.

\subsubsection{PTA}
\label{sec:PTA_orf}

\begin{figure}[!ht]     \centering
    \includegraphics[scale=0.27]{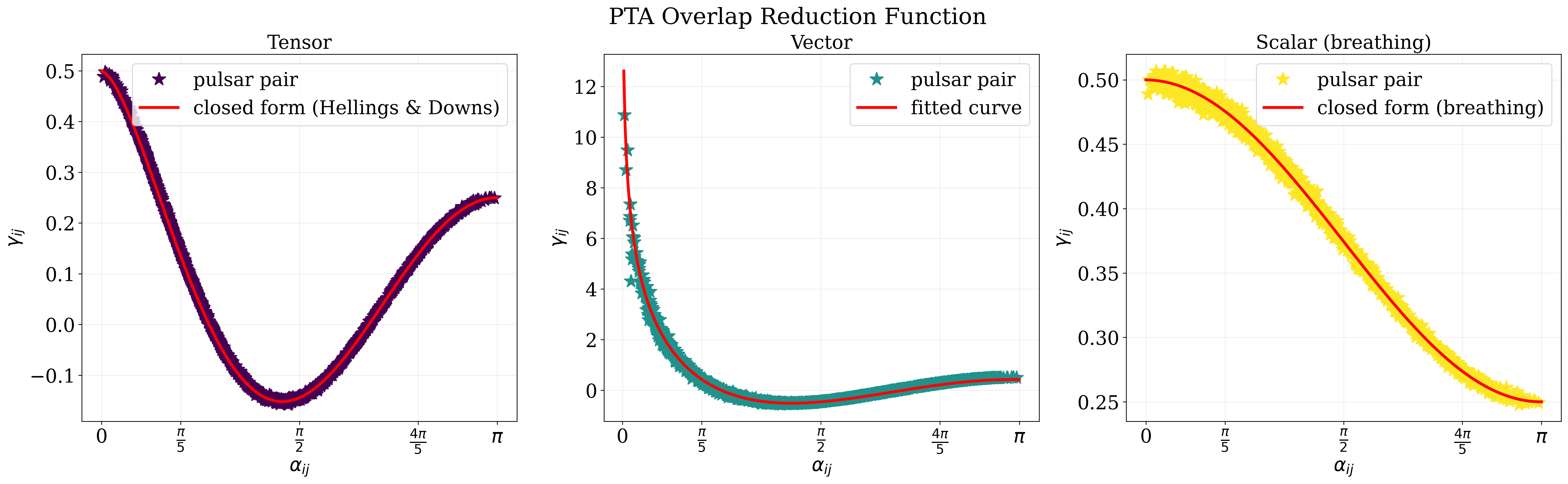}
    \caption{Overlap reduction function for the pair of pulsars in the PTA catalog as a function of their angular separation. Left: Tensor modes (purple stars) compared with the Hellings–Downs correlation curve (solid red line) from eq.~\ref{eq:hellingsdowns} (~\cite{Hellings:1983fr}). Center: vector modes (turquoise stars) compared with the approximate fit (solid red line) from eq.~\ref{eq:approximated_HD_vectors}. Right: scalar modes (only breathing polarization) (yellow stars) compared with the analytic curve (solid red line) from eq.~\ref{eq:hellingsdowns_scalars}~\cite{Lee_2008, Chamberlin:2011ev}.}
    \label{fig:HD_PTA}
\end{figure}

In Figure~\ref{fig:HD_PTA} we  show the overlap reduction function for the pair of pulsars in a PTA catalog\footnote{https://zenodo.org/records/14773896}~\cite{nanograv15yr} as a function of their angular separation at the fixed frequency of $f=10^{-8}\, \rm{Hz}$ \footnote{Note that for all the transverse polarization modes, namely tensors and breathing scalar, the overlap reduction functions can be seen as frequency independent, since the pulsars terms can be neglected~\cite{Mingarelli:2013dsa, Gair:2014rwa, Babak:2024yhu, Depta:2024ykq}. For vector polarization modes this is not the case in general. Close-by pulsars may experience frequency dependency since the pulsar terms have an impact~\cite{Chamberlin:2011ev, Lee_2008}. However, for pulsars which are separated by $\alpha_{ij}> \pi/10$ the behavior we show can be trusted. }. We see that for tensor modes the Hellings-Downs curve~\cite{Hellings:1983fr} (eq.~\eqref{eq:hellingsdowns}) is recovered and similarly for the scalar modes~\cite{Chamberlin:2011ev} (eq.~\eqref{eq:hellingsdowns_scalars}) \footnote{We specify here that in the case of scalar modes only the breathing mode is considered.}. In the case of vector modes instead, we can see that the pulsar pairs match the fitting function in eq.~\eqref{eq:approximated_HD_vectors}. Circular polarization here is not shown, since it is compatible with zero, as expected in literature~\cite{Kato:2015bye, Sato-Polito:2021efu, Cruz:2024esk}.
\subsection{Power Law Integrated Sensitivity Curves}
\subsubsection{LIGO-Virgo}

As mentioned in Section~\ref{PLS}, the sensitivity of an interferometer network to the background can be visually represented using the Power Law Integrated sensitivity curve (PLS)~\cite{Thrane:2013oya, Belgacem:2025oom}. This curve determines the minimum amplitude the SGWB signal must have at a given frequency to be detected, given a specific observation time $T$ and an SNR threshold $\rm SNR_{th}$.

\begin{figure}[t!]
    \centering
    \includegraphics[scale=0.45]{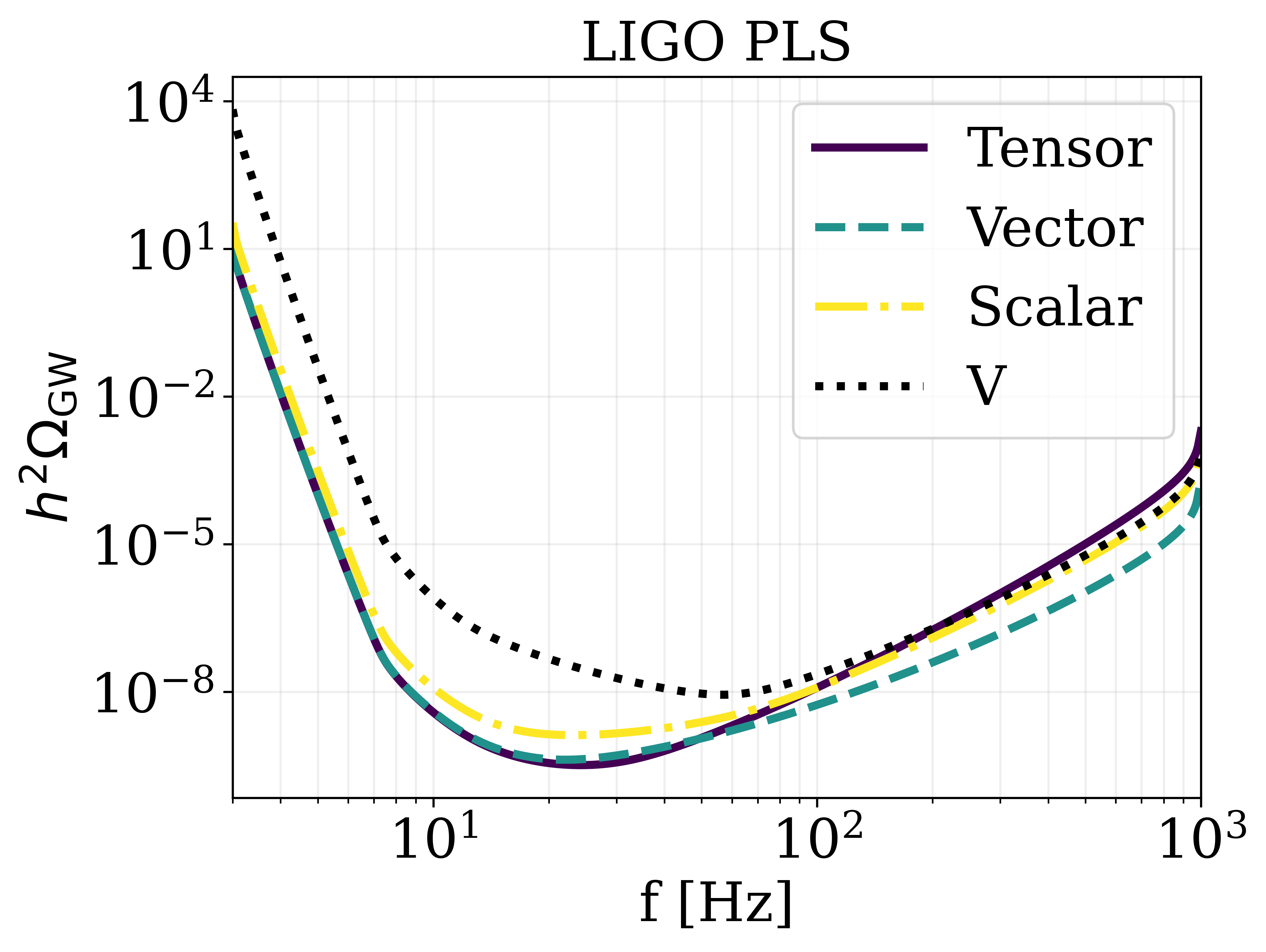}
    \caption{Power law integrated sensitivity curve for the two LIGO detectors (chosen as a reference example for 2G detectors). We choose an ${\rm SNR}_{ \rm th} = 1 $ and $T=1 {\rm yr}$. Purple solid line: tensor modes. Turquoise dashed line: vector modes. Yellow dot dashed line: scalar modes. Black dotted line: circular polarization mode.}
    \label{fig:LIGOpls}
\end{figure}

We start considering the two LIGO detectors at design sensitivity \footnote{https://dcc.ligo.org/LIGO-T1500293/public} and in Figure~\ref{fig:LIGOpls} we plot the PLS for the different polarization modes.
It can be noticed that, at low frequencies, the PLS is similar for tensor and vector polarization modes. However, at higher frequencies, the PLS for the vector modes shows greater sensitivity. This is because the overlap reduction function for tensor modes becomes smaller than that for the vector modes at higher frequencies, as illustrated in Figure~\ref{fig:LIGOorf}. Despite this, the minimum sensitivity is reached by tensor modes, as it can be seen in Table~\ref{tab:plsminLIGO}.
Circular polarization mode, instead, is almost 2 order of magnitude less sensitive than the other polarizations.
\begin{table}[!ht]
    \centering
    \begin{tabular}{ccccc}
    \multicolumn{5}{c}{LIGO} \\
    \hline
    \hline
                         & Tensor
                         & Vector & Scalar & V \\
                         \hline
        $h^2\Omega_{\rm GW}$ & $3.24 \times 10^{-10}$ & $4.17 \times 10^{-10}$ & $1.32\times 10^{-9}$ & $8.73 \times 10^{-9}$  \\
                          & {\scriptsize$@ \, 24 \, \rm Hz$} & {\scriptsize$@ \,22 \, \rm Hz$} & {\scriptsize$@ \,23 \, \rm Hz$} & {\scriptsize $@57 \, \rm Hz$}  \\
    \hline
    \end{tabular}
    \caption{Minimum detectable energy density spectrum (${\rm SNR_{th}=1}$ and $T=1{\rm yr}$) for a SGWB made of tensor, vector and scalar polarization modes separately.}
    \label{tab:plsminLIGO}
\end{table}
We examine the case where the SGWB consists of a combination of tensor, vector, and scalar polarization modes for a network made of the two LIGO detectors and Virgo.

\begin{figure}[t!]
    \centering
    \includegraphics[scale=0.45]{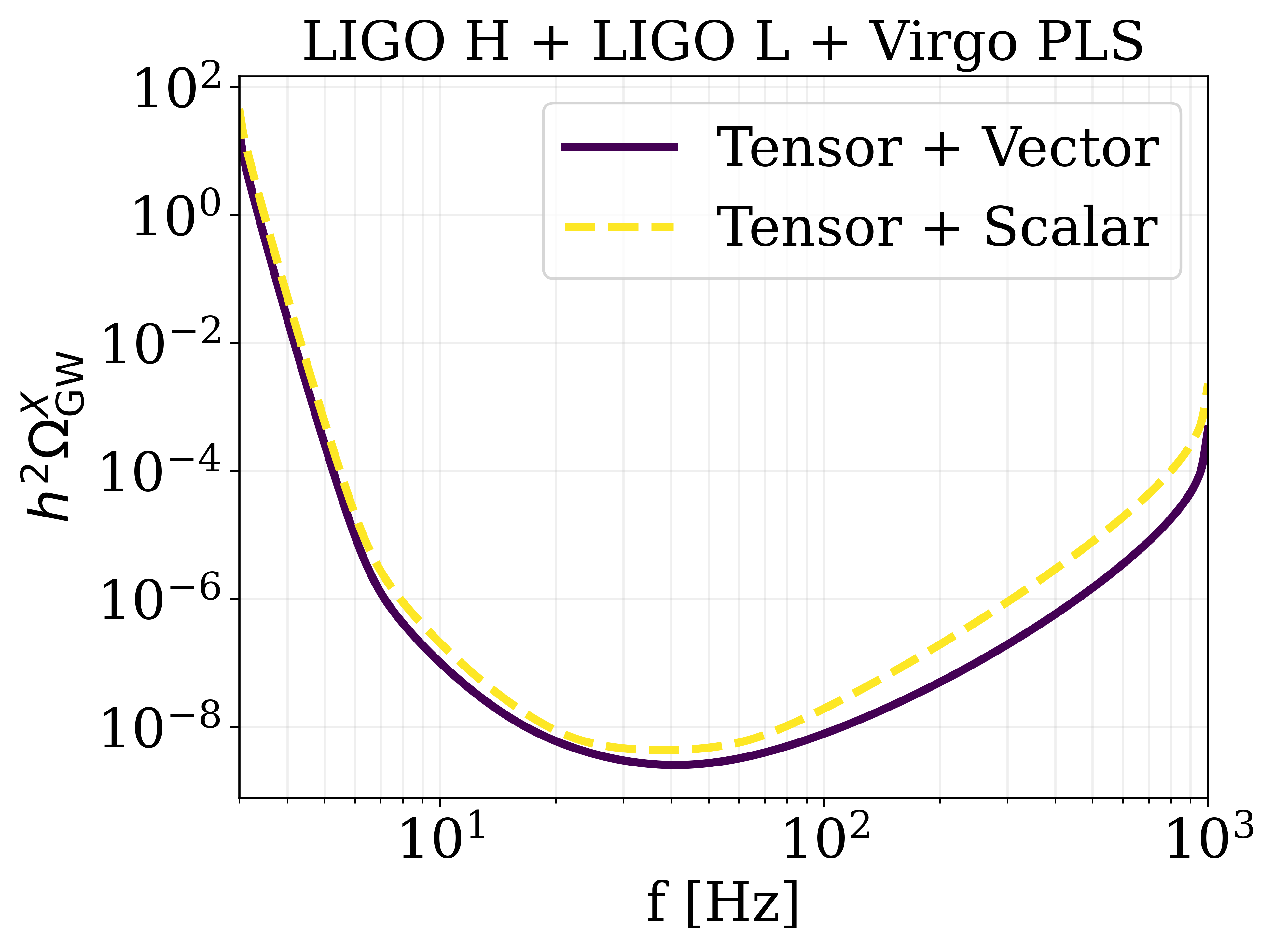}
    \includegraphics[scale=0.45]{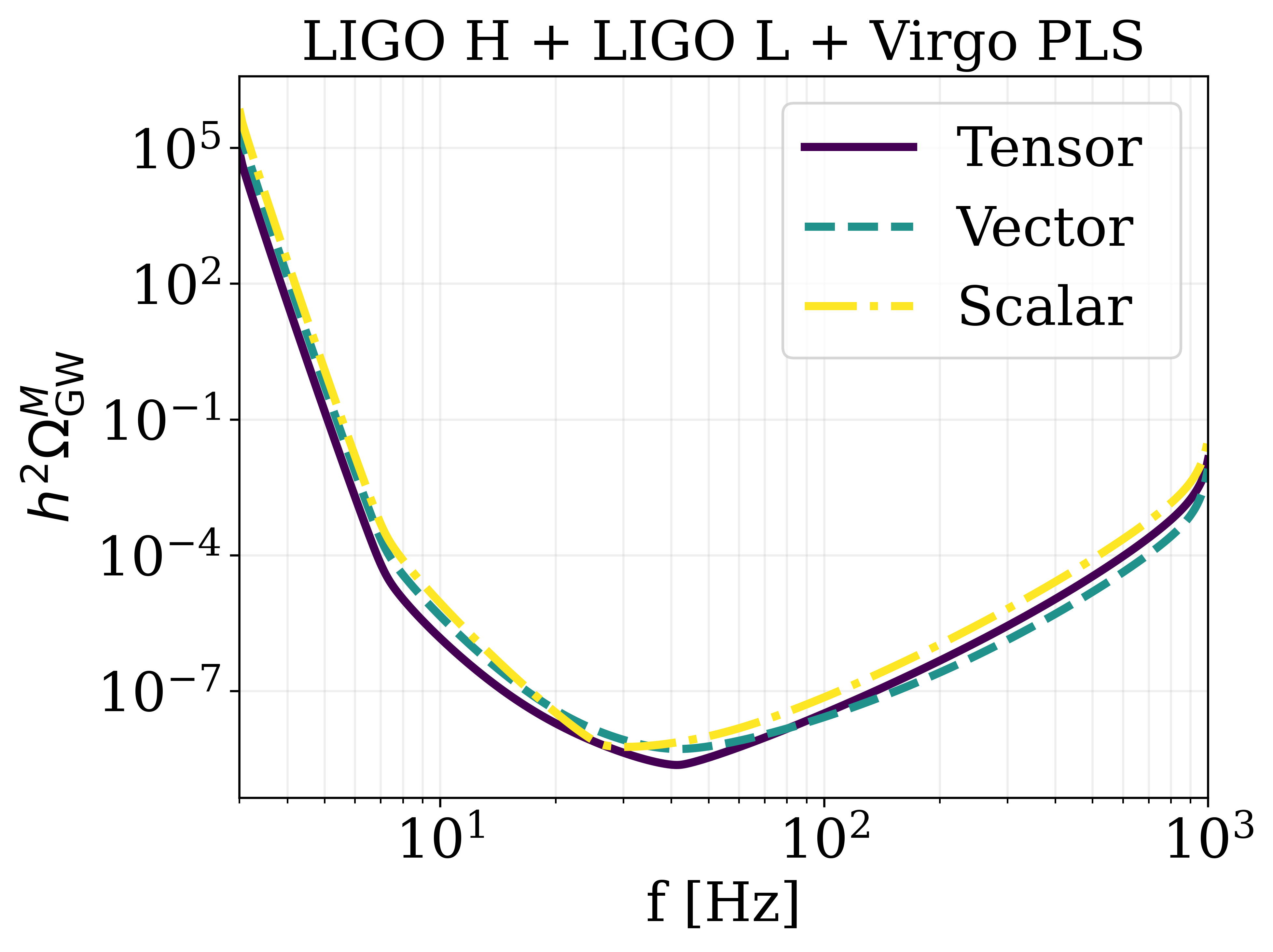}
    \label{fig:plsLIGOVirgoextrapol}
    \caption{Left: a SGWB made of tensor and an X polarization modes, with ${X = v, s}$. The solid purple line (yellow dashed line) represents the PLS for a SGWB made of vectors (scalars). Right: SGWB made of tensor, vector and scalar polarization modes, with with ${M = t,v, s}$. The purple solid line (turquoise dashed line - yellow dot dashed line) represents the PLS for a SGWB made of tensors (vectors - scalars). All the curves are obtained for ${\rm SNR_{\rm th}=1}$ and ${T = 1\rm \, yr}$. }
\end{figure}

First, we consider the scenario in which we aim to detect an X polarization mode (vector or scalar) other than the tensor one. By generating a PLS for this case, we can determine the minimum amplitude that the SGWB in the X polarization must have at a given frequency to be detectable by the detector. For the LIGO-Virgo network, we show our result in the left panel of Figure~\ref{fig:plsLIGOVirgoextrapol} and in Table~\ref{tab:plsminLIGOVirgo2pol}, where LIGO Hanford was chosen as the ``dominant'' detector~\cite{Amalberti:2021kzh}. We see in this latter case that the network is slightly more sensitive to vector polarization modes rather than the scalar ones. This can be easily interpreted by considering that the overlap reduction function for vector modes is overall a factor 3 higher than the scalar one.
\begin{table}[!ht]
    \centering
    \begin{tabular}{lc}
    \multicolumn{2}{c}{LIGO + Virgo} \\
    \hline\hline
    $h^2\Omega_{\rm GW}^v$ (Tensor + Vector) 
        & $2.53 \times 10^{-9}$ {\scriptsize@\,40.8\,Hz} \\
    $h^2\Omega_{\rm GW}^s$ (Tensor + Scalar) 
        & $4.29 \times 10^{-9}$ {\scriptsize@\,37.9\,Hz} \\
    \hline
    \end{tabular}
    \caption{Detectable energy density spectrum ($\mathrm{SNR_{th}}=1$ and $T=1\,\mathrm{yr}$) for a SGWB composed of tensor–vector modes and tensor–scalar modes separately.}
    \label{tab:plsminLIGOVirgo2pol}
\end{table}
Then, we assess the detector's sensitivity to each individual polarization component (tensor, vector, or scalar) within an SGWB that contains all polarization modes. For the LIGO-Virgo network, our results—shown in the right panel of Figure~\ref{fig:plsLIGOVirgoextrapol} and in Table~\ref{tab:plsminLIGOVirgo3pol}—demonstrate that the network is most sensitive to the tensor than to vector and scalar components of the SGWB, when all of them are present.
\begin{table}[!ht]
    \centering
    \begin{tabular}{lc}
    \multicolumn{2}{c}{LIGO + Virgo} \\
    \hline\hline
    $h^2\Omega_{\rm GW}^{t}$ 
        & $2.31 \times 10^{-9}$ {\scriptsize @\,41.3\,Hz} \\
    $h^2\Omega_{\rm GW}^{ v}$ 
        & $5.21 \times 10^{-9}$ {\scriptsize @\,41.8\,Hz} \\
    $h^2\Omega_{\rm GW}^{s}$ 
        & $5.79 \times 10^{-9}$ {\scriptsize @\,30.4\,Hz} \\
    \hline
    \end{tabular}
    \caption{Minimum detectable energy density spectrum ($\mathrm{SNR_{th}}=1$ and $  T=1\,\mathrm{yr}$) for a SGWB composed of tensor, vector, and scalar polarization modes.}
    \label{tab:plsminLIGOVirgo3pol}
\end{table}
\subsubsection{Einstein Telescope}
\label{sec:ETpls}

\begin{figure}[t!]
    \centering
    \includegraphics[scale=0.45]{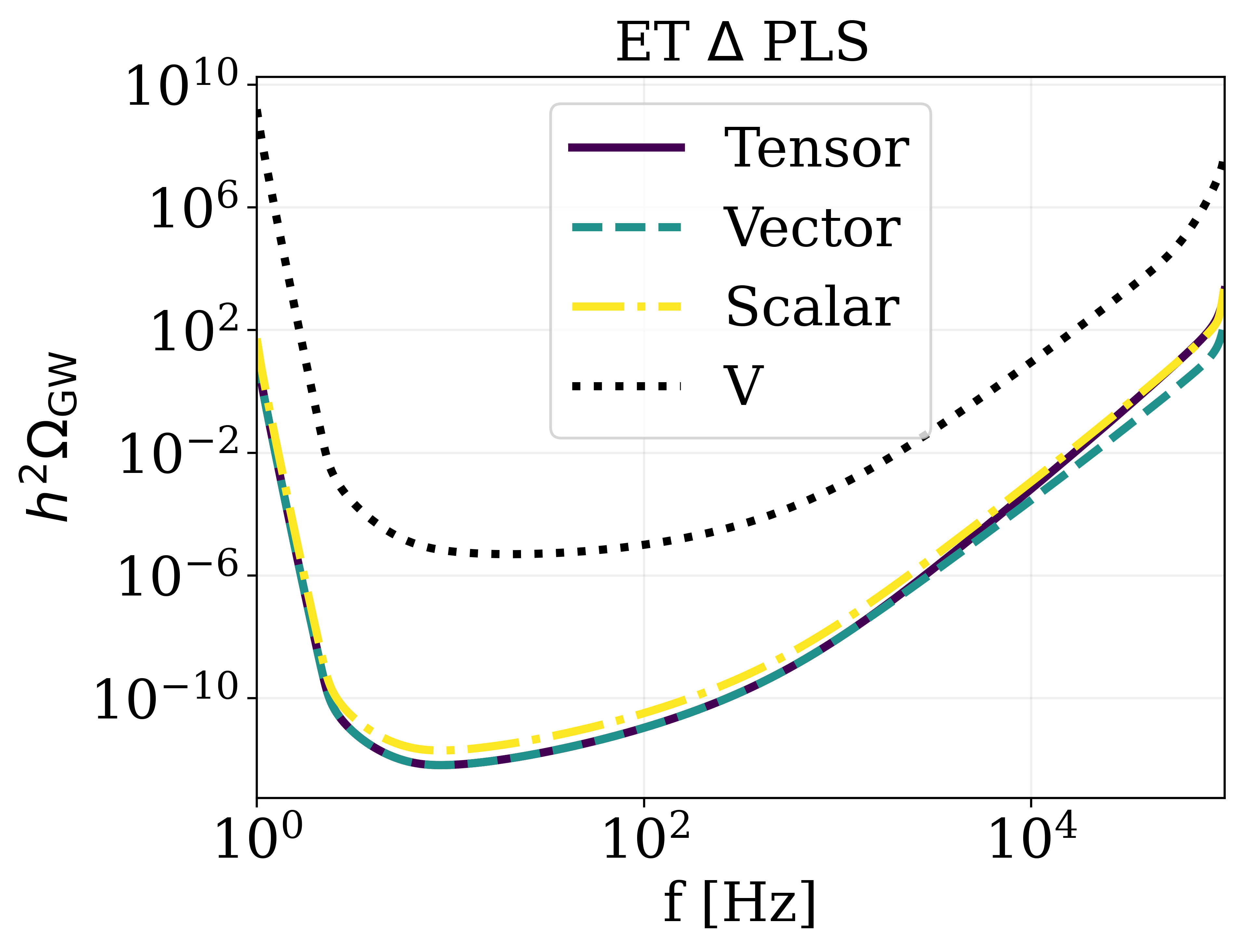}
    \includegraphics[scale=0.45]{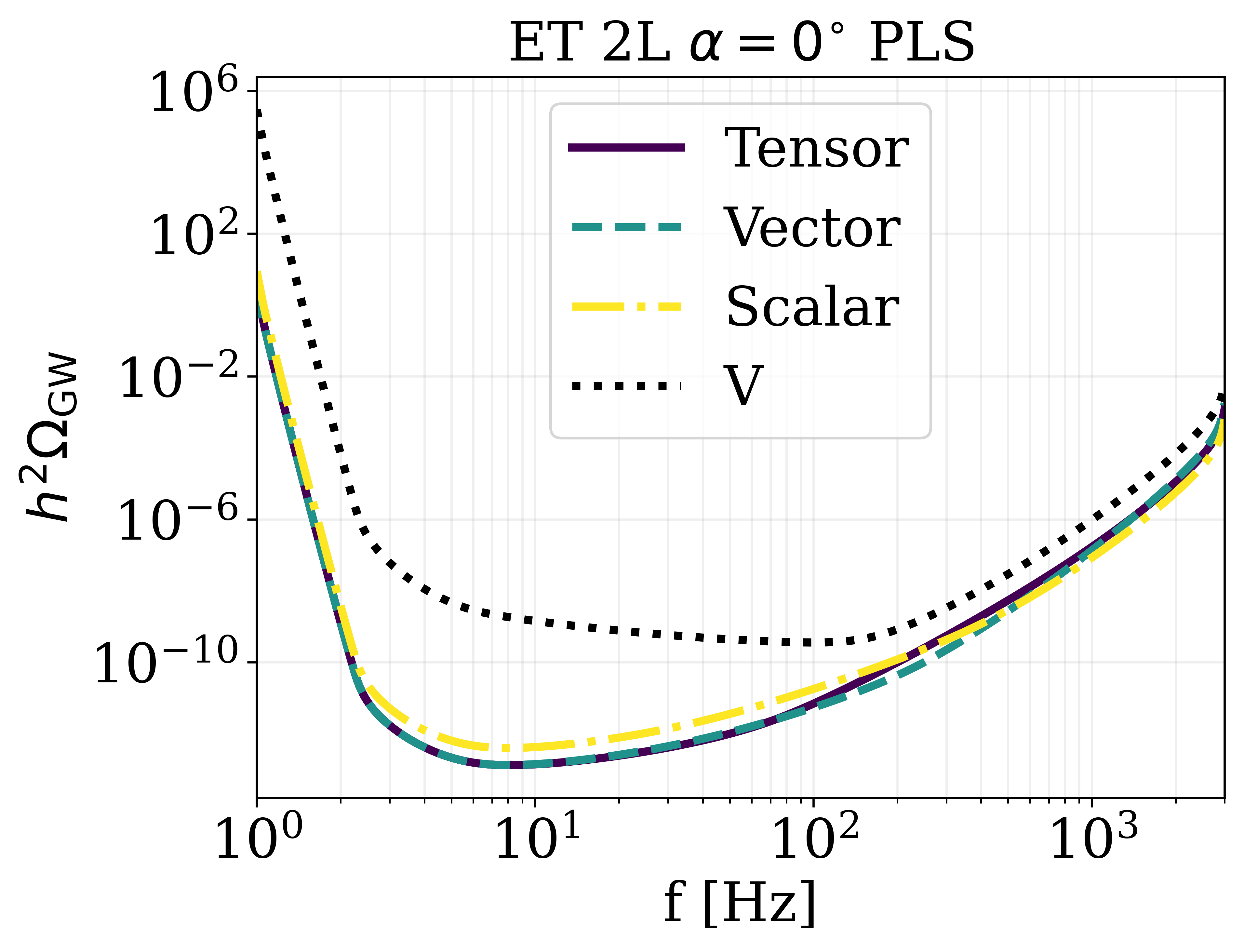}
    \includegraphics[scale=0.45]{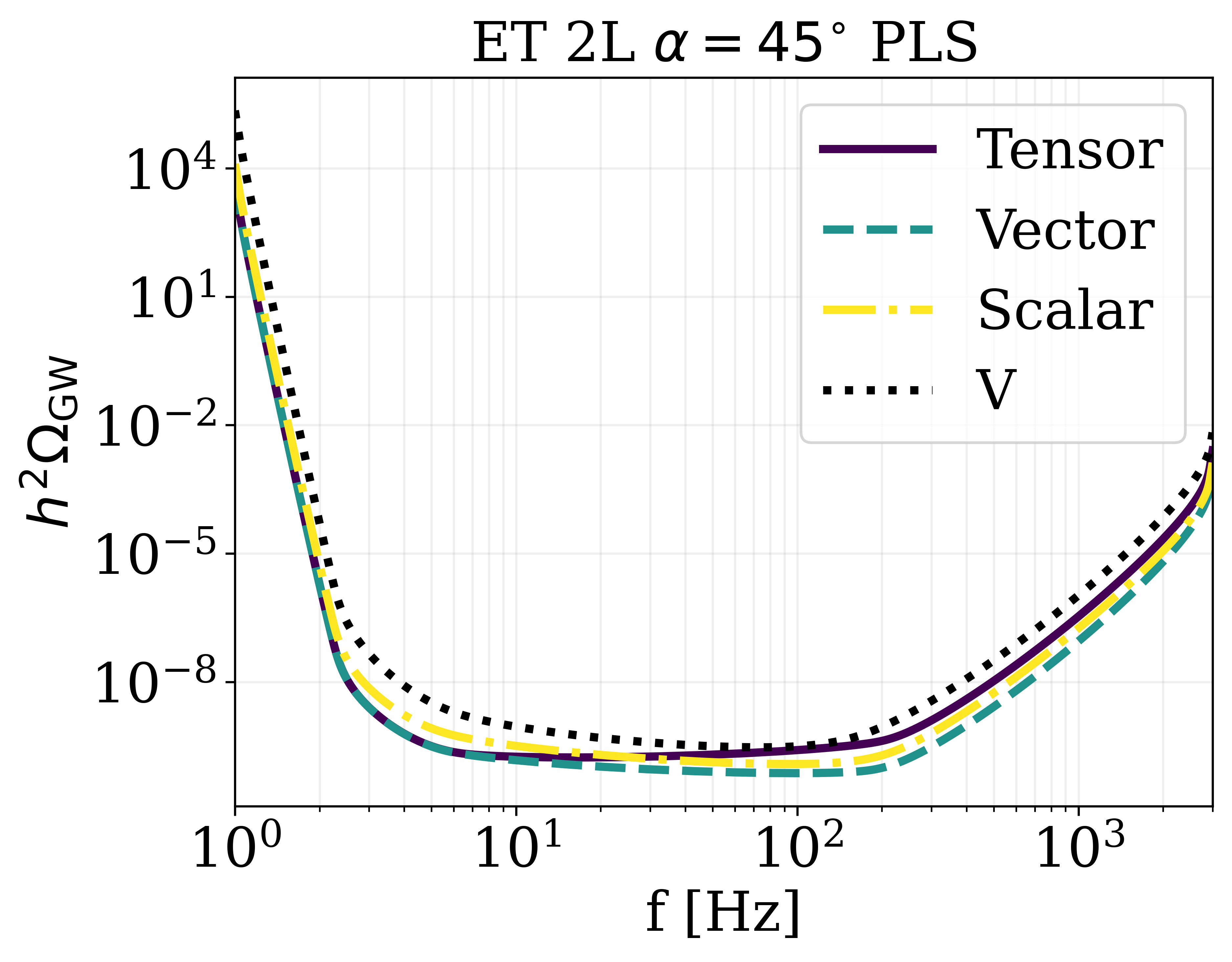}
    \caption{ET power law integrated sensitivity curves for different configurations analyzed. We choose an ${\rm SNR}_{ \rm th} = 1 $ and $T=1 {\rm yr}$. Upper left: triangular configuration. Upper right: 2L configuration aligned. Bottom center: 2L configuration misaligned of $\alpha=45^{\circ}$. Purple solid lines: tensor modes. Turquoise dashed lines: vector modes. Yellow dash-dotted lines: scalar modes. Black dashed lines: circular polarization mode.}
    \label{fig:ETpls}
\end{figure}

As mentioned above, ET design is still under discussion, so we derive here the PLS for an isotropic SGWB for different configurations and orientations. In Figure~\ref{fig:ETpls} we plot the PLS considering $T=1 \rm{yr}$ of observation time and ${\rm SNR_{th}=1}$ for the three different configuration \footnote{In the triangular configuration of ET, only the X and Y channels of the interferometers are included in the network analysis.}. We see from the plot and from Table~\ref{tab:plsminET} that the 2L aligned configuration is more sensitive to the background than the other two configurations for tensor, vector and scalar polarizations. The 2L misaligned case appears to be disfavored over the whole frequency range. This behavior is common to tensor, vector and scalar polarization modes, as they share very similar overlap reduction functions, which is the only thing that differs in them (see Figure~\ref{fig:ETorf}, right panel). On the contrary, the circular polarization mode shows a reasonable value only for the 2L configurations rather than the triangular one, while the misaligned configurations yield a slightly lower minimum detectable energy density spectrum.
\begin{table}[t!]
    \centering
    \begin{tabular}{ccccc}
    \multicolumn{5}{c}{ET}\\
    \hline\hline
    \multicolumn{5}{c}{$h^2\Omega_{\rm GW}$} \\
    \hline
                             & Tensor                 & Vector                  & Scalar                 & V \\
    $\Delta$                  & $6.54\times10^{-13}$   & $6.54\times10^{-13}$    & $1.96\times10^{-12}$   & $4.90\times10^{-6}$  \\
                             & {\scriptsize@\,8.9\,Hz} & {\scriptsize@\,8.9\,Hz} & {\scriptsize@\,8.9\,Hz} & {\scriptsize@\,20.5\,Hz}   \\
    2L $\alpha=0^\circ$       & $1.31\times10^{-13}$   & $1.32\times10^{-13}$    & $3.96\times10^{-13}$   & $3.56\times10^{-10}$  \\
                             & {\scriptsize@\,8.0\,Hz} & {\scriptsize@\,8.0\,Hz} & {\scriptsize@\,8.0\,Hz} & {\scriptsize@\,100.3\,Hz}   \\
    2L $\alpha=45^\circ$      & $1.71\times10^{-10}$   & $7.34\times10^{-11}$    & $1.19\times10^{-10}$   & $2.95\times10^{-10}$  \\
                             & {\scriptsize@\,16.3\,Hz} & {\scriptsize@\,97.9\,Hz} & {\scriptsize@\,93.3\,Hz} & {\scriptsize@\,74.0\,Hz}   \\
    \hline
    \end{tabular}
    \caption{Minimum detectable energy density spectrum ($\mathrm{SNR_{th}}=1$ and $T=1\,\mathrm{yr}$) for a SGWB composed of tensor, vector, scalar, or circular polarization modes under three different ET configurations.}
    \label{tab:plsminET}
\end{table}
Now, we discuss the case where the SGWB is made of the combination of tensor, vector and scalar polarization modes for a network made of ET (in its different configurations) and Cosmic Explorer (CE)\footnote{Placed in the LIGO Hanford site with 40 km arms.}.

\begin{figure}[t!]
    \centering
    \includegraphics[scale=0.45]{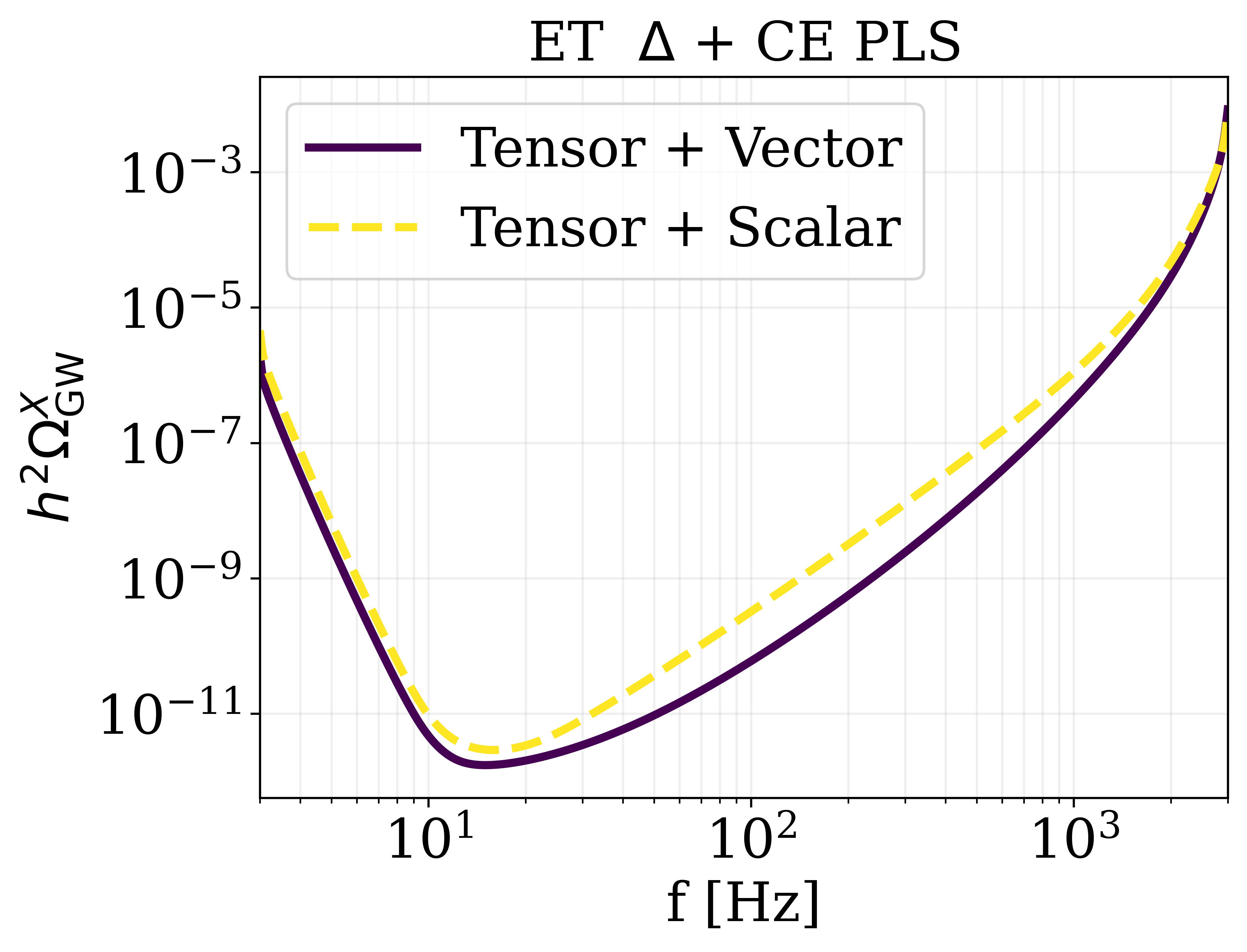}
    \includegraphics[scale=0.45]{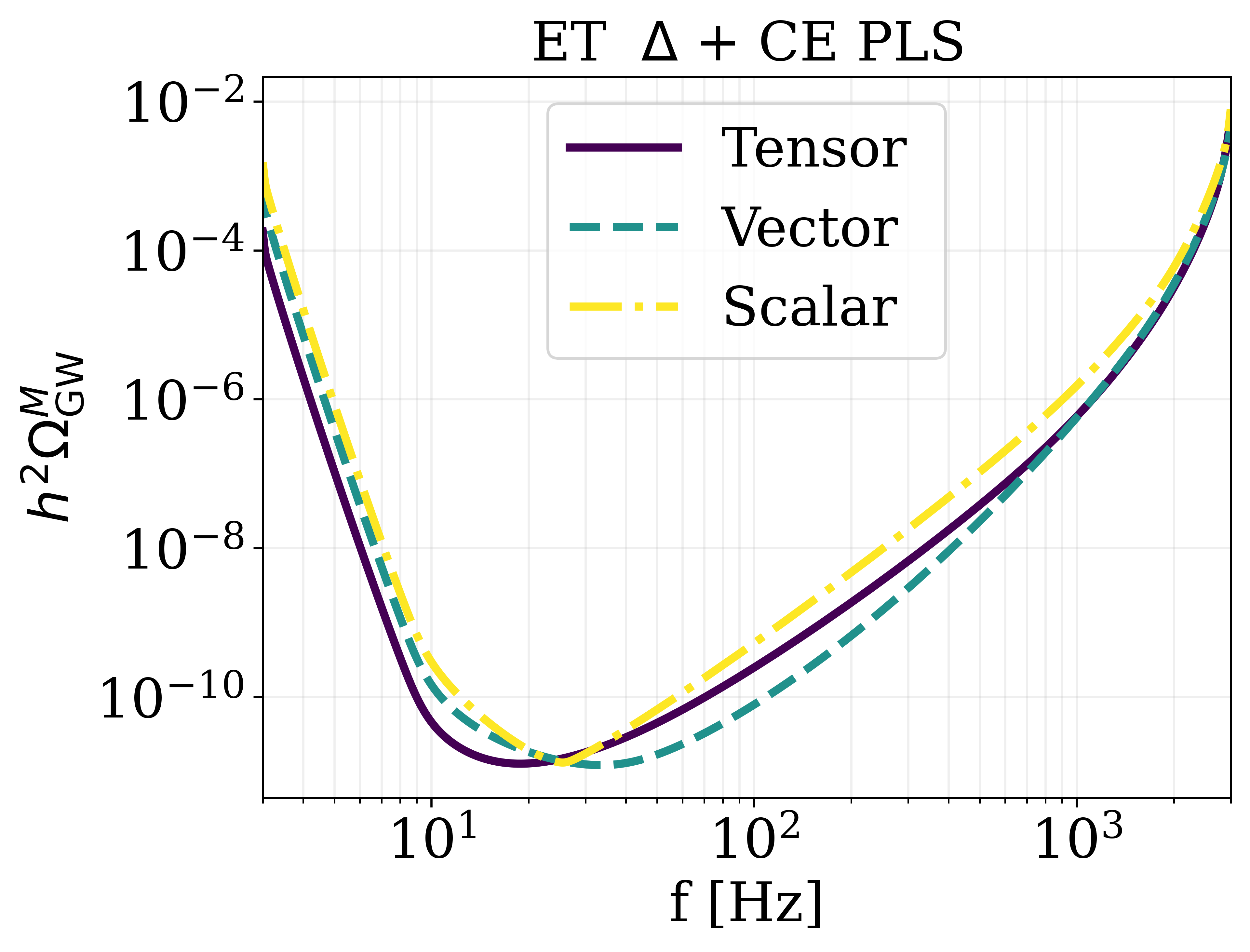}
    \includegraphics[scale=0.45]{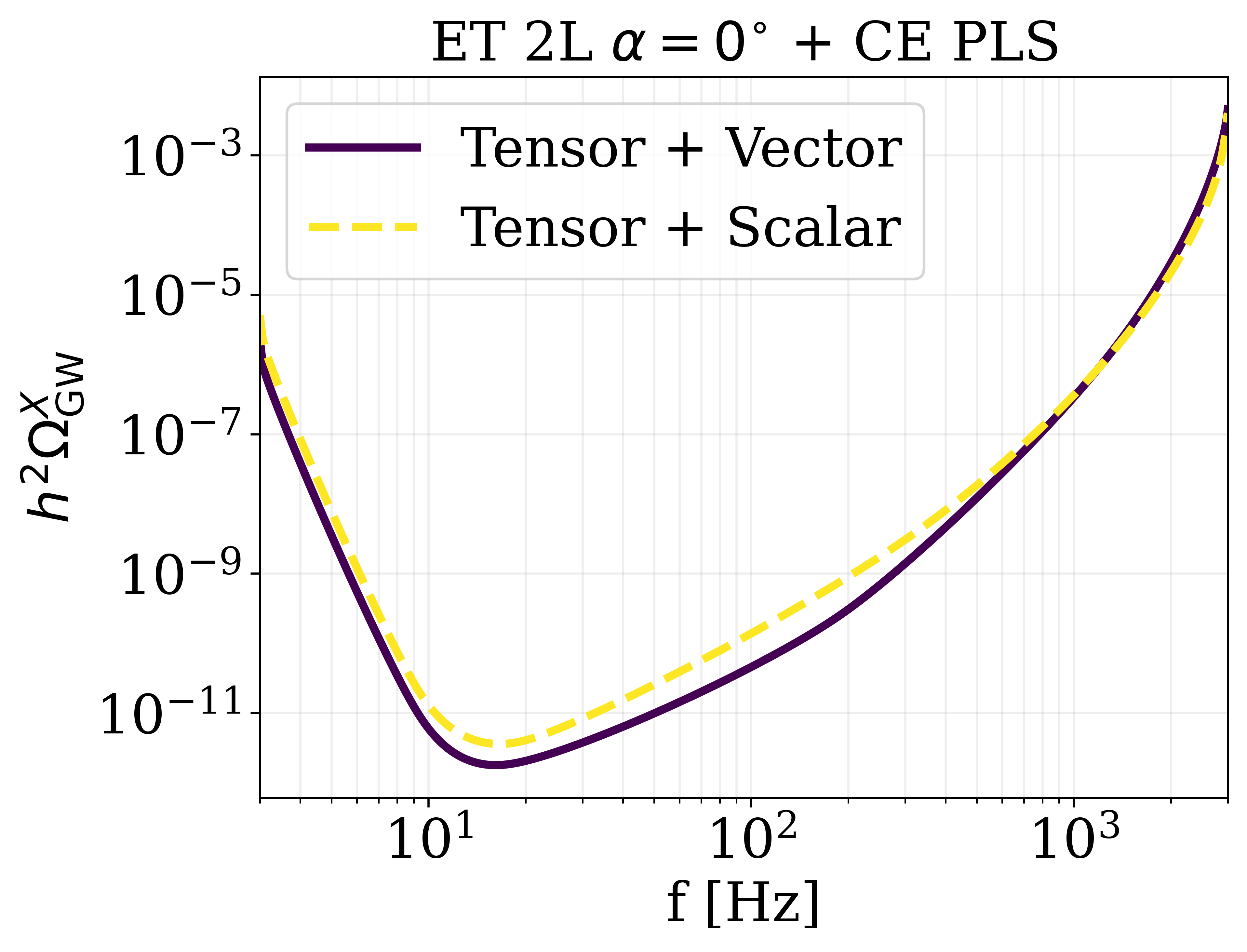}
    \includegraphics[scale=0.45]{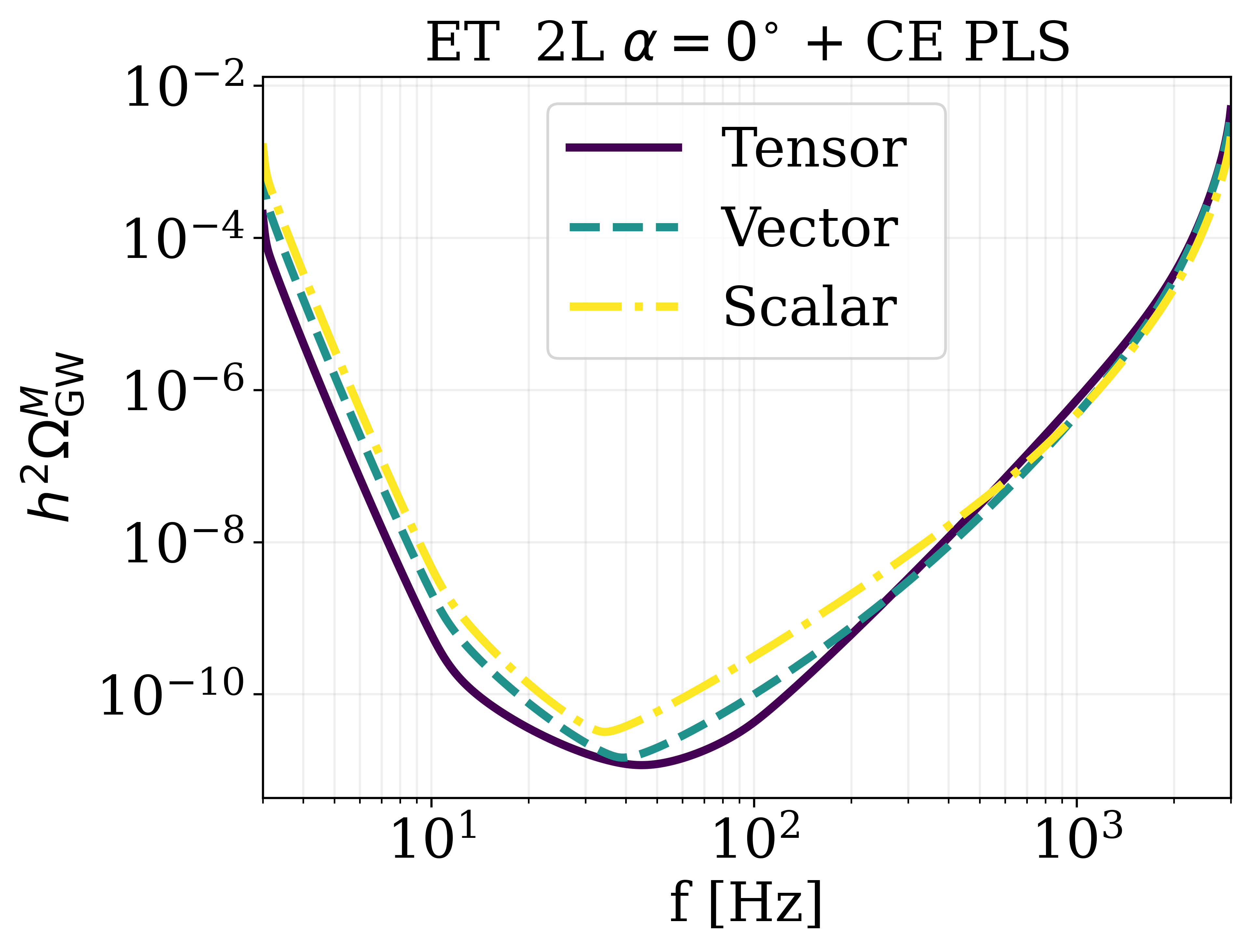}
    \includegraphics[scale=0.45]{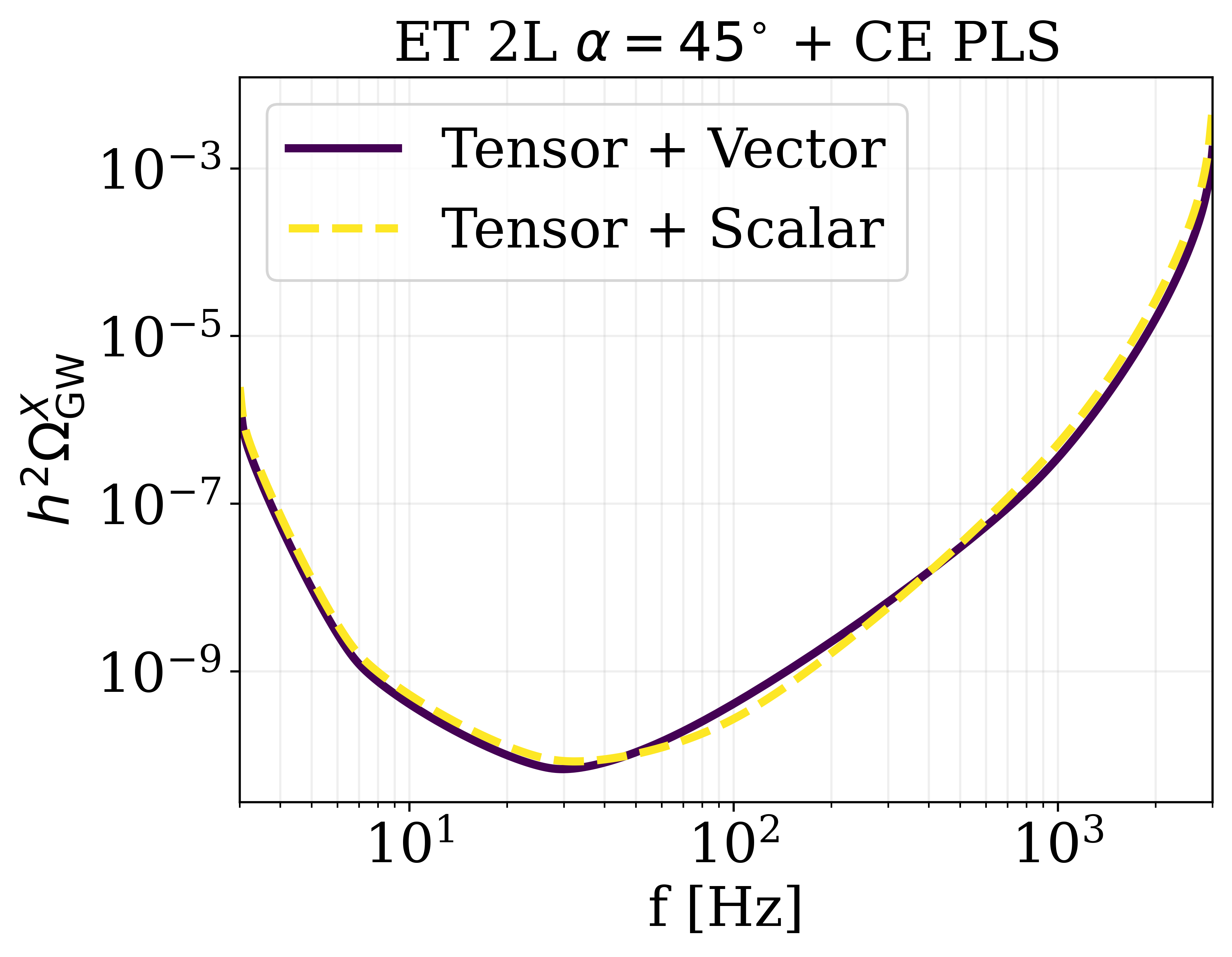}
    \includegraphics[scale=0.45]{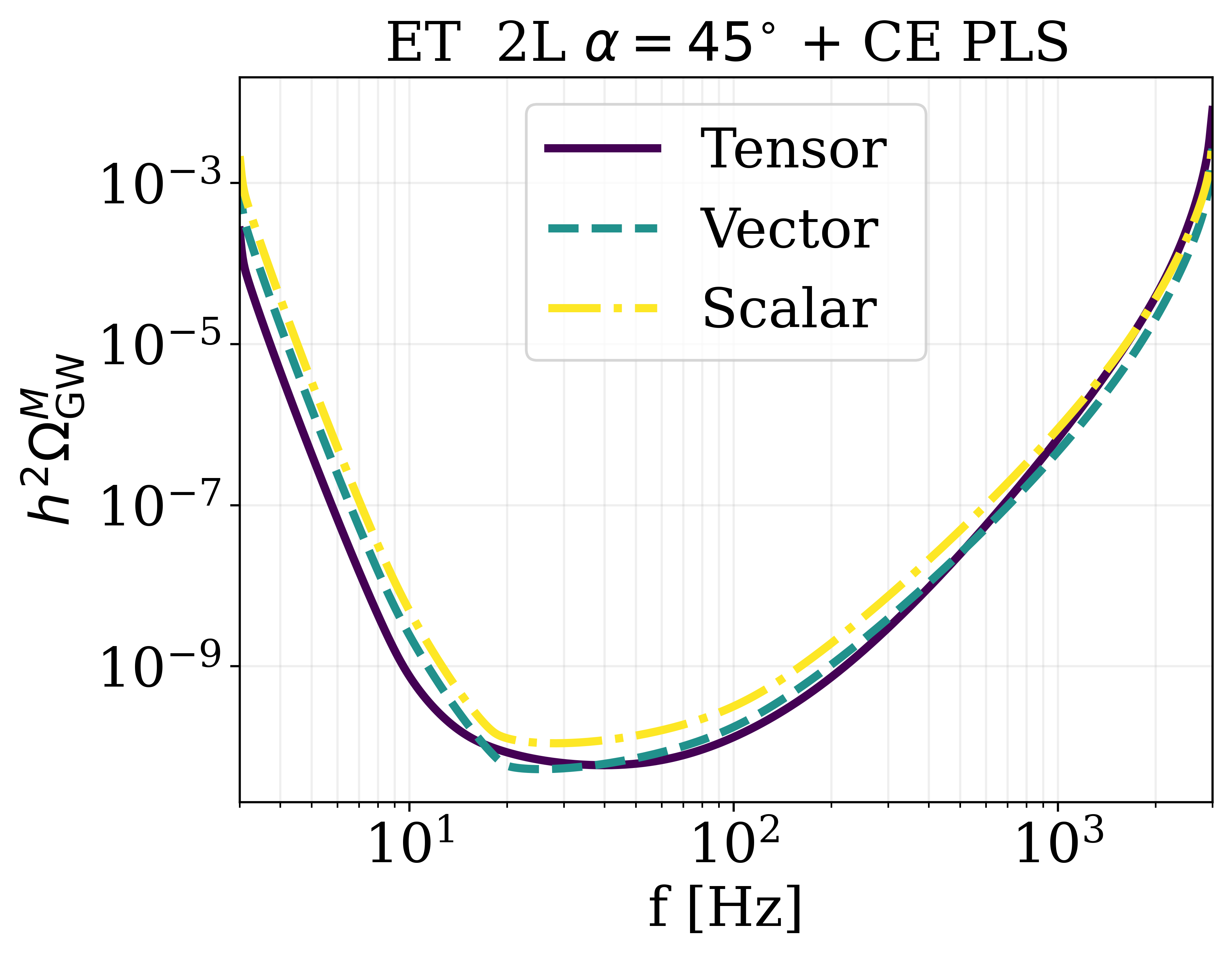}
    \caption{Left column: a SGWB made of tensor and an X polarization modes, with ${\rm X = v, s}$. The solid purple line (yellow dashed line) represents the sensitivity curve to a SGWB made of vectors (scalars). Right column: a SGWB made of tensor, vector and scalar polarization modes, with with ${\rm M = t,v, s}$. The purple solid line (turquoise dashed line - yellow dot dashed line) represents the sensitivity curve to a SGWB made of tensors (vectors - scalars). All the curves are obtained for ${\rm SNR_{\rm th}=1}$ and ${T = \rm 1\,  yr}$. }
    \label{fig:plsETCEextrapol}
\end{figure}

First, we analyze the case where we want to be sensitive to an X polarization mode (vector or scalar) other than the tensor one. By producing a PLS in this latter case, we are able to assess, at a given frequency, the minimum amplitude that the SGWB of the X polarization need to have in order to be visible by the instrument.
For the ET-CE network, we show our result in the left column of Figure~\ref{fig:plsETCEextrapol} and in Table~\ref{tab:plsminETCE2pol}. We notice that the network is slightly more sensitive to vector polarization modes rather than scalar ones for all the configurations under consideration. 
\begin{table}[!ht]
    \centering
    \begin{tabular}{cccc}
                                          & ET $\Delta$ + CE        & ET 2L $\alpha=0^{\circ}$ + CE &  ET 2L $\alpha=45^{\circ}$ + CE\\
    \hline  
    \hline
      $h^2\Omega_{\rm GW}^v$ Tensor + Vector  &  $1.73 \times 10^{-12}$ & $1.79 \times 10^{-12}$        &$6.81 \times 10^{-11}$\\
                                              &  {\scriptsize$@ \, 15.0 \, \rm Hz $} & {\scriptsize$@ \, 16.1  \, \rm Hz $}        & {\scriptsize$@ \, 29.7 \, \rm Hz $}\\
      $h^2\Omega_{\rm GW}^s$Tensor + Scalar   &  $2.89 \times 10^{-12}$ & $3.60 \times 10^{-12}$        &$8.41 \times 10^{-11}$\\
                                              & {\scriptsize$@ \, 16.0 \, \rm Hz $} & {\scriptsize$@ \, 16.7  \, \rm Hz $}        &{\scriptsize$@ \, 32.4 \, \rm Hz $}\\
      \hline
    \end{tabular}
    \caption{Minimum detectable energy density (${\rm SNR_{th}=1}$ and $T=1{\rm yr}$) for a SGWB made of tensor - vector modes and tensor - scalar modes separately.}
    \label{tab:plsminETCE2pol}
\end{table}
Secondly, we analyze the detectors sensitivity to a specific contribution of the SGWB from a given polarization mode, assuming an SGWB that includes all polarization modes. For the ET-CE network, our results, shown in the right column of Figure~\ref{fig:plsETCEextrapol} and in Table~\ref{tab:plsminETCE3pol}, demonstrate that, for tensor polarization modes, the best network is ET in its 2L aligned configuration and CE, for vector and scalar polarization modes the best one is the ET in its triangular configuration and CE.

We stress that the ET-CE network’s varying sensitivities to tensor, vector, and scalar-polarized backgrounds arise solely from its geometric configuration, not from any inherent features of the backgrounds themselves. These results could change if the relative positions and orientations of the detectors were different. \\


\begin{table}[th!]
    \centering
    \begin{tabular}{cccc}
                               & ET $\Delta$ + CE        & ET 2L $\alpha=0^{\circ}$ + CE &  ET 2L $\alpha=45^{\circ}$ + CE\\
    \hline 
    \hline
      $h^2\Omega_{\rm GW}^{t}$   &  $1.27 \times 10^{-11}$ & $1.17 \times 10^{-11}$ & $5.89 \times 10^{-11}$\\
                               &  {\scriptsize$@ \, 18.9 \, \rm Hz $} & {\scriptsize$@ \, 44.8 \, \rm Hz $} & {\scriptsize$@ \, 39..8 \, \rm Hz $}\\
      $h^2\Omega_{\rm GW}^{v}$   &  $1.22 \times 10^{-11}$ & $1.47 \times 10^{-11}$ & $5.26 \times 10^{-11}$\\
                               &  {\scriptsize$@ \, 33.5 \, \rm Hz $} & {\scriptsize$@ \, 39.3 \, \rm Hz $} & {\scriptsize$@ \, 25.4 \, \rm Hz $}\\
      $h^2\Omega_{\rm GW}^{s}$   &  $1.32 \times 10^{-11}$ & $3.20 \times 10^{-11}$ & $1.10 \times 10^{-10}$\\
                               &  {\scriptsize$@ \, 25.6 \, \rm Hz $} & {\scriptsize$@ \, 34.4 \, \rm Hz $}& {\scriptsize$@ \, 28.2 \, \rm Hz $}\\
      \hline
    \end{tabular}
    \caption{Detectable energy density ($\rm SNR_{th}=1$ and $T=1 {\rm yr}$ ) for a SGWB made of tensor vector and scalar polarization modes.}
    \label{tab:plsminETCE3pol}
\end{table}

\subsubsection{LISA}

\begin{figure}[t!]
    \centering
    \includegraphics[scale=0.45]{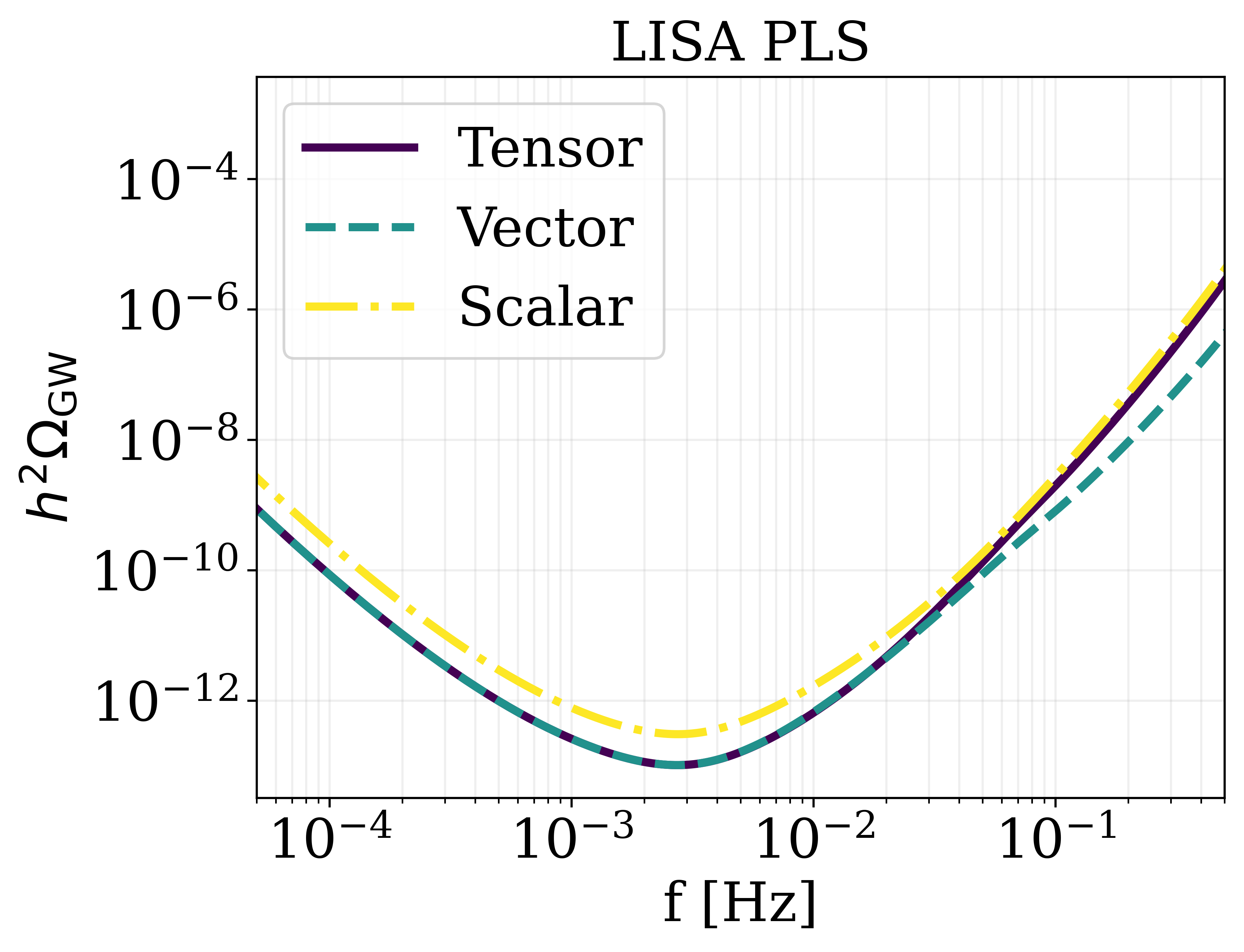}
    \caption{LISA power law integrated sensitivity curves for tensor (purple line), scalar (dashed dotted yellow line) and vector (dashed turquoise line) modes. We choose ${\rm SNR_{\rm th} = 10}$ and $T=3{\rm\,  yrs}$.}
    \label{fig:LISApls}
\end{figure}

Now we turn to the space-based LISA detector, and we compute the PLS for tensor, vector, and scalar modes. The results are presented in Figure~\ref{fig:LISApls}. In this case, the PLS is reweighted across the three TDI channels, namely we sum over the A, E and T channels. The sensitivity appears to be essentially the same for tensor and vector modes, since their overlap reduction functions are equal for $f \lesssim 10^{-2} \, \rm Hz$ (see Figure \ref{fig:LISAorf_AET}), while for $f \gtrsim 10^{-2} \, \rm Hz$ vector overlap reduction function is slightly higher for both the A (E) and T channel. On the contrary, scalar polarization modes exhibit the standard $1/3$ reduction factor in sensitivity (see Table \ref{tab:plsminLISA}) due to their overlap reduction function normalization.

Circular polarization mode is not considered here, since the overlap reduction function is compatible with zero (see Section~\ref{sec:overlapLISA}) and therefore, the network is insensitive to parity violating signals for isotropic SGWB.

\begin{table}[!h]
    \centering
    \begin{tabular}{cccc}
    \multicolumn{4}{c}{LISA} \\
    \hline
    \hline
                         & Tensor & Vector & Scalar \\
                         \hline
        $h^2\Omega_{\rm GW}$ & $1.02 \times 10^{-13}$ & $1.02 \times 10^{-13}$ & $3.03\times 10^{-13}$  \\
        & {\scriptsize$@ \, 2.74 \times 10^{-3}\, \rm Hz$}  & {\scriptsize$@ \, 2.74 \times 10^{-3}\, \rm Hz$} & {\scriptsize$@ \, 2.74 \times 10^{-3}\, \rm Hz$}  \\
    \hline
    \end{tabular}
    \caption{Minimum detectable energy density (${\rm SNR_{th}}=10$ and $T=3 {\rm yrs}$) for a SGWB made of tensor, vector or scalar polarization modes separately.}
    \label{tab:plsminLISA}
\end{table}

\subsubsection{PTA}

\begin{figure}[t!]
    \centering
    \includegraphics[scale=0.45]{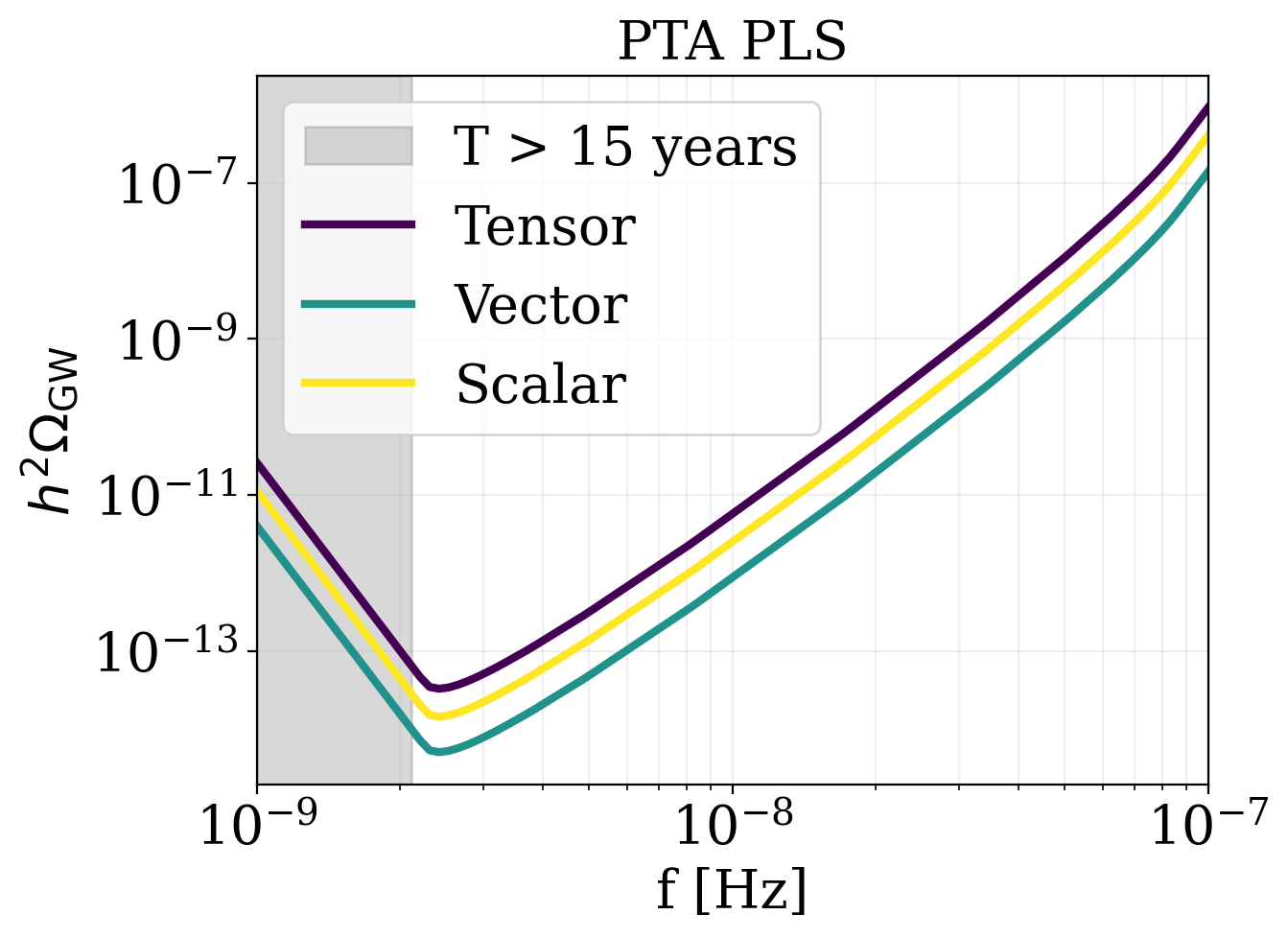}
    \caption{PLS computed from a PTA catalog. We choose ${\rm SNR_{th} = 1}$ and $T=15{\rm\,  yrs}$. Purple line: tensor modes. Turquoise line: vector modes. Yellow line: scalar modes.}
    \label{fig:plsPTA}
\end{figure}

Finally, we consider the case of PTA. Also in this case we derive the PLS, considering a catalog of 60 pulsars. We plot our results in Figure~\ref{fig:plsPTA} and we report the minimum detectable energy density spectrum in Table~\ref{tab:plsminPTA}. 
Here, we consider only white noise for each pulsar in the catalog. Under such assumptions, we can clearly see that the array is most sensitive to vector modes. This can be easily interpreted considering Figure~\ref{fig:HD_PTA}, where we see that the overlap reduction function for vector modes is enhanced with respect to the other modes considered. 

The PLS for tensor modes in our analysis lies below that reported in~\cite{Babak:2024yhu} for the Square Kilometer Array (SKA)~\cite{Janssen:2014dka}. This difference can probably be attributed to two factors: our analysis assumes a 15-year observation time, while theirs uses 10 years, and we consider an array of 60 pulsars compared to their 50. Overall, the results remain reasonably consistent: we find a minimum sensitivity of order $10^{-14}$, compared to their $10^{-13}$. Given the assumptions on pulsar noise and the uniform 15-year observation time in our analysis, our approach appears robust.

\begin{table}[!h]
    \centering
    \begin{tabular}{cccc}
    \multicolumn{4}{c}{PTA (60 pulsars)} \\
    \hline
    \hline
                         & Tensor & Vector & Scalar \\
                         \hline
        $h^2\Omega_{\rm GW}$ & $3.49 \times 10^{-14}$ & $5.24 \times 10^{-15}$ & $1.53\times 10^{-14}$  \\
        & {\scriptsize$@ \, 2.11 \times 10^{-9} \, \rm Hz$} &  {\scriptsize$ @ \, 2.11 \times 10^{-9} \, \rm Hz$} &  {\scriptsize$ @ \, 2.11 \times 10^{-9} \, \rm Hz$}  \\
    \hline
    \end{tabular}
    \caption{Minimum detectable energy density (${\rm SNR_{th}}=1$ and $T=15 {\rm yrs}$) for a SGWB made of tensor, vector and scalar polarization modes respectively.}
    \label{tab:plsminPTA}
\end{table}

\subsection{Angular Power Law Integrated Sensitivity Curves}
\subsubsection{LIGO-Virgo}

\begin{figure}[t!]
    \centering
    \includegraphics[scale=0.45]{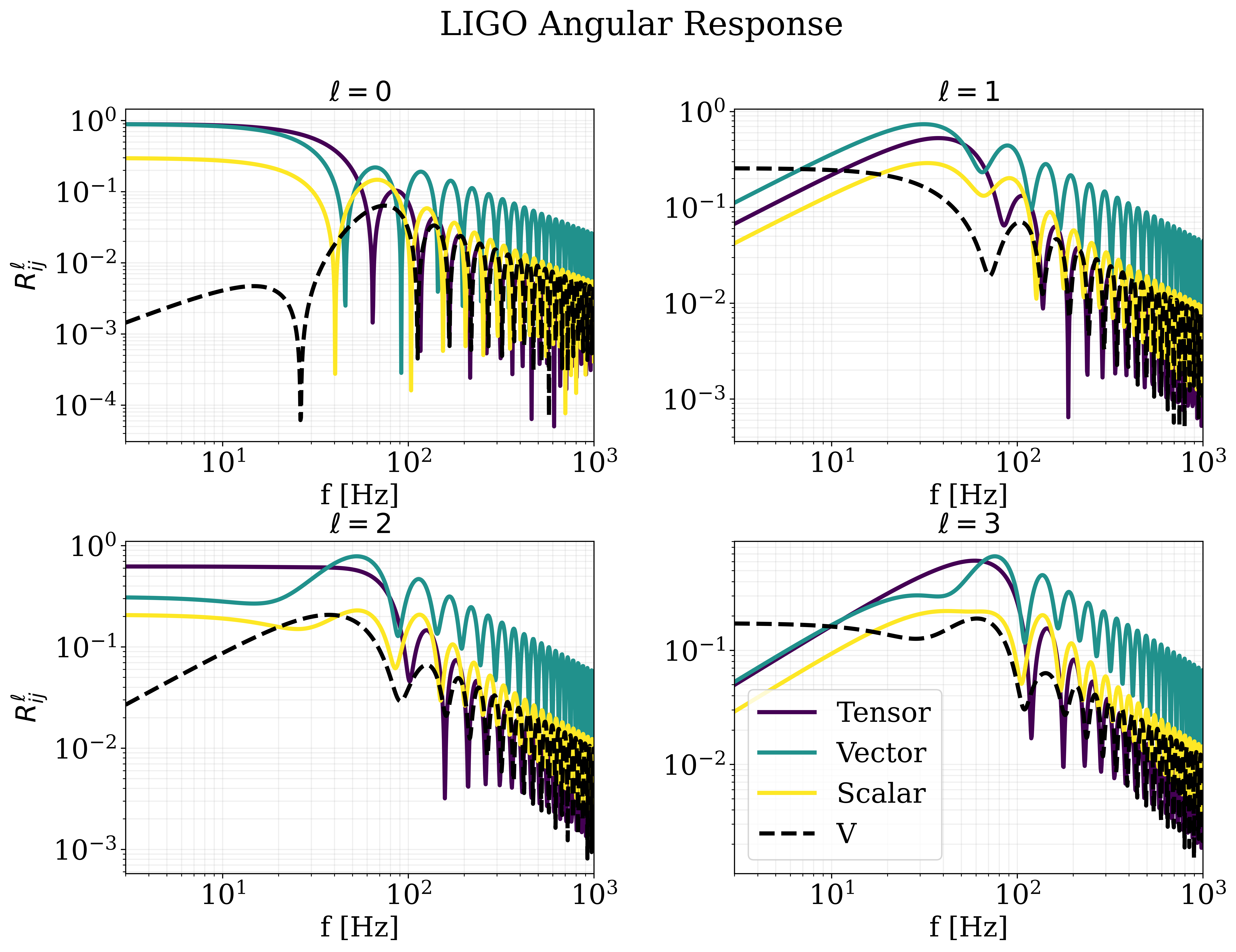}
    \caption{Angular response function for LIGO for $\ell = 0, 1, 2, 3$ in the triangular configuration. Purple solid lines: tensor modes.  Turquoise solid lines: vector modes. Yellow solid lines: scalar modes. Black dashed lines: circular polarization mode.}
    \label{fig:LIGOarf}
\end{figure}

Now we move to the angular response function defined in eq.\eqref{R_ell}, for different polarization modes. In Figure~\ref{fig:LIGOarf} we plot the response function for different $\ell$ multipoles for the LIGO Hanford - LIGO Livingston network, using the formalism presented in Section~\ref{angularresponse}. 

For tensor, vector, and scalar polarization modes, the even multipoles ($\ell=0,2$) exhibit a plateau at lower frequencies, followed by a sharp oscillatory decline. Instead, the odd multipoles ($\ell=1,3$) show a steep rise that transitions into an oscillatory decay at higher frequencies, resembling a broken power-law behavior. For the circular polarization mode, on the other hand, the even multipoles ($\ell=0,2$) display a broken power-law–like trend, while the odd multipoles ($\ell=1,3$) exhibit a plateau at low frequencies, followed by an oscillatory drop.

In a similar manner to the PLS, we aim to assess the sensitivity of the detector network to anisotropies. Specifically, we seek to determine the required amplitude of the SGWB monopole needed to detect an anisotropy of order $\ell$, given its amplitude, as presented in Section~\ref{sensitivity to multipoles}.

\begin{figure}[!ht]
    \centering
    \includegraphics[scale=0.45]{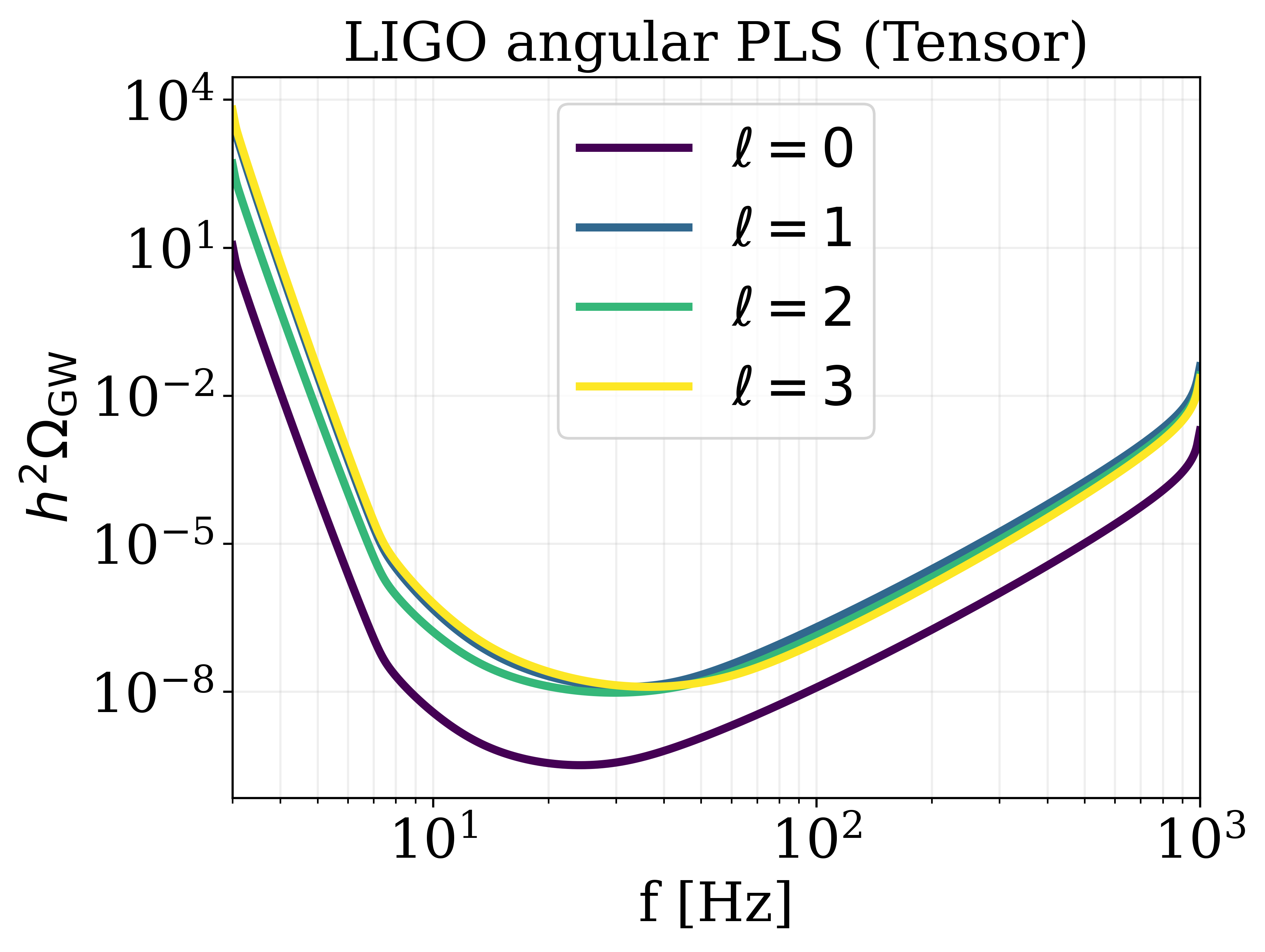}
    \includegraphics[scale=0.45]{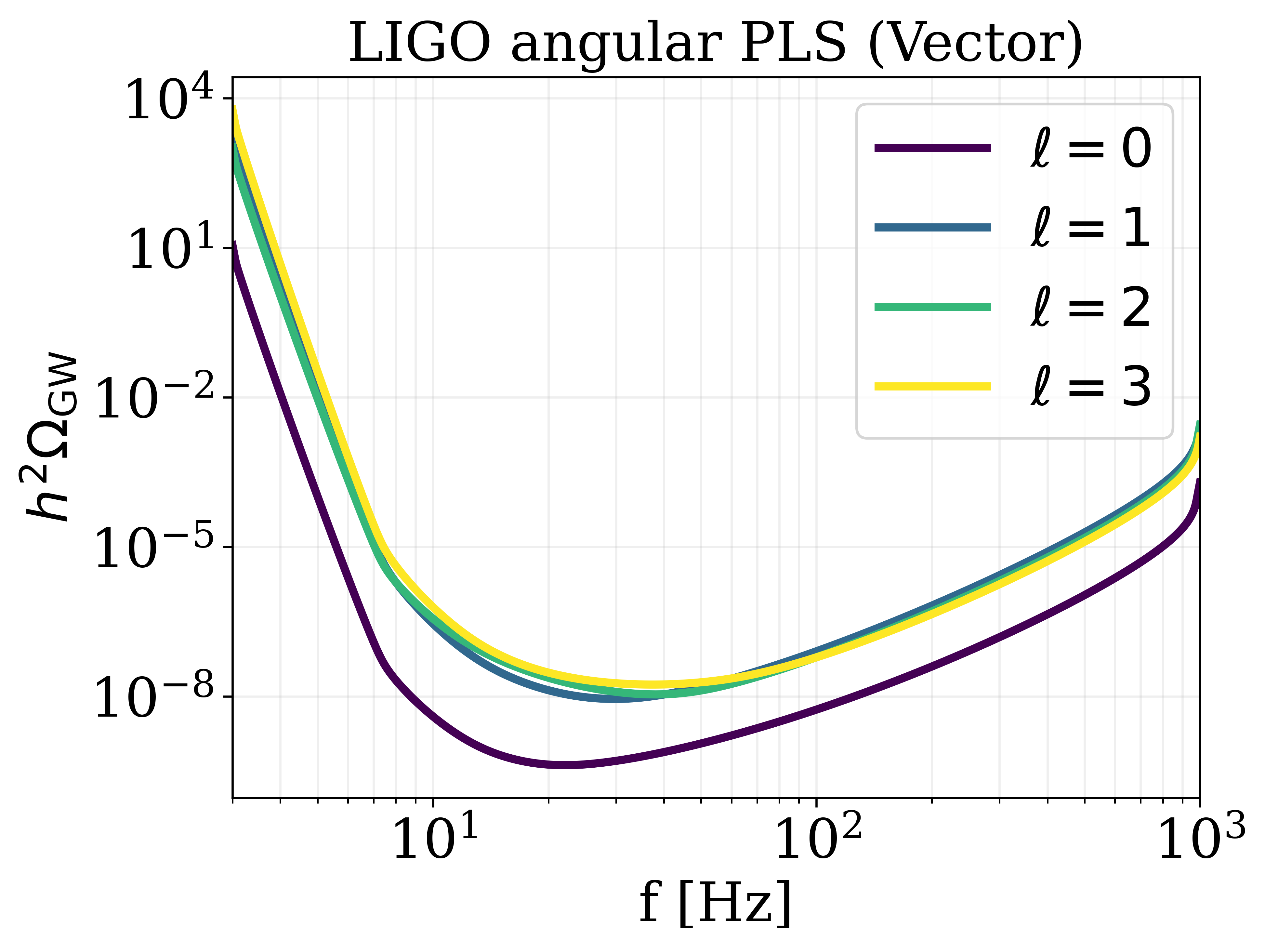}
    \includegraphics[scale=0.45]{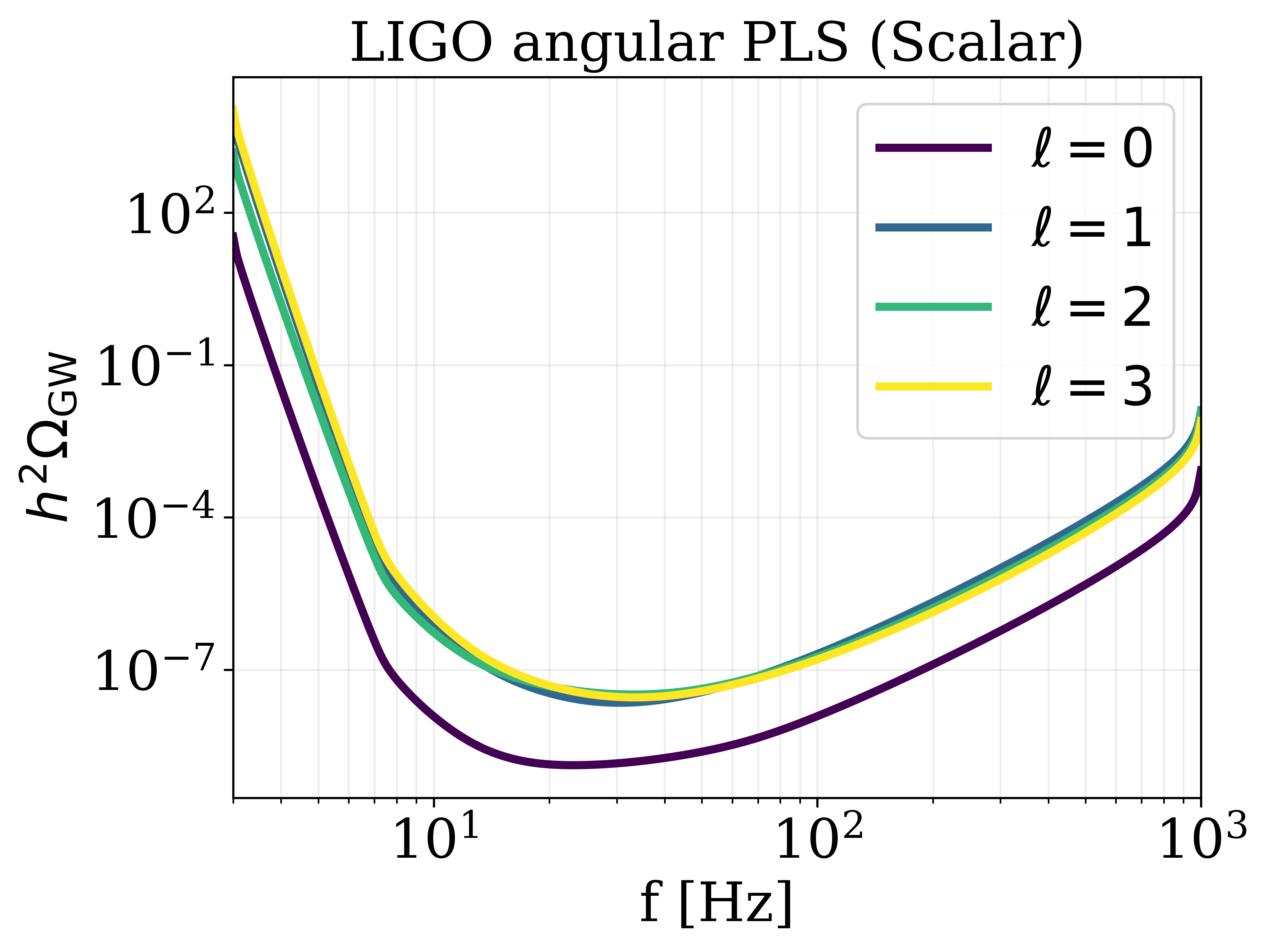}
    \includegraphics[scale=0.45]{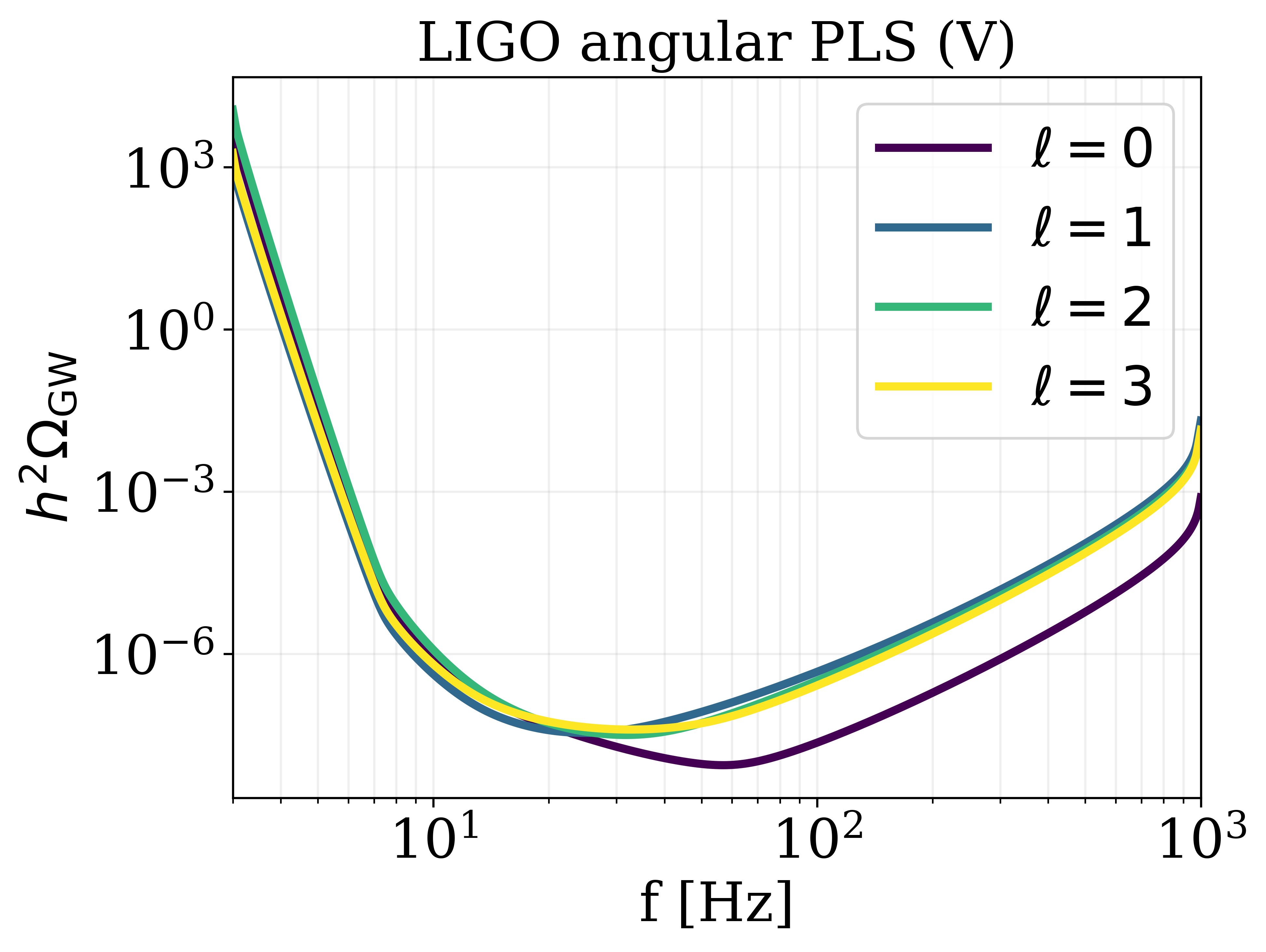}
    \caption{Angular power law integrated sensitivity curves for the LIGO pair. We set ${\rm \langle SNR\rangle}_{\ell, \rm th} = 1$, $T= 1 {\rm yr}$, $C_{\ell = 1, 2, 3}^{\rm GW}=10^{-3}$.}
    \label{fig:LIGOapls}
\end{figure}

As shown in Figure~\ref{fig:LIGOapls} and Table \ref{tab:aplsminLIGO}, for the LIGO detectors, for all the polarization modes considered, the three multipoles $\ell=1, 2,3$ set nearly the same monopole value required for detecting the anisotropy, given the same anisotropy contribution of $C_{\ell}^{ \rm GW}=10^{-3}$. However, for tensor modes, the detectors exhibit a slightly better sensitivity to the quadrupole ($\ell=2$) rather than to the dipole.
While the Advanced LIGO network’s varying sensitivities arise largely from its geometric configuration, it is important to note that these angular sensitivities can also strongly depend on the specific model considered, since they depend on the angular power spectrum.

\begin{table}[!ht]
    \centering
    \begin{tabular}{ccccc}
    \multicolumn{5}{c}{LIGO} \\
    \hline
    \hline
    \multicolumn{5}{c}{$h^2\Omega_{\rm GW}$} \\
    \hline
                         & Tensor & Vector & Scalar & V\\
                         \hline
        $\ell=1$ & $1.24 \times 10^{-8}$ & $8.90 \times 10^{-9}$ & $2.24\times 10^{-8}$   & $3.46\times 10^{-8}$ \\
                 & {\scriptsize$@ \, 31.6 \, \rm Hz$} & {\scriptsize$@ \, 29.8 \, \rm Hz$} & {\scriptsize$@ \, 30.5 \, \rm Hz$}  & {\scriptsize$@ \, 24.8 \, \rm Hz$} \\
        $\ell=2$ & $9.55 \times 10^{-9}$ & $1.10 \times 10^{-8}$ & $3.26\times 10^{-8}$  & $3.16\times 10^{-8}$ \\
                 & {\scriptsize$@ \, 29.8 \, \rm Hz$} & {\scriptsize$@ \, 38.1 \, \rm Hz$} & {\scriptsize$@ \, 33.1 \, \rm Hz$}  & {\scriptsize$@ \, 31.6 \, \rm Hz$} \\
        $\ell=3$ & $1.25 \times 10^{-8}$ & $1.72\times 10^{-8}$  & $2.85\times 10^{-8}$  & $3.98\times 10^{-8}$\\
                 & {\scriptsize$@ \, 36.6 \, \rm Hz$} & {\scriptsize$@ \, 37.4 \, \rm Hz$} & {\scriptsize$@ \, 33.5 \, \rm Hz$}  & {\scriptsize$@ \, 32.7 \, \rm Hz$} \\
    \hline
    \end{tabular}
    \caption{Minimum energy density spectrum amplitude required (${\rm \langle SNR\rangle}_{\ell, \rm th}=1$, $T=1{\rm yr}$ and $C_{\ell=1, 2, 3}^{ \rm GW}=10^{-3}$ ) for a SGWB made of tensor, vector, scalar or circular polarization modes.}
    \label{tab:aplsminLIGO}
\end{table}

\subsubsection{Einstein Telescope}

\begin{figure}[t!]
    \centering
    \includegraphics[scale=0.45]{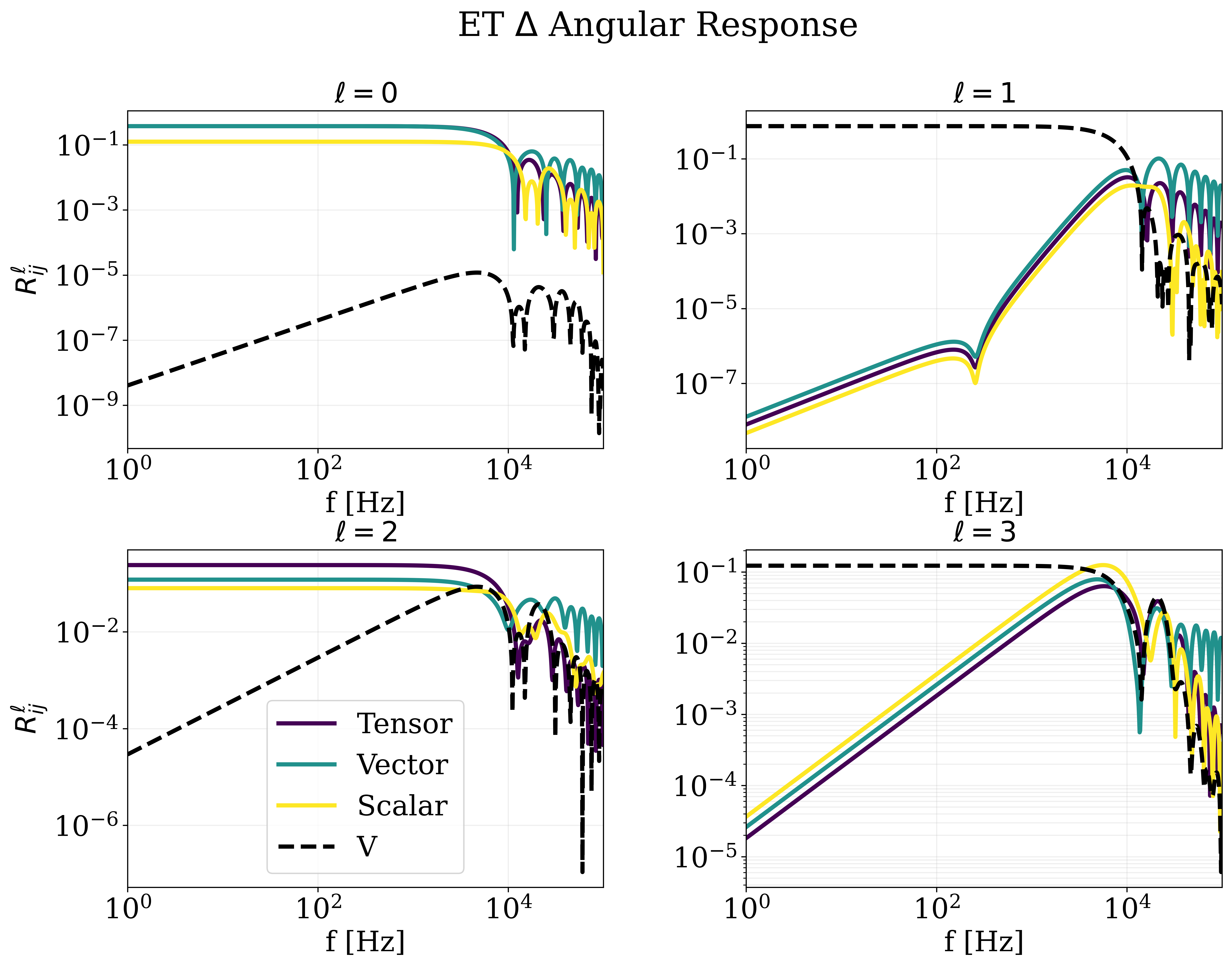}
    \caption{Angular response function for ET for $\ell = 0, 1, 2, 3$ in the triangular configuration. Purple solid lines: tensor modes.  Turquoise solid lines: vector modes. Yellow solid lines: scalar modes. Black dashed lines: circular polarization mode.}
    \label{fig:ET_Rell}
\end{figure}

\begin{figure}[t!]
    \centering
    \includegraphics[scale=0.45]{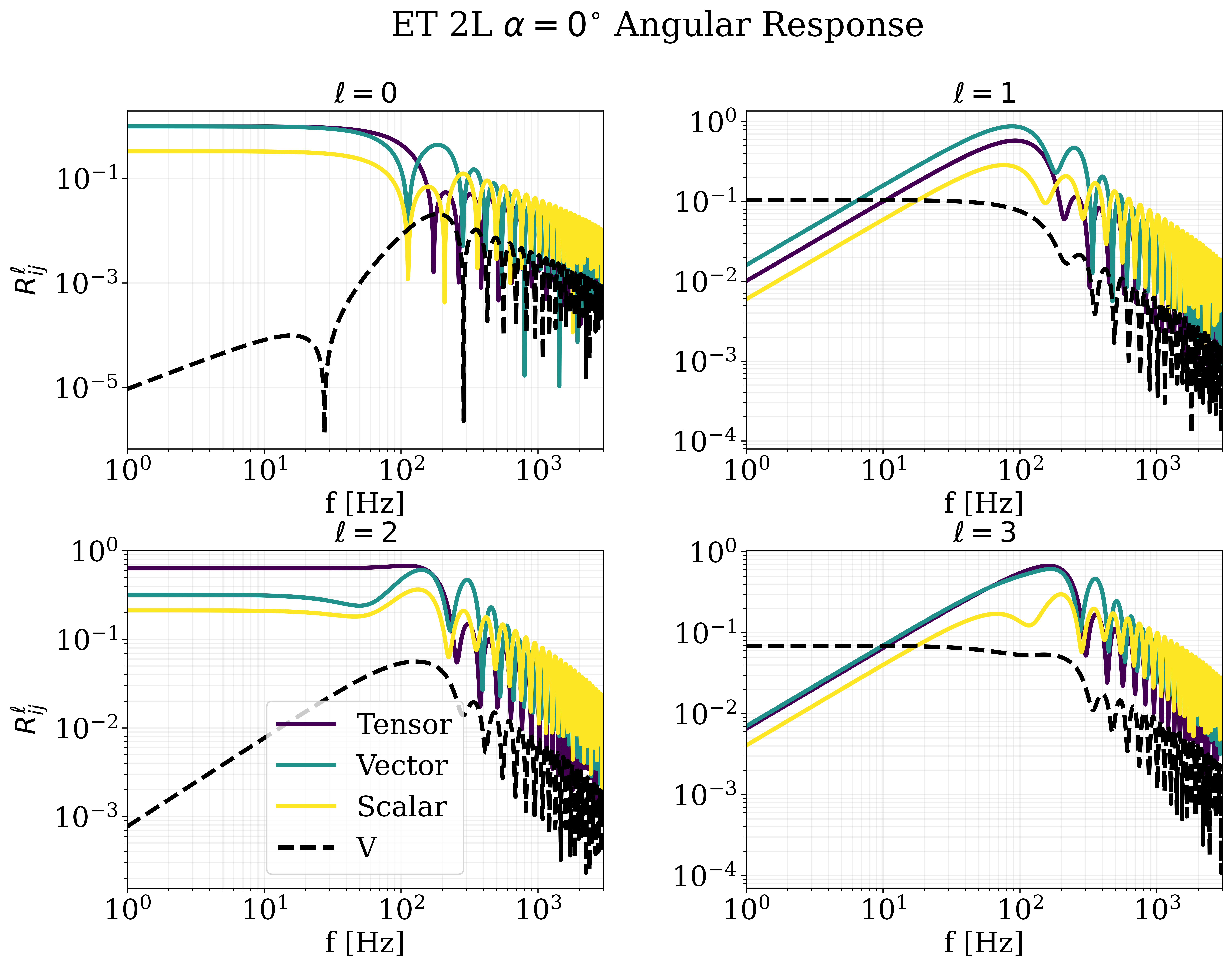}
    \caption{Angular response function for ET for $\ell = 0, 1, 2, 3$ in the 2L configuration aligned. Purple solid lines: tensor modes. Turquoise solid lines: vector modes. Yellow solid lines: scalar modes. Black dashed lines: circular polarization mode. }
    \label{fig:ET2L0_Rell}
\end{figure}

\begin{figure}[!ht]
    \centering
    \includegraphics[scale=0.45]{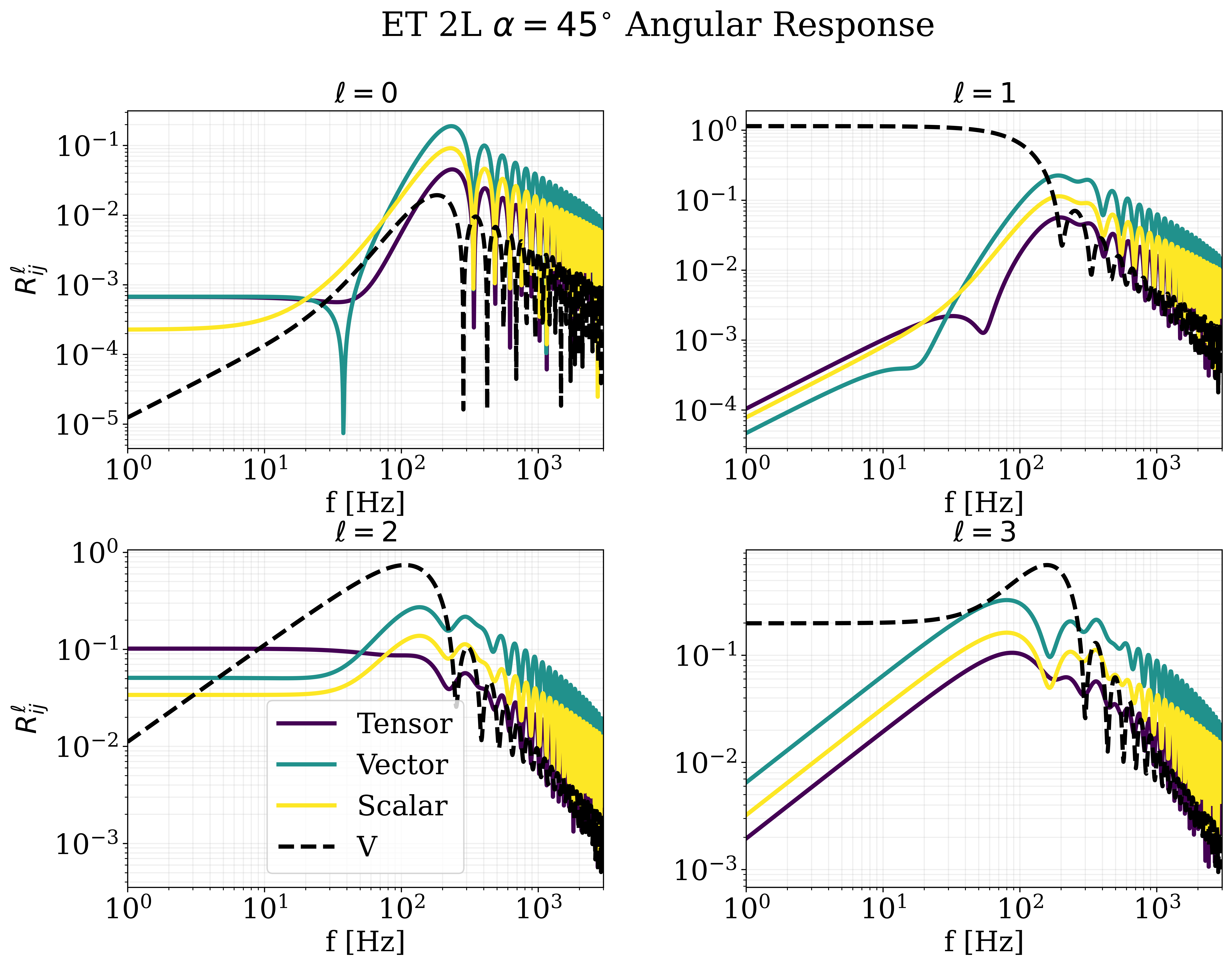}
    \caption{Angular response function for ET for $\ell = 0, 1, 2, 3$ in the 2L configuration misaligned of $\alpha=45^{\circ}$. Purple solid lines: tensor modes. Turquoise solid lines: vector modes. Yellow solid lines: scalar modes. Black dashed lines: circular polarization mode.}
    \label{fig:ET2L45_Rell}
\end{figure}

In Figure~\ref{fig:ET_Rell},~\ref{fig:ET2L0_Rell},~\ref{fig:ET2L45_Rell} we present the angular response functions for the different $\ell$ multipoles across the various configurations considered for ET.

First, let us discuss the angular response for ET in its triangular configuration \footnote{As before in Section \ref{sec:ETpls}, only the X and Y channels are considered for ET in the triangular setup.} in Figure~\ref{fig:ET_Rell}.
For tensor, vector and scalar modes, even multipoles ($\ell=0,2$) show a plateau in the lower frequency range and then a sudden oscillatory drop at $f\sim 10^{3} \, \rm Hz$.
Instead, odd multipoles ($\ell=1, 3$) shows a steep rise and then the usual oscillatory drop at higher frequencies; again, the behavior changes after $f\sim 10^{3} \, \rm Hz$. This behavior is common to tensor, vector and scalar polarization modes; however it appears that, for $\ell=3$ the scalar modes response is higher than for the other two polarization modes. 
For the circular polarization mode, instead, the behavior is the opposite. In the case $\ell=0,2$ we have first a steep rise and then an oscillatory drop after $f\sim 10^{3} \, \rm Hz$, while odd multipoles ($\ell = 1, 3$) show a plateau at lower frequency and than a sudden oscillatory drop at $f\sim 10^{3} \, \rm Hz$.

Now, let us now discuss the angular response for ET in the 2L aligned configuration in Figure~\ref{fig:ET2L0_Rell}. 
As in the previous case, for tensor, vector and scalar polarization modes, even multipoles ($\ell=0,2$) show a plateau in the lower frequency range and then a sudden oscillatory drop , while odd multipoles ($\ell=1, 3$) exhibit a steep rise and then a fall at higher frequencies in an almost broken power law behavior. 
For the circular polarization mode, the behavior is inverted, as in ET in the triangular configuration. What is different here is the frequency at which the behavior changes. In the case of the 2L configuration aligned, the behavior changes at $f\sim 10^{2} \, \rm Hz$. This is because the two detectors are farther apart ($|\Delta \vec{X}| \sim 1165 \,\rm km$,~\cite{Branchesi:2023mws}) than in the triangular case ($|\Delta \vec{X}| \sim 10 \,\rm km$).

Lastly, let us discuss the angular response for ET in the 2L misaligned configuration in Figure~\ref{fig:ET2L45_Rell}.
Here, for tensor, vector and scalar polarization modes, $\ell=0$ shows a low plateau in the lower frequency range, a sudden rise at $f\sim 10^{2} \, \rm  Hz$ and then an oscillatory drop, $\ell=2$ has the same behavior only for vector and scalar modes, while tensor ones only exhibit a plateau and then the oscillatory drop. Instead, odd multipoles show the same behavior as before in the $\alpha = 0^{\circ}$ case.
For the circular polarization mode, instead, even multipoles show a power-law behavior at low frequencies followed by the oscillatory drop at $f \sim 100 \, \rm Hz$, while odd multipoles exhibit a plateau at low frequencies.

\begin{figure}[!ht]
    \centering
    \includegraphics[scale=0.45]{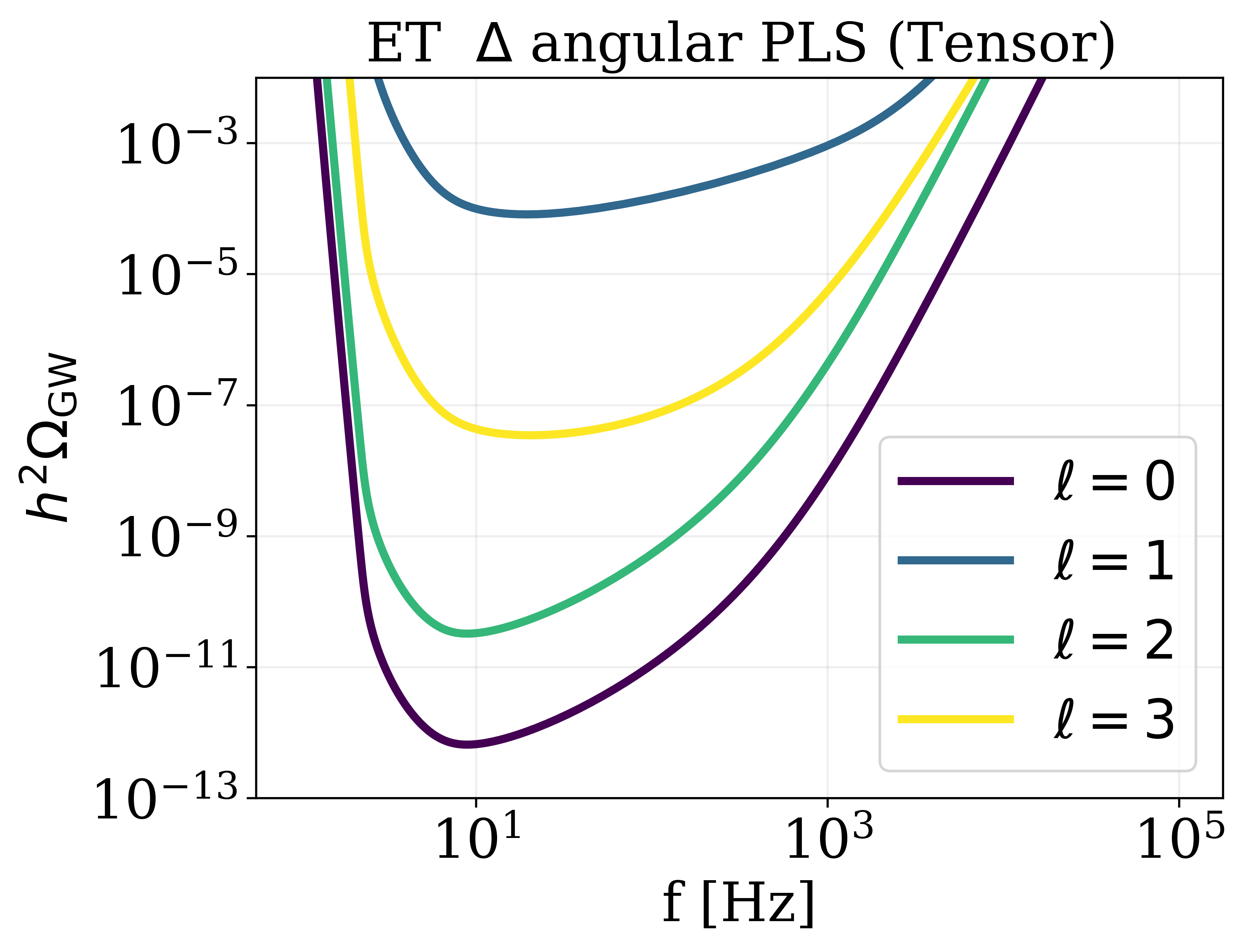}
    \includegraphics[scale=0.45]{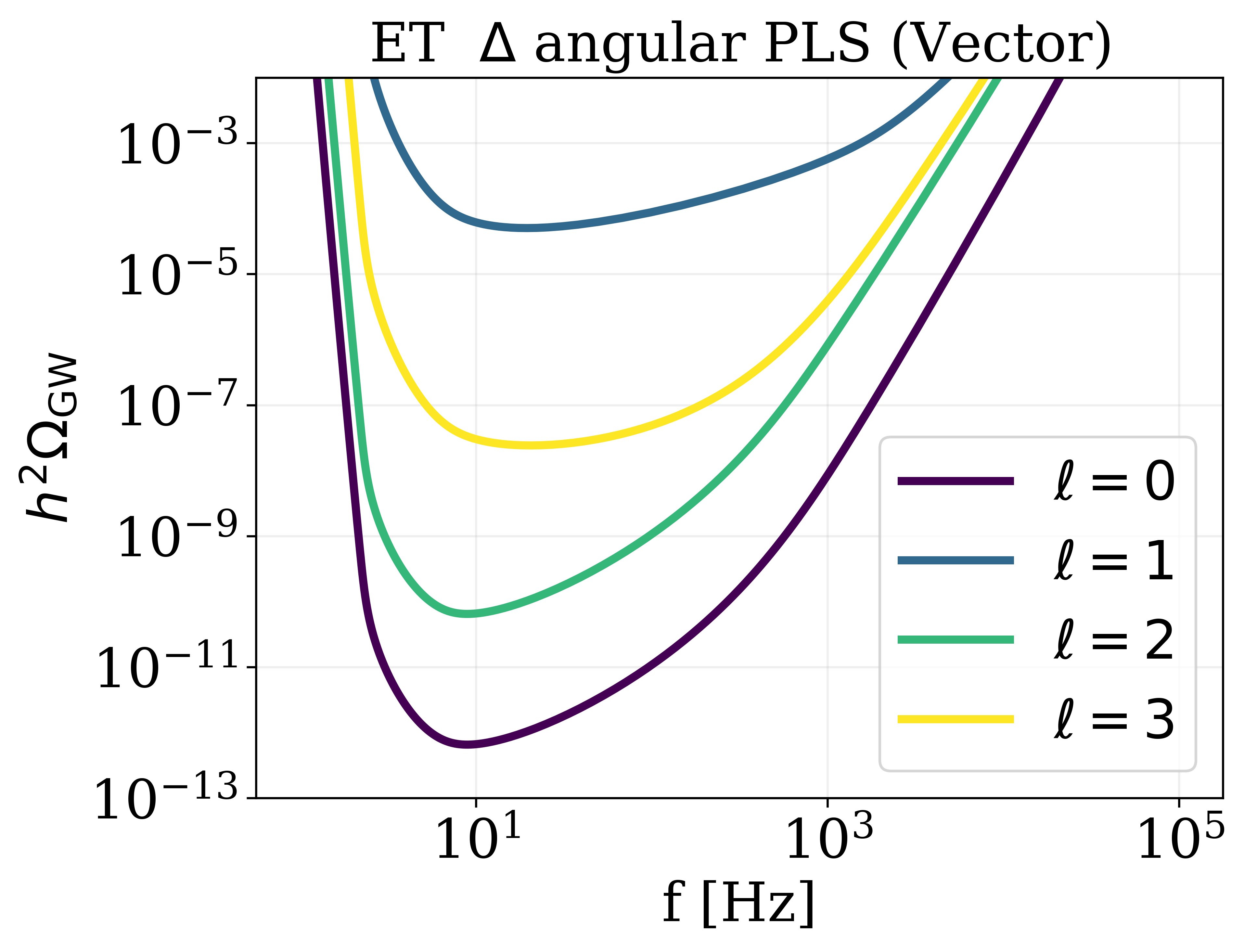}
    \includegraphics[scale=0.45]{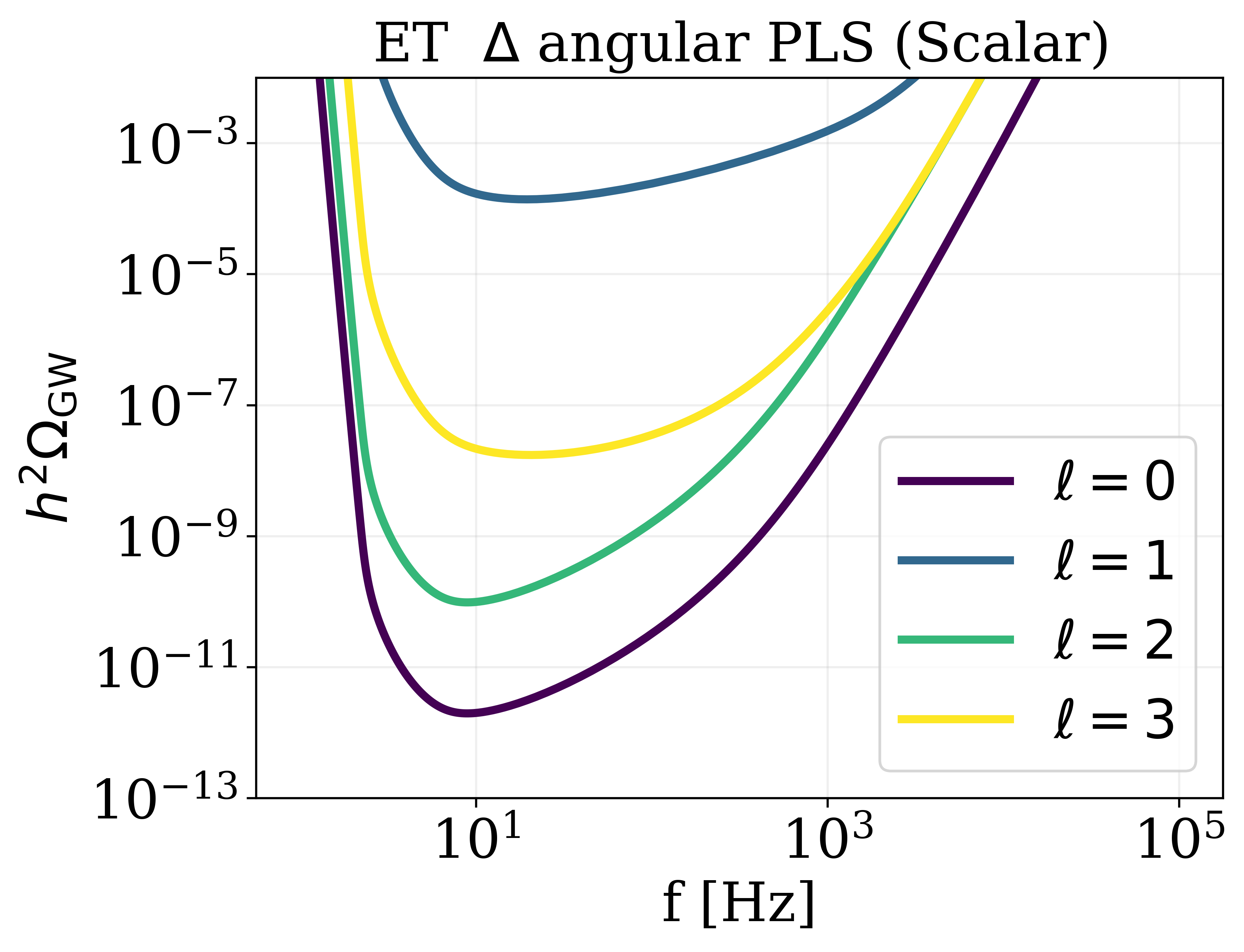}
    \includegraphics[scale=0.45]{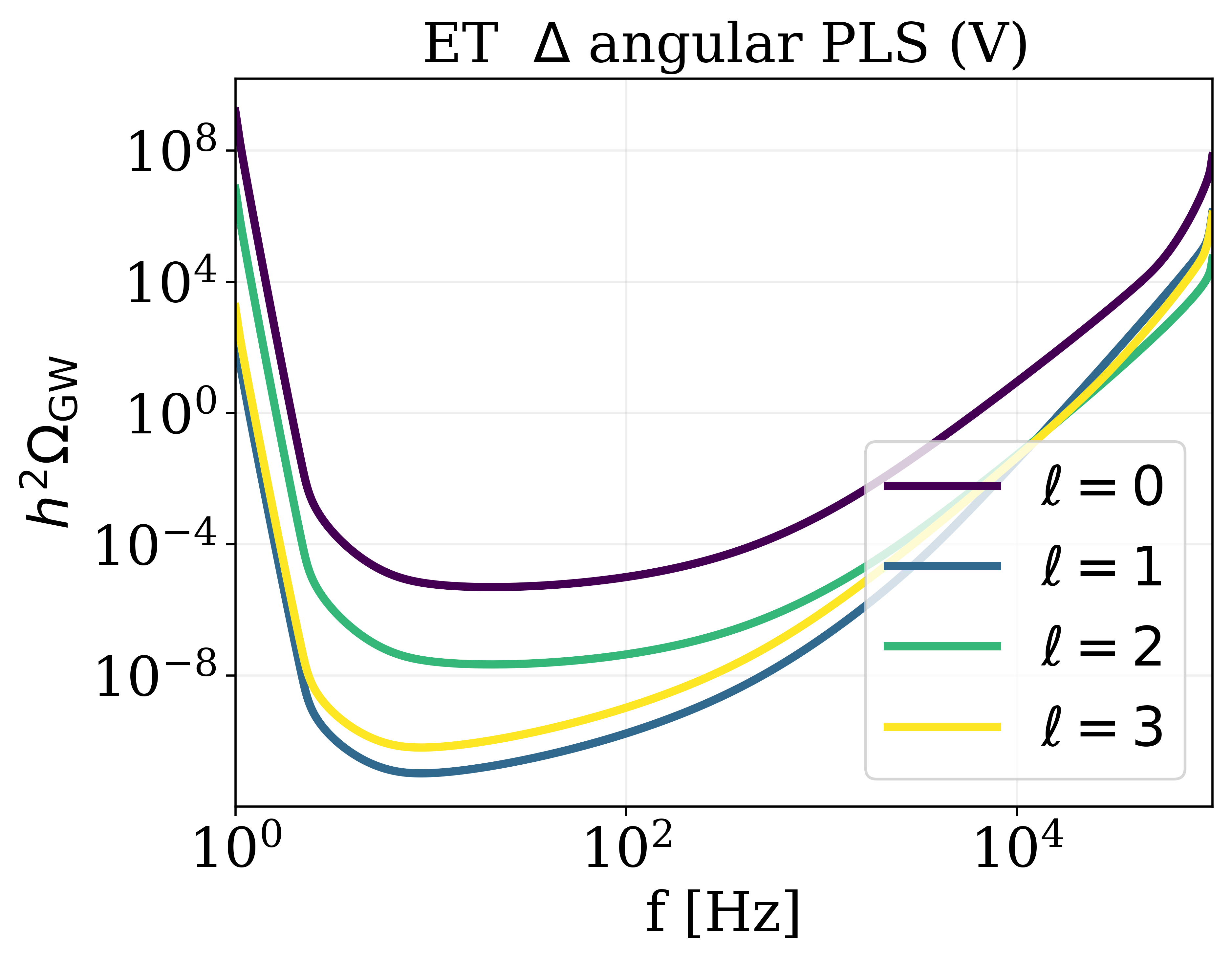}
    \caption{Angular power law integrated sensitivity curves for ET for $\ell = 0, 1, 2, 3$ in the triangular configuration. We set $\langle\rm SNR \rangle_{\ell, th}=1$, $T= 1 {\rm yr}$, $C_{\ell = 1, 2, 3}^{GW}=10^{-3}$.}
    \label{fig:ETtri_Nell}
\end{figure}

\begin{figure}[!ht]
    \centering
    \includegraphics[scale=0.45]{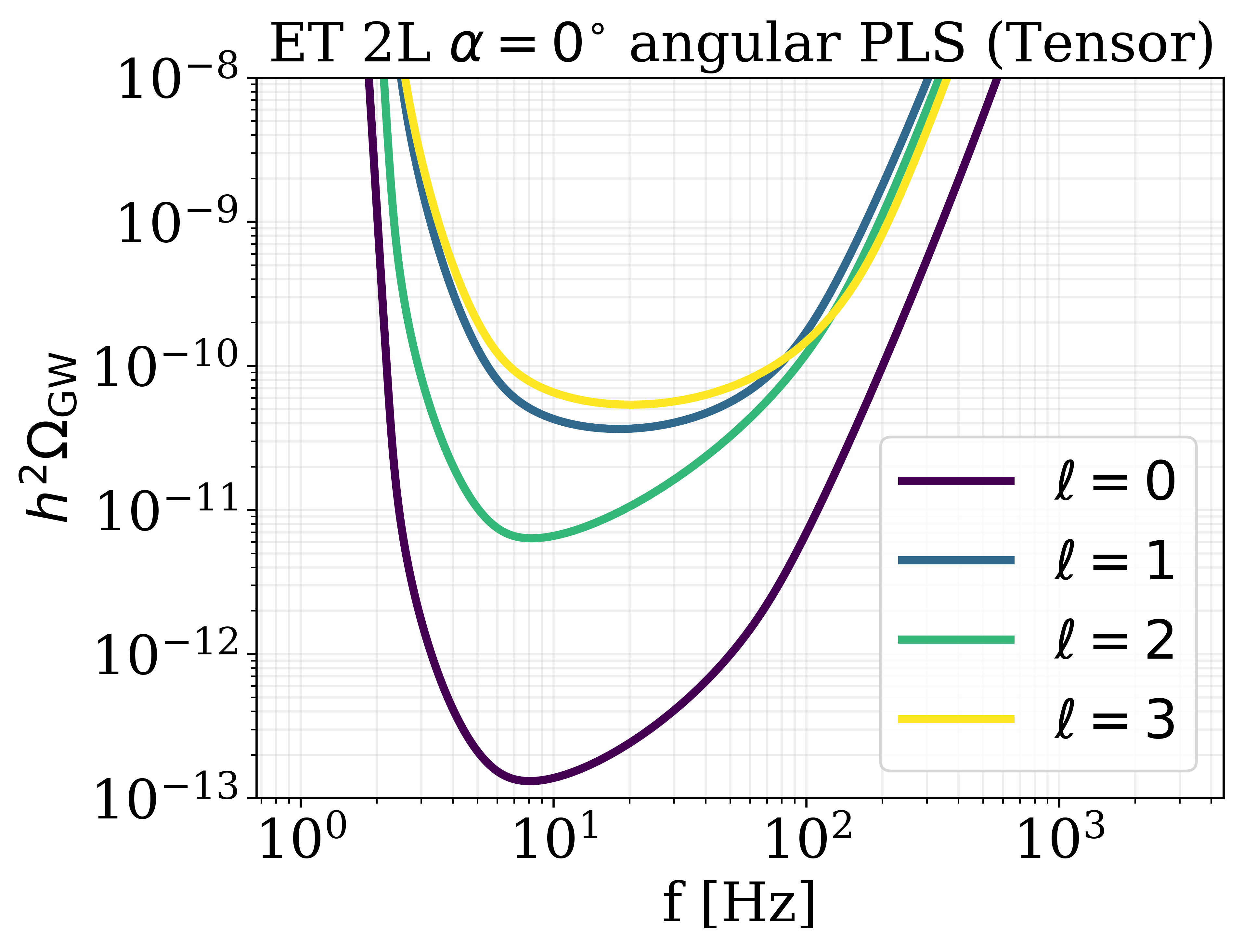}
    \includegraphics[scale=0.45]{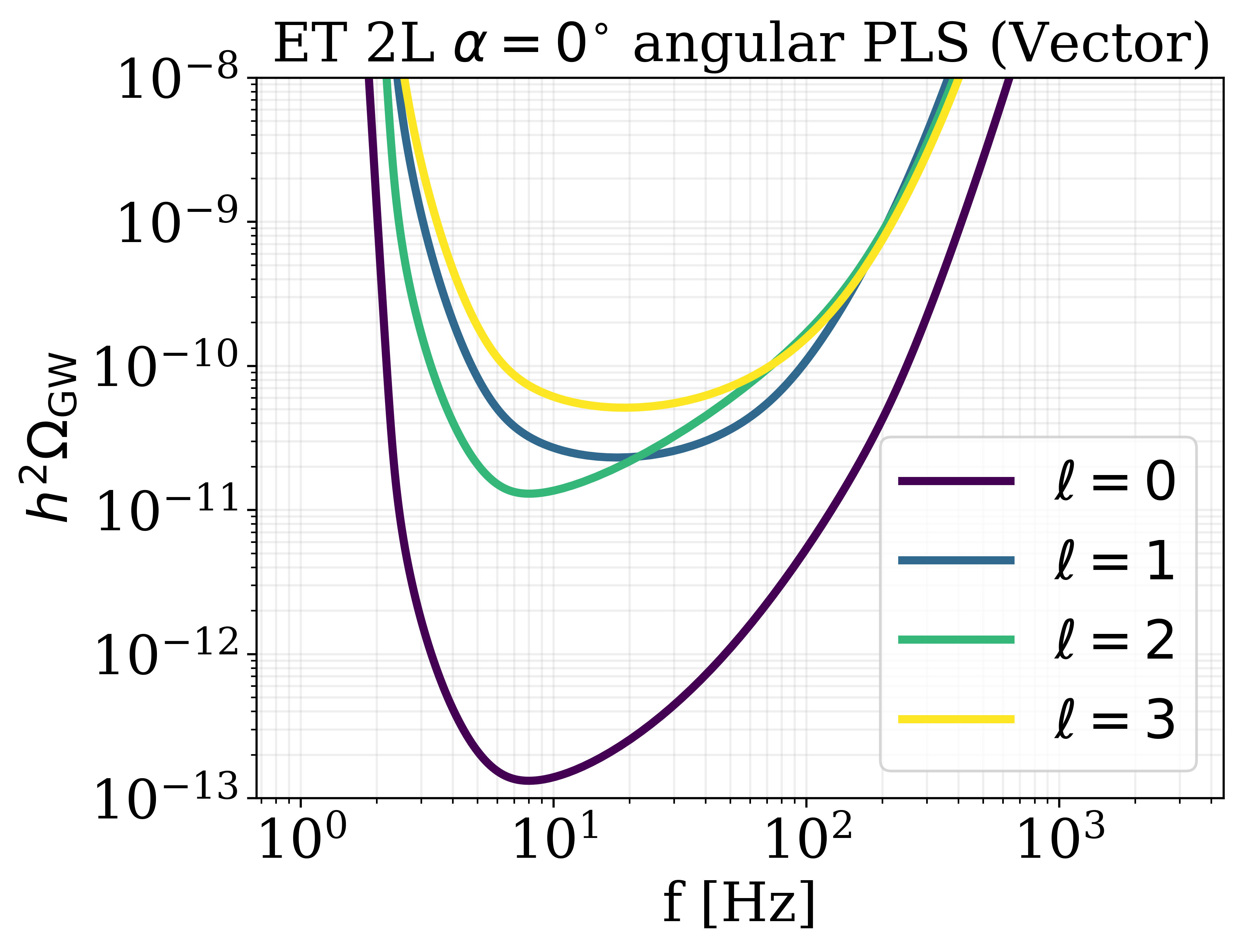}
    \includegraphics[scale=0.45]{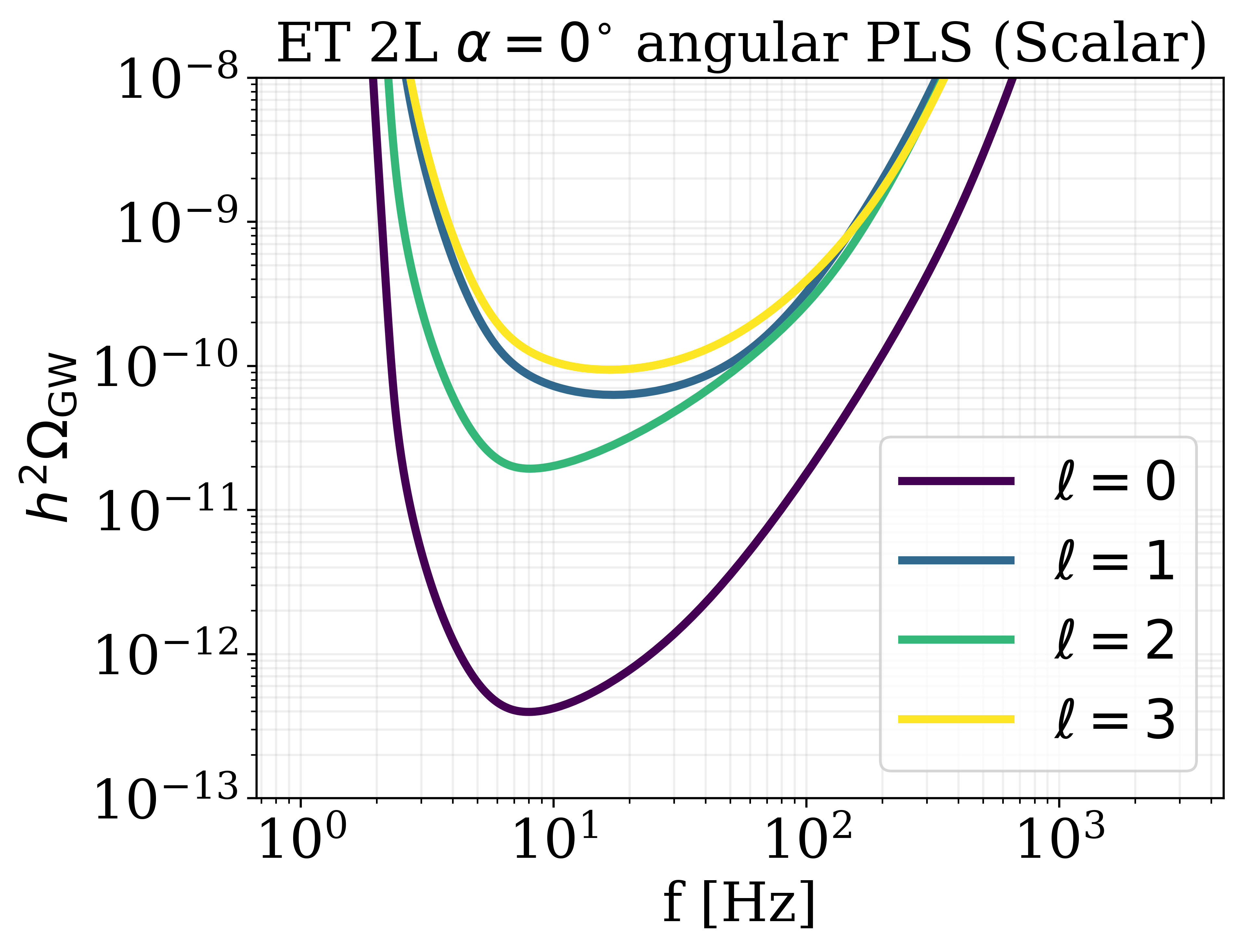}
    \includegraphics[scale=0.45]{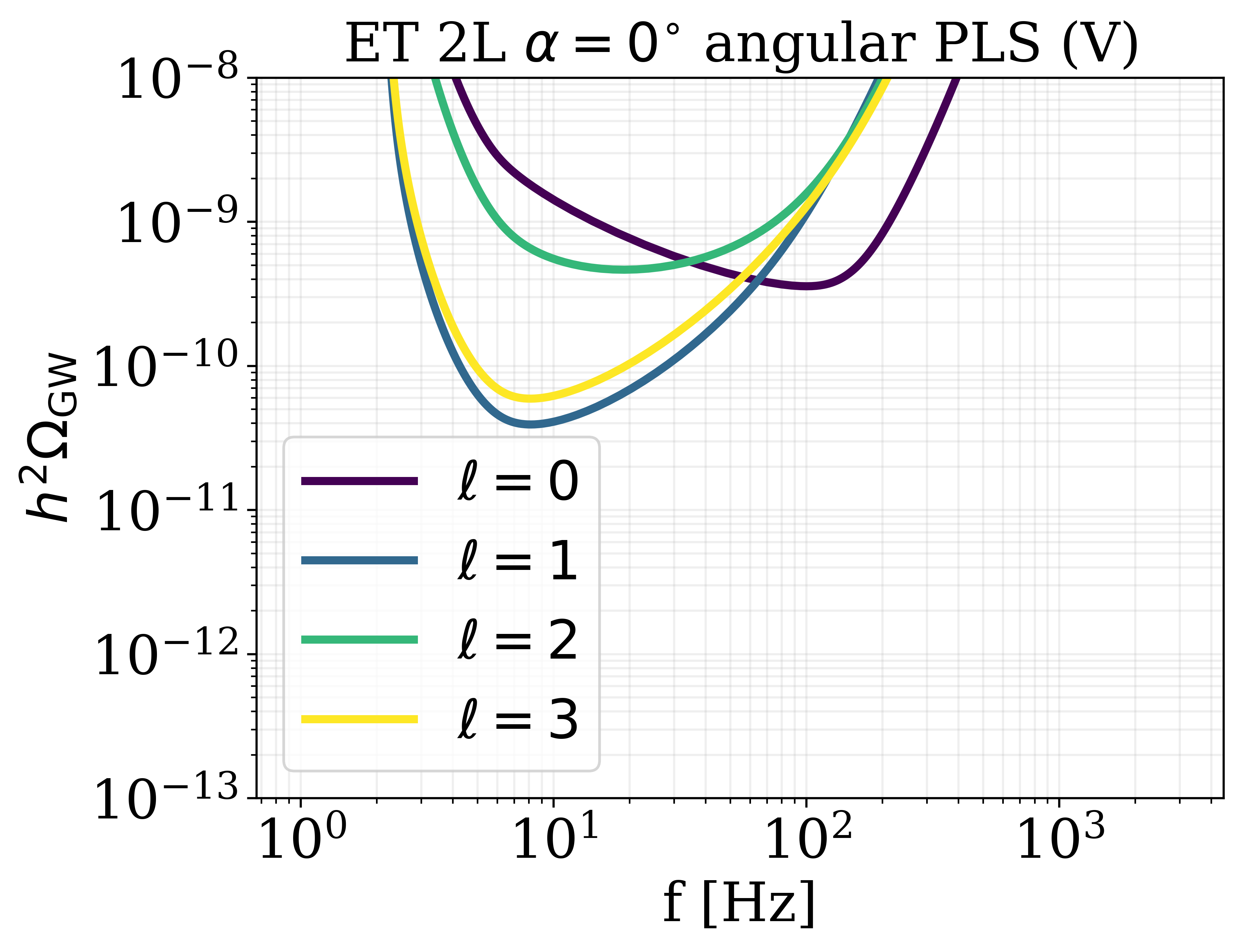}
    \caption{Angular power law integrated sensitivity curves for ET for $\ell = 0, 1, 2, 3$ in the 2L aligned configuration. We set $\langle\rm SNR \rangle_{\ell, th}=1$, $T= 1 {\rm yr}$, $C_{\ell = 1, 2, 3}^{GW}=10^{-3}$.}
    \label{fig:ET2L0_Nell}
\end{figure}

\begin{figure}[!ht]
    \centering
    \includegraphics[scale=0.45]{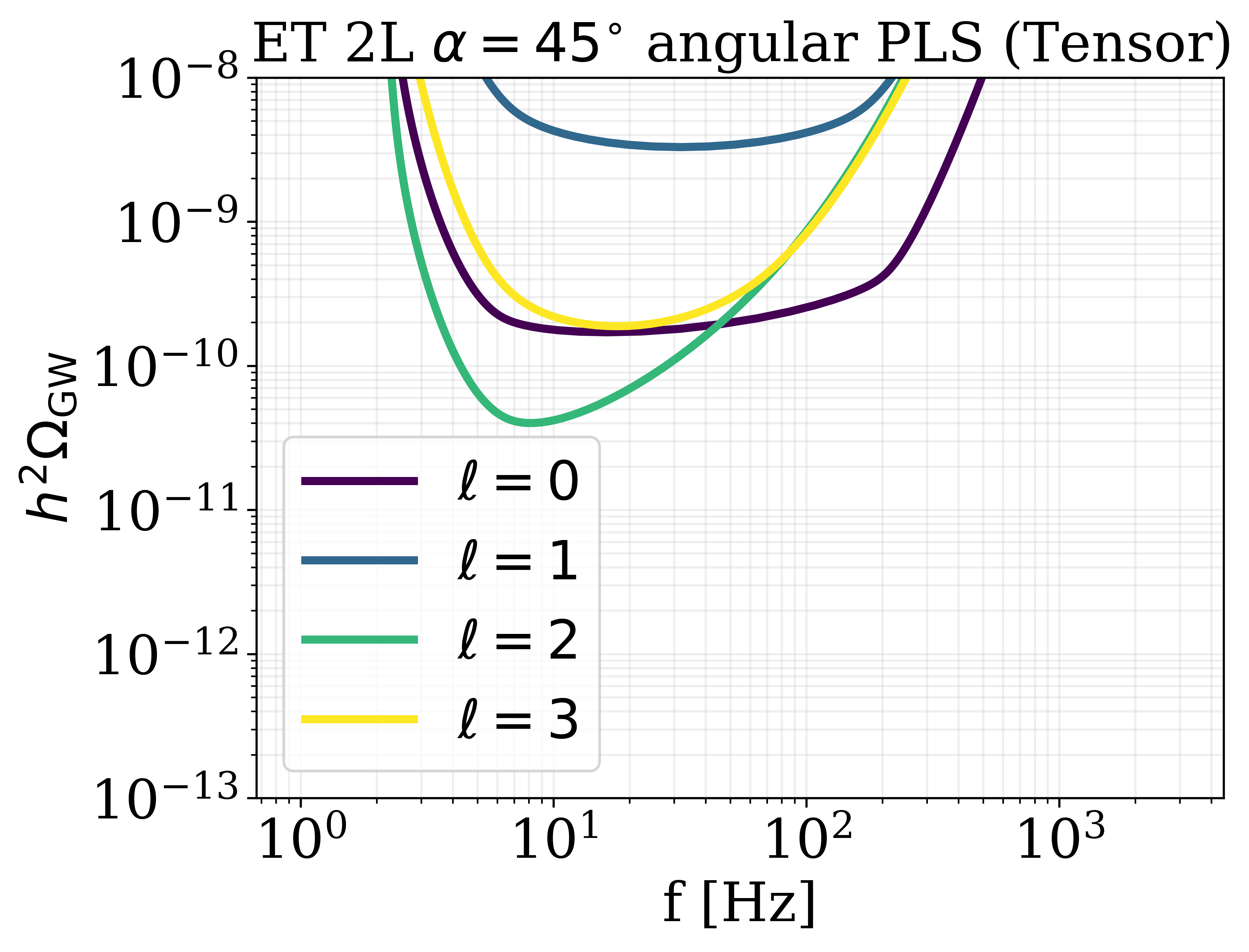}
    \includegraphics[scale=0.45]{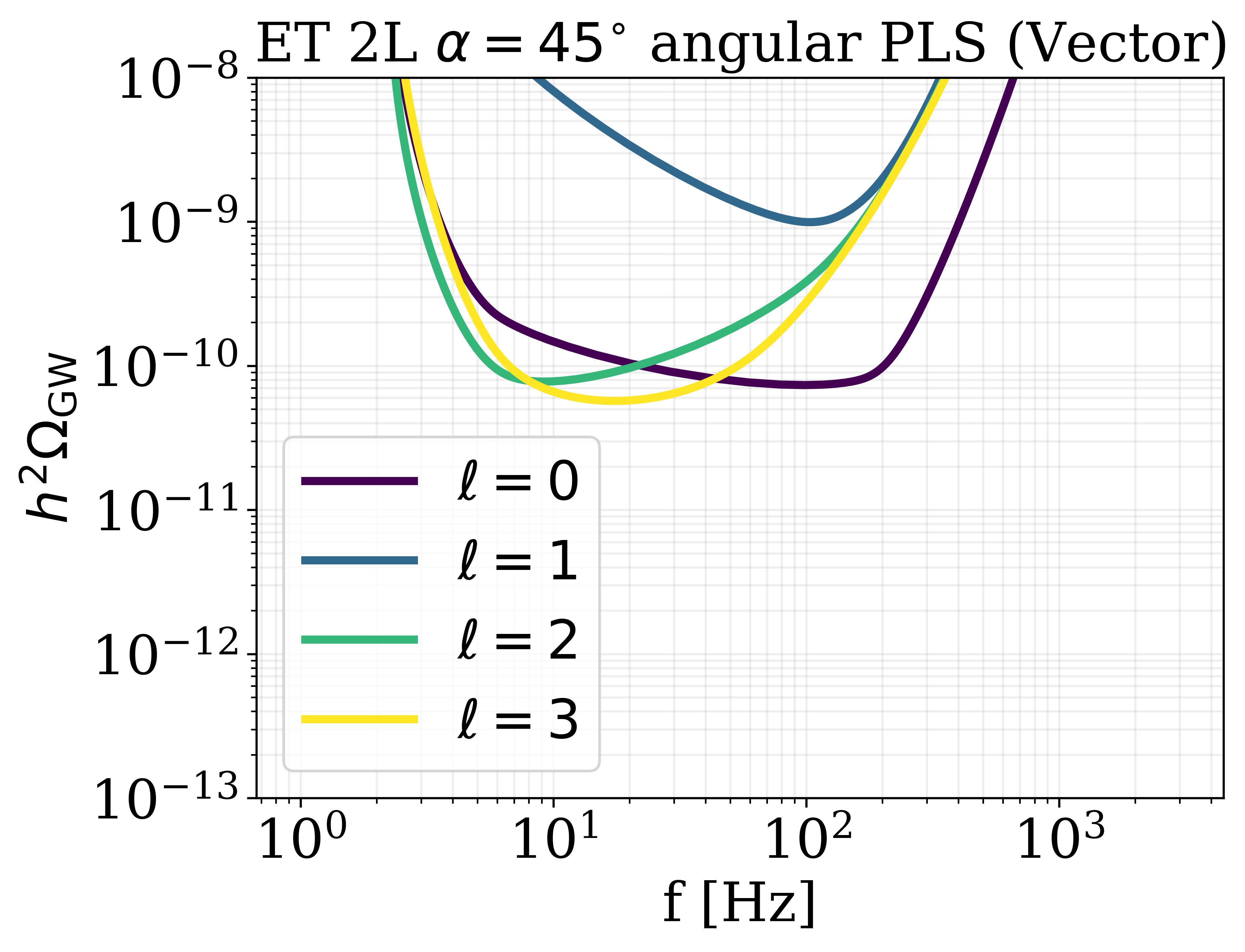}
    \includegraphics[scale=0.45]{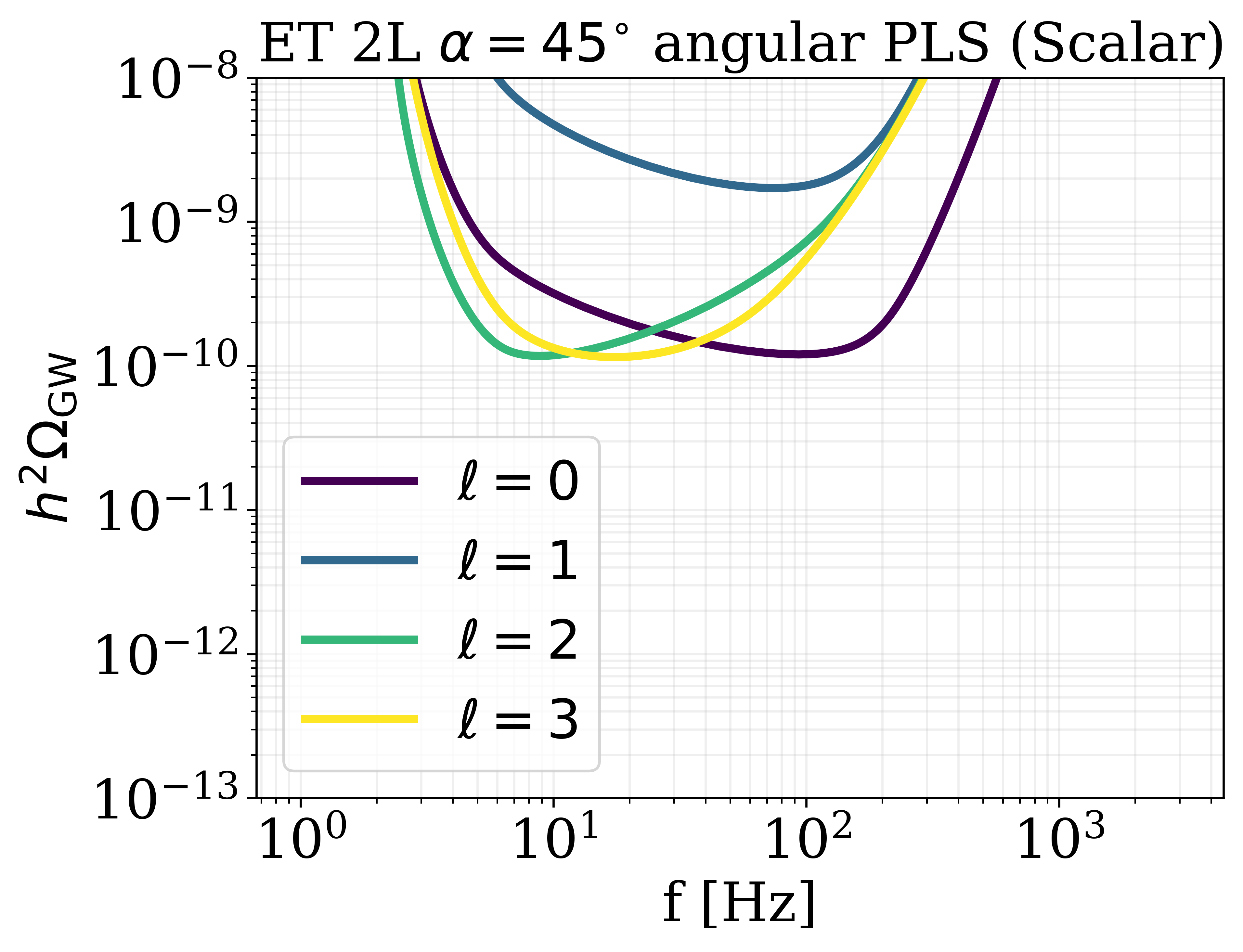}
    \includegraphics[scale=0.45]{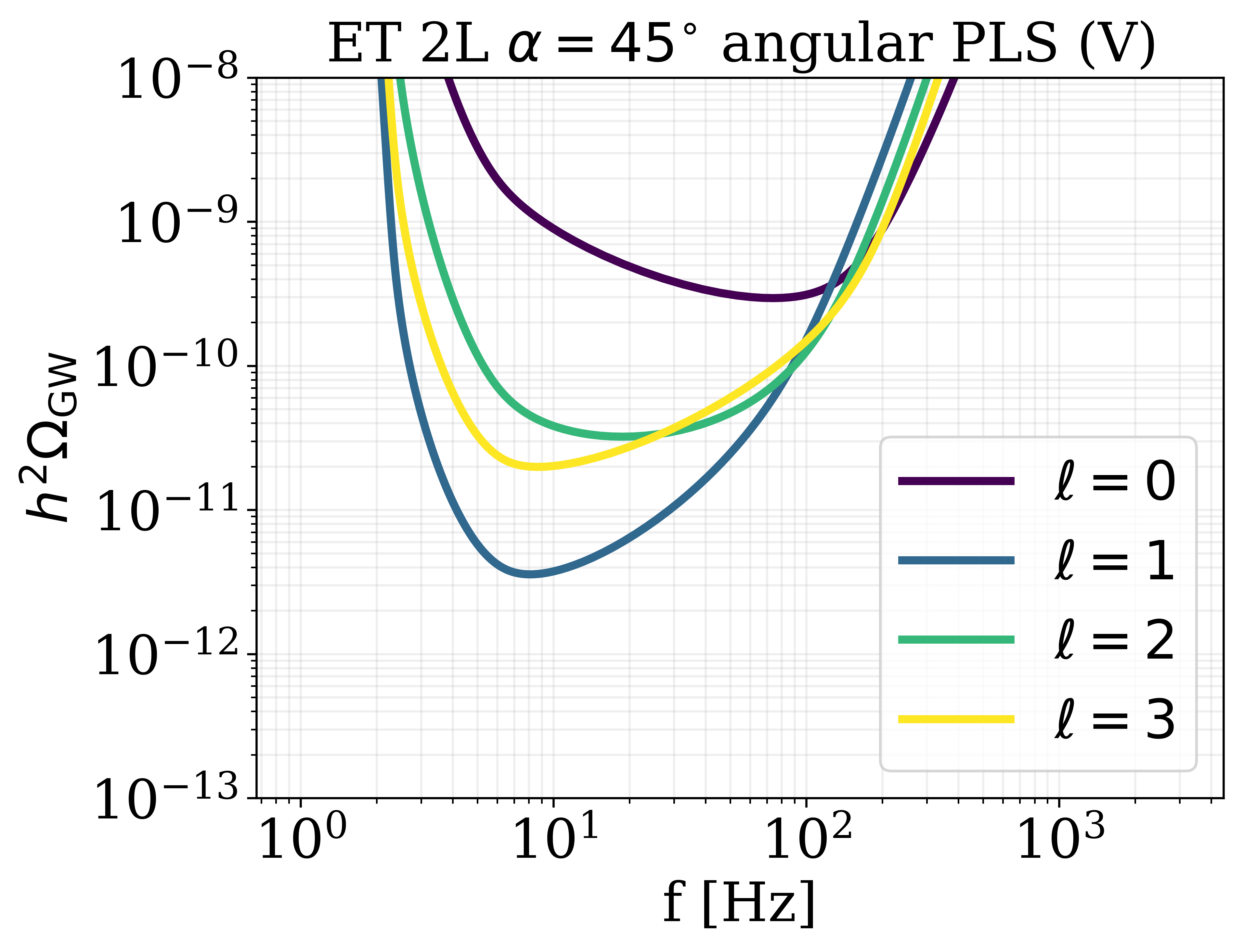}
    \caption{Angular power law integrated sensitivity curves for ET for $\ell = 0, 1, 2, 3$ in the 2L misaligned configuration. We set $\langle\rm SNR \rangle_{\ell, th}=1$, $T= 1 {\rm yr}$, $C_{\ell = 1, 2, 3}^{\rm GW}=10^{-3}$.}
    \label{fig:ET2L45_Nell}
\end{figure}

Having discussed the angular response for ET in all its configurations, we can now look at the sensitivity required for the monopole to spot anisotropies at a given multipole $\ell$, as in Figure~\ref{fig:ETtri_Nell},~\ref{fig:ET2L0_Nell},~\ref{fig:ET2L45_Nell}. As in Section~\ref{sensitivity to multipoles}, we set a reasonable value for $C_{\ell}^{\rm GW}$ that a specific mechanism could produce. Given this, the sensitivity curve indicates the required monopole value to detect that level of anisotropy with $\langle\rm SNR \rangle_{\ell, threshold}$ and a given observation time $T$.

For the triangular configuration case, we see that large amplitudes of the monopole are needed to detect the dipole, with respect to the octupole or quadrupole. Specifically, odd multipoles are disfavored in the search. This holds for tensor, vector and scalar polarization modes. For circular polarization mode instead, we find that the dipole and the octupole are actually favored in the search, compared to even multipoles. This can be easily understood considering the angular response functions in Figure~\ref{fig:ET_Rell}. This behavior is common to all the polarization modes, as it can be seen in Figure~\ref{fig:ETtri_Nell} and Table~\ref{tab:aplsminETtr}.

\begin{table}[!ht]
    \centering
    \begin{tabular}{ccccc}
    \multicolumn{5}{c}{ET $\Delta$} \\
    \hline
    \hline
    \multicolumn{5}{c}{$h^2\Omega_{\rm GW}$} \\
    \hline
                 & Tensor                & Vector                & Scalar                & V \\
    \hline
        $\ell=1$ & $8.14 \times 10^{-5}$  & $5.03 \times 10^{-5}$  & $1.38\times 10^{-4}$   & $1.03\times 10^{-11}$\\
                 & {\scriptsize$@ \, 19.3 \, \rm Hz$} & {\scriptsize$@ \, 19.3 \, \rm Hz$} & {\scriptsize$@ \, 19.3 \, \rm Hz$}  & {\scriptsize$@ \, 8.9 \, \rm Hz$} \\
        $\ell=2$ & $3.24 \times 10^{-11}$ & $6.47 \times 10^{-11}$ & $9.71\times 10^{-11}$  & $2.15\times 10^{-8}$\\
                 & {\scriptsize$@ \, 8.9 \, \rm Hz$} & {\scriptsize$@ \, 8.93\, \rm Hz$} & {\scriptsize$@ \, 8.9 \, \rm Hz$}  & {\scriptsize$@ \, 20.5 \, \rm Hz$} \\
        $\ell=3$ & $3.45 \times 10^{-8}$  & $2.42 \times 10^{-8}$  & $1.73\times 10^{-8}$   & $6.33\times 10^{-11}$\\
                 & {\scriptsize$@ \, 20.47 \, \rm Hz$} & {\scriptsize$@ \, 20.5 \, \rm Hz$} & {\scriptsize$@ \, 20.5 \, \rm Hz$}  & {\scriptsize$@ \, 8.9 \, \rm Hz$} \\
    \hline
    \end{tabular}
    \caption{Minimum energy density spectrum amplitude required ($\langle\rm SNR \rangle_{\ell, th}=1$, $T=1 {\rm yr}$ and  $C_{\ell=1,2,3}=10^{-3}$) for a SGWB made of tensor, vector, scalar or circular polarization modes to spot for anisotropies.}
    \label{tab:aplsminETtr}
\end{table}

In the 2L aligned configuration, we see that for tensor modes the behavior is analogous to the triangular case, while instead vector polarization modes show that the monopole energy density required to spot for the dipole and the quadrupole is comparable, despite having a different spectral shape, as it can be seen from Figure~\ref{fig:ET2L0_Nell} and Table~\ref{tab:aplsminET2L0}.

\begin{table}[!ht]
    \centering
    \begin{tabular}{ccccc}
    \multicolumn{5}{c}{ET 2L $\alpha=0^{\circ}$} \\
    \hline
    \hline
    \multicolumn{5}{c}{$h^2\Omega_{\rm GW}$} \\
    \hline
                 & Tensor                & Vector               & Scalar               & V \\
    \hline
        $\ell=1$ & $3.64 \times 10^{-11}$  & $2.31 \times 10^{-11}$    & $6.27\times 10^{-11}$    & $3.91\times 10^{-11}$  \\
                 & {\scriptsize$@ \, 18.1 \, \rm Hz$} & {\scriptsize$@ \, 17.9 \, \rm Hz$} & {\scriptsize$@ \, 17.2 \, \rm Hz$}  & {\scriptsize$@ \, 8.1 \, \rm Hz$} \\
        $\ell=2$ & $6.35 \times 10^{-12}$  & $1.29 \times 10^{-11}$    & $1.93\times 10^{-11}$    & $4.64\times 10^{-10}$  \\
                 & {\scriptsize$@ \, 8.2 \, \rm Hz$} & {\scriptsize$@ \, 8.0 \, \rm Hz$} & {\scriptsize$@ \, 8.1 \, \rm Hz$}  & {\scriptsize$@ \, 18.9 \, \rm Hz$} \\
        $\ell=3$ & $5.36 \times 10^{-11}$  & $5.11 \times 10^{-11}$    & $9.37\times 10^{-11}$    & $5.91\times 10^{-11}$  \\
                 & {\scriptsize$@ \, 20.2 \, \rm Hz$} & {\scriptsize$@ \, 19.1 \, \rm Hz$} & {\scriptsize$@ \, 16.7 \, \rm Hz$}  & {\scriptsize$@ \, 8.1 \, \rm Hz$} \\
    \hline
    \end{tabular}
    \caption{Minimum energy density spectrum amplitude required ($\langle\rm SNR \rangle_{\ell, th}=1$, $T=1{\rm yr}$ and  $C_{\ell=1,2,3}=10^{-3}$) for a SGWB made of tensor, vector, scalar or circular polarization modes to spot for anisotropies.}
    \label{tab:aplsminET2L0}
\end{table}

Lastly, in the 2L misaligned configuration, as shown in Figure~\ref{fig:ET2L45_Nell} and Table~\ref{tab:aplsminET2L45}, we observe that, among the multipoles considered, the dipole generally requires a higher monopole amplitude to be detectable across all polarization modes. However, for tensor modes, the network is more sensitive to the quadrupole, assuming $C^{\ell=2}_{\rm GW} = 10^{-3}$, than to other multipoles. This trend, although less pronounced, is also present in the vector and scalar polarization modes. In the case of circular polarization modes, the network appears to be more sensitive to anisotropies, under the given assumptions, than to the monopole.

\begin{table}[!ht]
    \centering
    \begin{tabular}{ccccc}
    \multicolumn{5}{c}{ET 2L $\alpha=45^{\circ}$} \\
    \hline
    \hline
    \multicolumn{5}{c}{$h^2\Omega_{\rm GW}$} \\
    \hline
                 & Tensor               & Vector               & Scalar                & V \\
    \hline
        $\ell=1$ & $3.29 \times 10^{-9}$   & $9.91 \times 10^{-10}$     & $1.71\times 10^{-9}$    & $3.57\times 10^{-12}$ \\
                 & {\scriptsize$@ \, 31.6 \, \rm Hz$} 
                 & {\scriptsize$@ \, 103.6 \, \rm Hz$} 
                 & {\scriptsize$@ \, 74.6 \, \rm Hz$}  
                 & {\scriptsize$@ \, 8.1 \, \rm Hz$} \\
        $\ell=2$ & $4.00 \times 10^{-11}$  & $7.76 \times 10^{-11}$    & $1.17\times 10^{-10}$     & $3.21\times 10^{-11}$  \\
                 & {\scriptsize$@ \, 8.1 \, \rm Hz$} & {\scriptsize$@ \, 9.2 \, \rm Hz$} & {\scriptsize$@ \, 8.8 \, \rm Hz$}  & {\scriptsize$@ \, 18.8 \, \rm Hz$} \\
        $\ell=3$ & $1.88 \times 10^{-10}$  & $5.69 \times 10^{-11}$    & $1.15\times 10^{-10}$     & $1.98 \times 10^{-11}$  \\
                 & {\scriptsize$@ \, 17.8 \, \rm Hz$} & {\scriptsize$@ \, 17.5 \, \rm Hz$} & {\scriptsize$@ \, 17.5 \, \rm Hz$}  & {\scriptsize$@ \, 8.7 \, \rm Hz$} \\
    \hline
    \end{tabular}
    \caption{Minimum energy density spectrum amplitude required ($\langle\rm SNR \rangle_{\ell, th}=1$, $T=1{\rm yr}$ and  $C_{\ell=1,2,3}=10^{-3}$) for a SGWB made of tensor, vector, scalar or circular polarization modes to spot for anisotropies.}
    \label{tab:aplsminET2L45}
\end{table}

\subsubsection{LISA}

\begin{figure}[t!]
    \centering
    \includegraphics[scale=0.45]{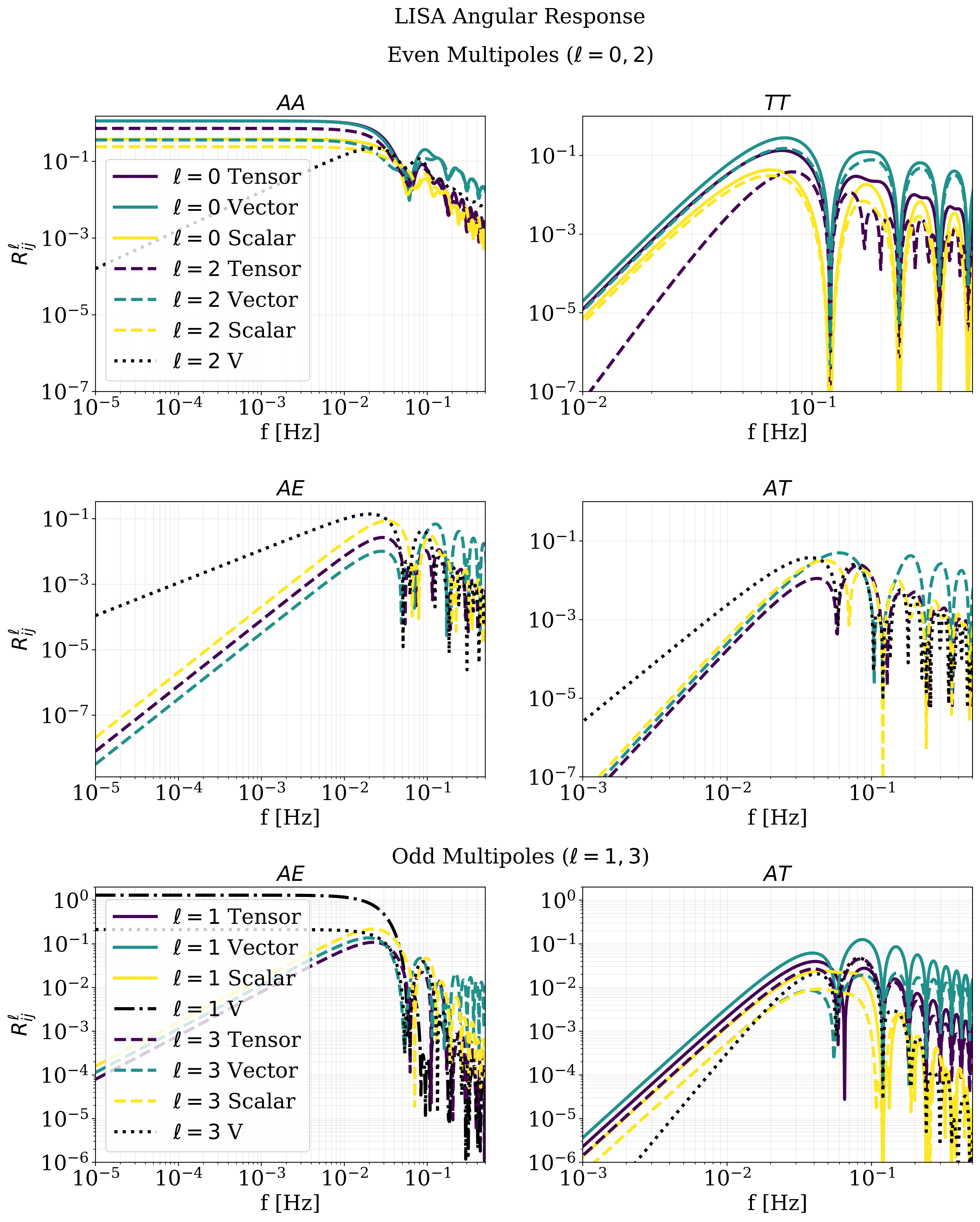}
    \caption{LISA angular response function in the AET basis for the different polarization modes for $\ell = 0, 2$ (even modes) on top and $\ell = 1, 3$ (odd modes) on the bottom. Purple lines: tensor modes. Turquoise lines: vector modes. Yellow lines: scalar modes. Black lines: circular polarization. }
    \label{fig:LISA_Rl}
\end{figure}

In Figure~\ref{fig:LISA_Rl}, we show the angular response function for each polarization mode, computed for LISA in the AET basis. The first four multipoles are plotted in two groups—even ($\ell = 0, 2$) and odd ($\ell = 1, 3$)—and within each group, results are shown for the AA, TT, AE, and AT channels.\footnote{We do not show the EE and ET channels, as they are equal to AA and AT channels, respectively.}.

Aside from an overall normalization constant, our results agree with those in~\cite{LISACosmologyWorkingGroup:2022kbp} for tensor modes. This normalization constant, however, does not affect the sensitivity, as it is correctly accounted for in the definition of eq.~\eqref{mean_SNR}.

It is important to emphasize that the basis is no longer diagonal when considering multipoles higher than the monopole.

For even multipoles, the AA channel exhibits a plateau followed by a sudden onset of oscillatory suppression around $f \sim 10^{-2}, \rm Hz$, consistently across tensor, vector, and scalar polarization modes. The circular polarization mode for $\ell = 0$ in this channel is consistent with zero, as discussed in Section~\ref{sec:overlapLISA}. For $\ell = 2$, on the other hand, the response shows an approximate broken power-law behavior, again featuring an oscillatory drop near $f \sim 10^{-2}, \rm Hz$. In contrast, the TT, AT, and AE channels exhibit a broken power-law behavior for tensor, vector and scalar polarization modes. Circular polarization mode exhibits the broken power law behavior for AE and AT channels, while the TT channel is insensitive to it for $\ell=0,2$.

For odd multipoles, only the AE and AT channels are non-vanishing. These channels again show a broken power-law behavior with the oscillatory drop for tensor, vector, and scalar polarization modes. In the case of circular polarization, the AE channel shows a low-frequency plateau followed by a drop after $f \sim 10^{-2}, \rm Hz$ for both $\ell = 1$ and $\ell = 3$. For the AT channel, this behavior is present for $\ell = 3$, while for $\ell = 1$, the amplitude remains below $10^{-16}$, making it effectively null.

\begin{figure}[!ht]
    \centering
    \includegraphics[scale=0.45]{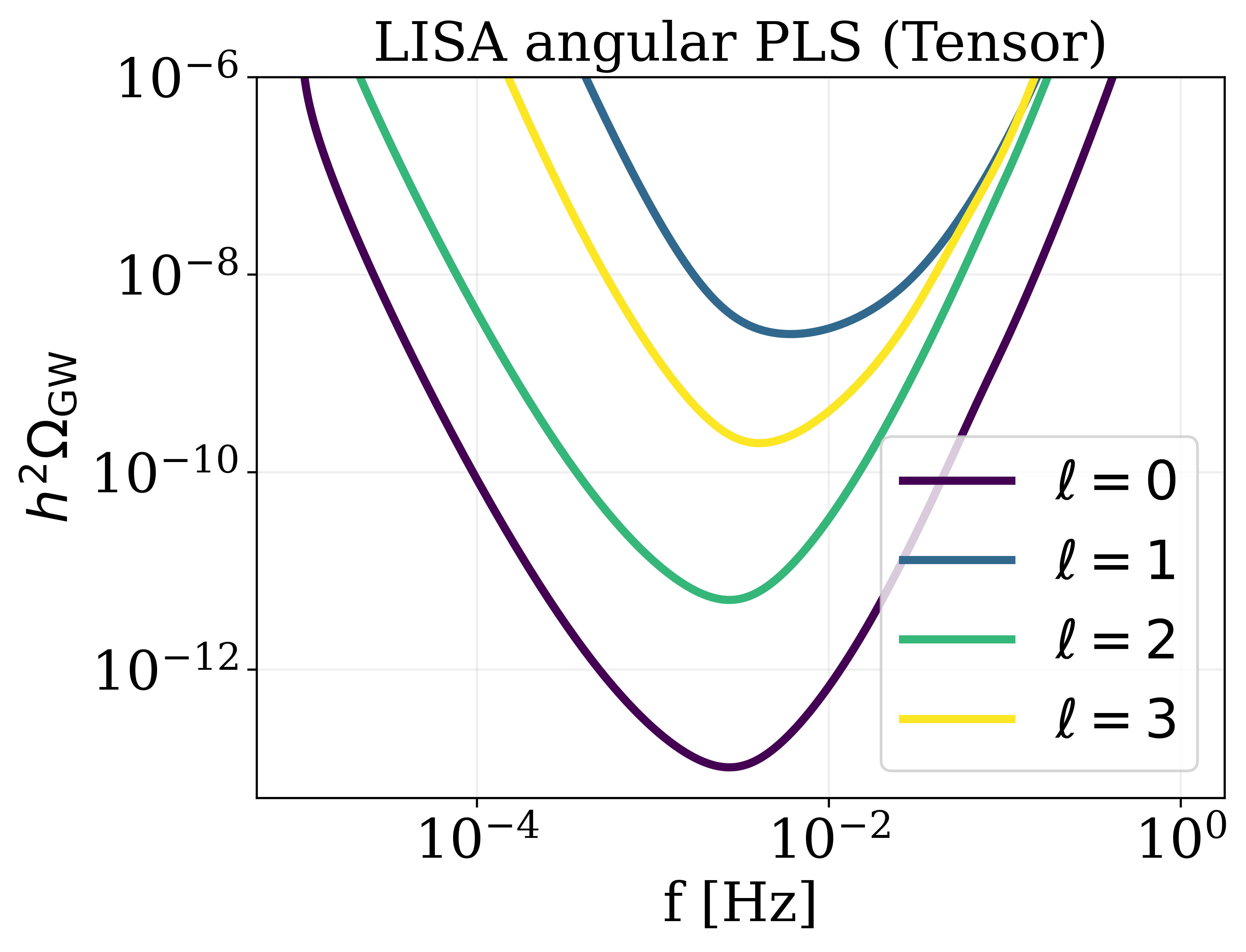}
    \includegraphics[scale=0.45]{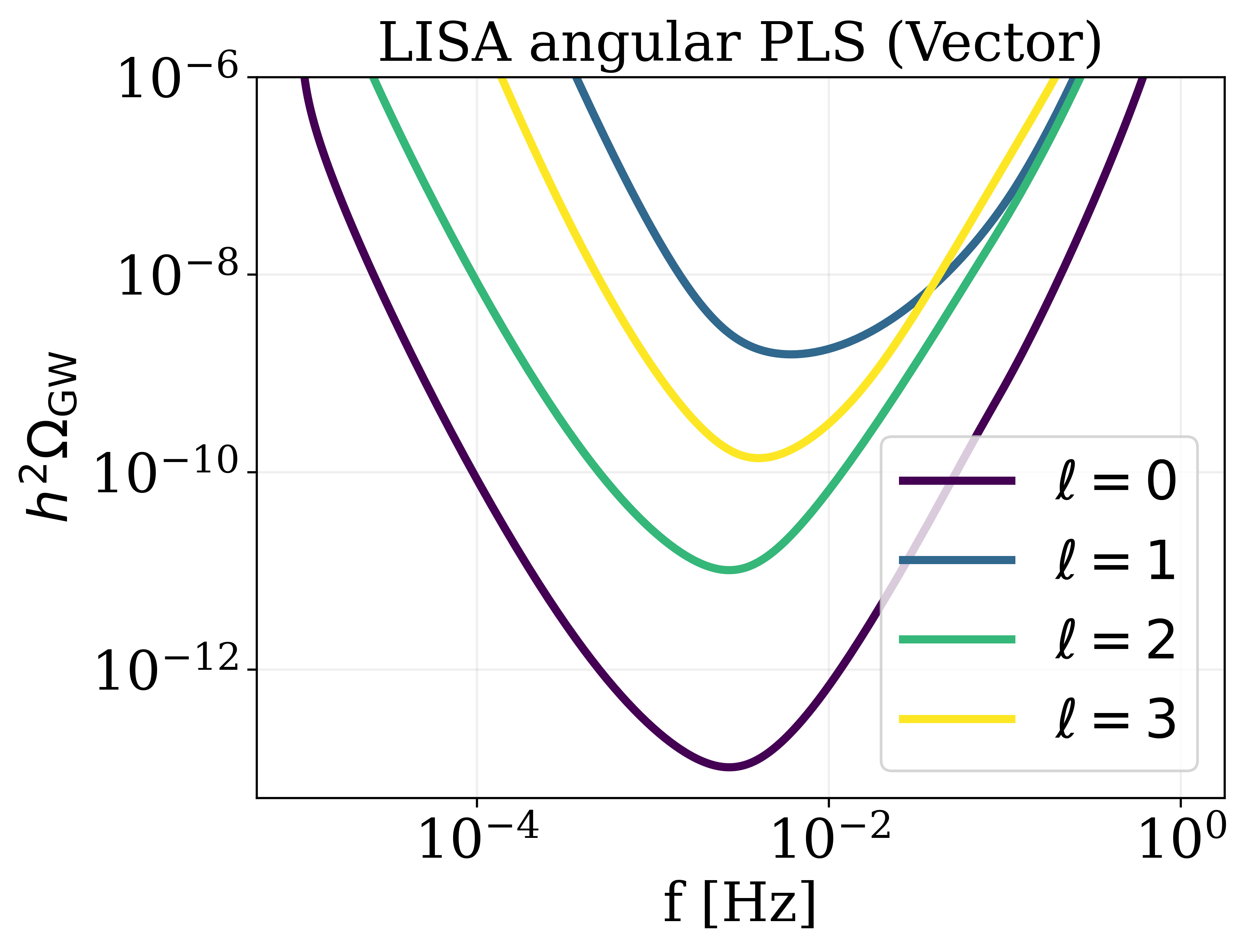}
    \includegraphics[scale=0.45]{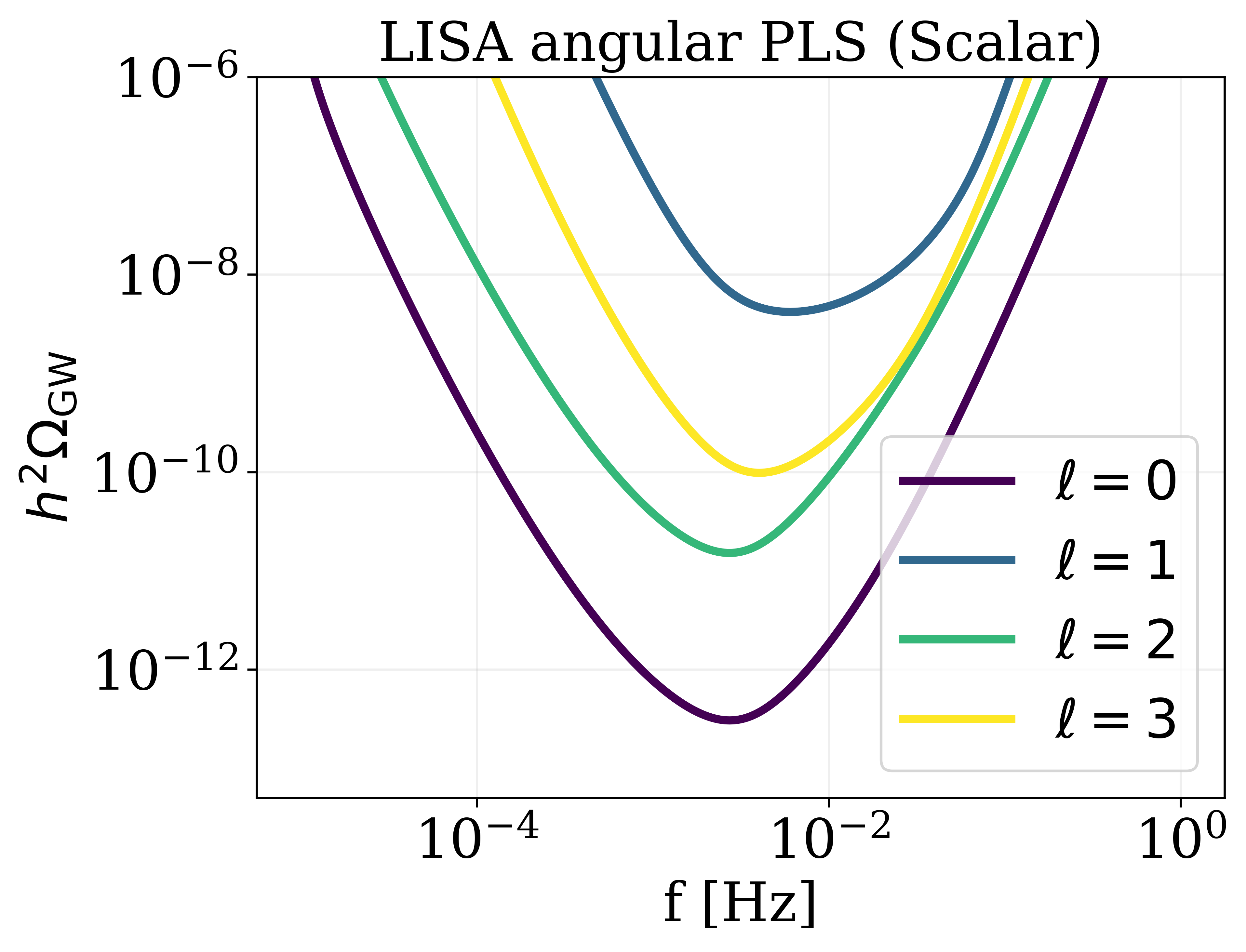}
    \includegraphics[scale=0.45]{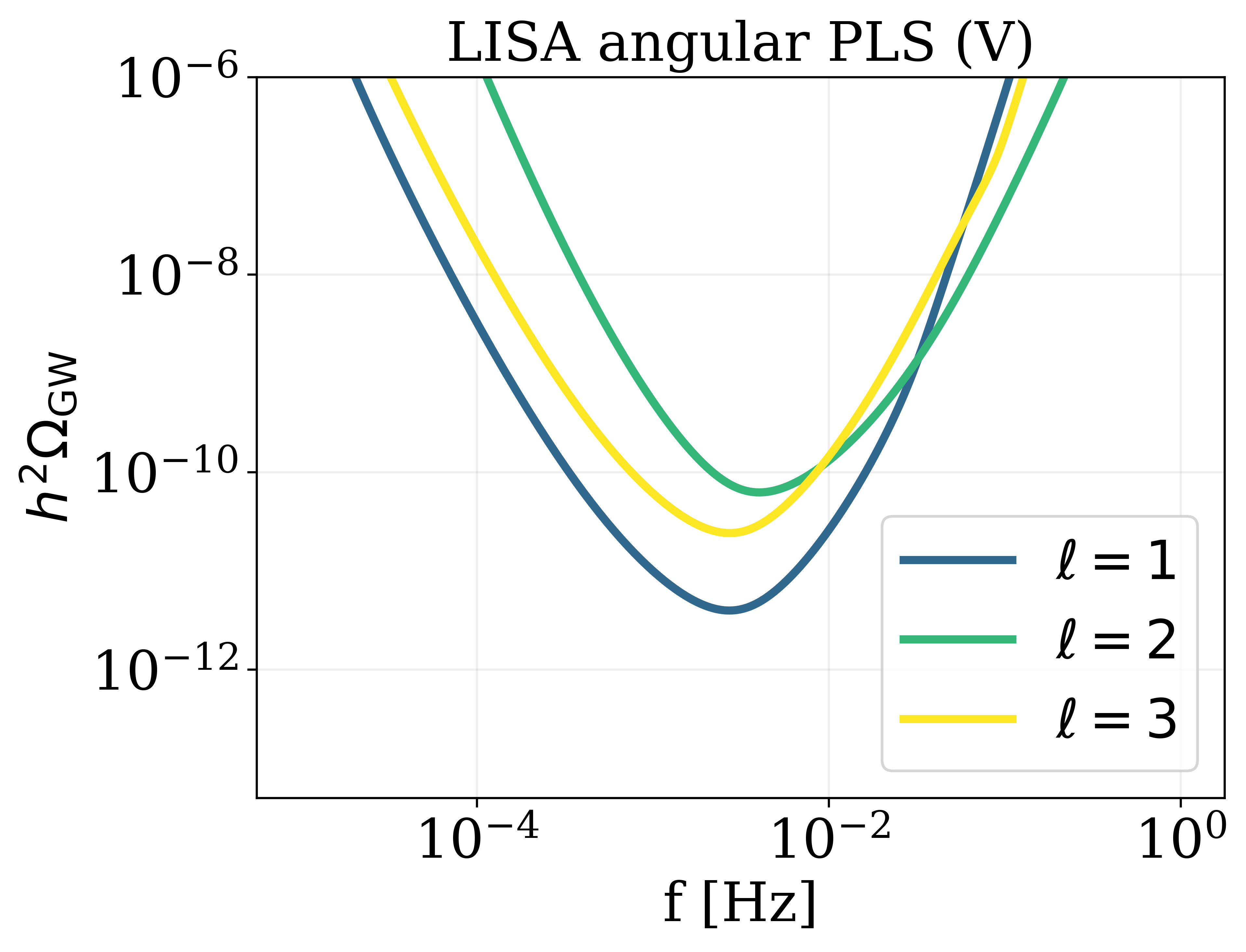}
    \caption{Angular sensitivity for LISA for $\ell = 0, 1, 2, 3$. We set$\langle\rm SNR \rangle_{\ell, th}=10$, $T= 3 {\rm yrs}$, $C_{\ell = 1, 2, 3}^{GW}=10^{-3}$.}
    \label{fig:LISA_Nell}
\end{figure}

As a next step, we present the APLS for LISA \footnote{In this case, the sum is performed over all possible channel combinations corresponding to each multipole.}. The results for different polarizations modes are shown in Figure~\ref{fig:LISA_Nell}. Assuming $C_{\ell = 1, 2, 3}^{\rm GW} = 10^{-3}$, we observe that the detector exhibits greater sensitivity to even multipoles for tensor, vector, and scalar polarization modes. Among these polarizations and the multipoles considered, the dipole appears to be the one to which LISA is less sensitive, as illustrated in Table~\ref{tab:aplsminLISA} for isotropic SGWB. In contrast, for the circular polarization mode, LISA requires a lower monopole amplitude to detect the dipole compared to the other two multipoles, under such assumptions.

\begin{table}[!ht]
    \centering
    \begin{tabular}{ccccc}
    \multicolumn{5}{c}{LISA} \\
    \hline
    \hline
    \multicolumn{5}{c}{$h^2\Omega_{\rm GW}$} \\
    \hline
                 & Tensor                & Vector               & Scalar                     & V  \\
    \hline
        $\ell=1$ & $2.49 \times 10^{-9}$   & $1.55 \times 10^{-9}$     & $4.17\times 10^{-9}$   & $3.95 \times 10^{-12}$ \\
                 & {\scriptsize$@ \, 6.06 \times 10^{-3} \, \rm Hz$} & {\scriptsize$@ \, 6.13 \times 10^{-3}  \, \rm Hz$} & {\scriptsize$@ \, 6.06 \times 10^{-3}  \, \rm Hz$} & {\scriptsize$@ \, 2.74 \times 10^{-3}  \, \rm Hz$} \\
        $\ell=2$ & $5.05 \times 10^{-12}$  & $1.01 \times 10^{-11}$    & $1.51\times 10^{-11}$  & $6.21 \times 10^{-11}$     \\
                 &{\scriptsize$@ \, 2.74 \times 10^{-3} \, \rm Hz$} 
                 &{\scriptsize$@ \, 2.71 \times 10^{-3}  \, \rm Hz$} &{\scriptsize$@ \, 2.74 \times 10^{-3}  \, \rm Hz$} &{\scriptsize$@ \, 4.05 \times 10^{-3}  \, \rm Hz$} \\
        $\ell=3$ & $1.96 \times 10^{-10}$  & $1.38 \times 10^{-10}$    & $9.80\times 10^{-11}$  & $2.41 \times 10^{-11}$     \\
                 &{\scriptsize$@ \, 4.01 \times 10^{-3} \, \rm Hz$}
                 &{\scriptsize$@ \, 4.01 \times 10^{-3}  \, \rm Hz$} &{\scriptsize$@ \, 4.01 \times 10^{-3}  \, \rm Hz$}  &{\scriptsize$@ \, 2.74 \times 10^{-3}  \, \rm Hz$} \\
    \hline
    \end{tabular}
    \caption{Minimum energy density spectrum amplitude required ($\langle\rm SNR \rangle_{\ell, th}=10$, $T= 3 {\rm yrs}$ and  $C_{\ell=1,2,3}^{\rm GW}=10^{-3}$) for a SGWB made of tensor, vector, scalar or circular polarization modes to detect anisotropies.}
    \label{tab:aplsminLISA}
\end{table}

\subsubsection{PTA}
\label{sec:PTAanisotropies}

\begin{figure}[H] 
    \centering
    \includegraphics[scale=0.45]{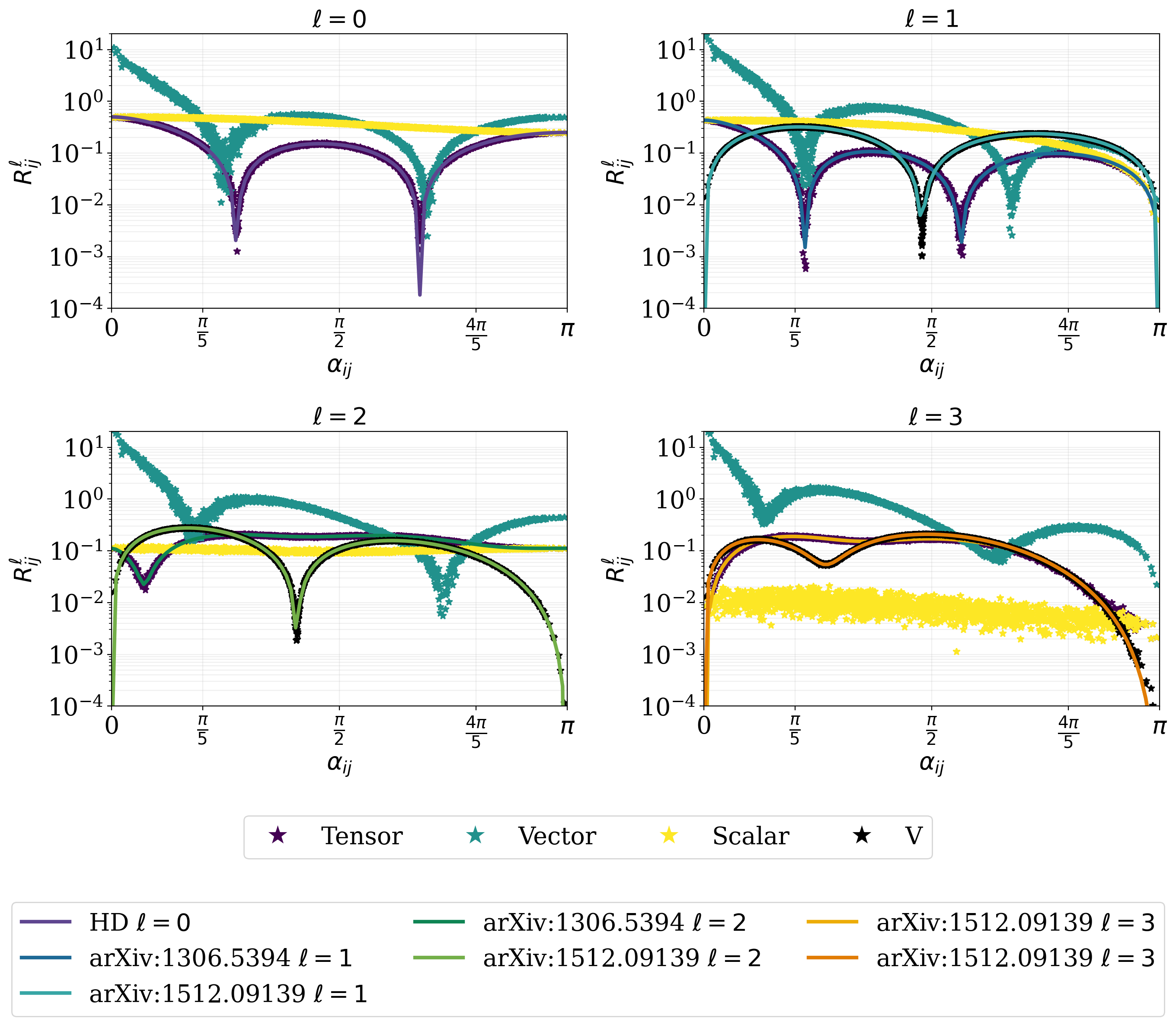}
    \caption{Angular response for the pair of pulsars in the catalog considered for different multipoles. For $\ell=0$ our results match the Hellings-Downs curve~\cite{Hellings:1983fr}. For $\ell=1,2$, in the case of tensor polarization, our results match~\cite{Mingarelli:2013dsa} (given our prescription in eq. \eqref{R_ell}). For $\ell=3$ for tensor and $\ell=1,2,3$ for circular polarization our results match with~\cite{Kato:2015bye} (given our prescription in eq.\eqref{R_ell}.}
    \label{fig:HDell}
\end{figure}

Finally, we present the angular response function for the PTA, computed from the same pulsar catalog. In Figure~\ref{fig:HDell}, the response is plotted against the separation angle between pulsar pairs at a fixed frequency  of $f = 10^{-8}\,\mathrm{Hz}$ \footnote{We again point out that this has an impact only on the vector modes among the ones considered in the analysis, as pointed out in Section \ref{sec:PTA_orf}. For all the transverse polarization modes, the angular overlap reduction function and therefore, the angular responses, can be assumed independent of the frequency.}, analogous to the overlap reduction function shown in Figure~\ref{fig:HD_PTA}.

For completeness, we compare our results with the analytical expressions presented in~\cite{Mingarelli:2013dsa} (as well as in~\cite{Gair:2014rwa}) and~\cite{Kato:2015bye}. Although the two calculations are performed in different reference frames — ours in the so-called cosmic rest frame and those in ~\cite{Mingarelli:2013dsa, Gair:2014rwa, Kato:2015bye, Sato-Polito:2021efu} in the computational frame — they agree when considering eq.~\eqref{R_ell}. This agreement arises because eq.~\eqref{R_ell} is invariant under rotations, and since transforming from one frame to the other involves a rotation, the results obtained in the two frames must coincide.

We note that our approach differs from~\cite{Kato:2015bye, Mingarelli:2013dsa, Gair:2014rwa, Sato-Polito:2021efu} in two ways: we perform the evaluation numerically in the cosmic rest frame rather than the usual computational frame~\cite{Mingarelli:2013dsa}, and we include the pulsar terms. The choice of including the pulsar terms come from the fact that when longitudinal polarization modes are considered (as for instance, vector polarization mode), these terms cannot be neglected \cite{Chamberlin:2011ev, Lee_2008}. Therefore, for the sake of consistency, we include them for all the polarization modes considered. 

We obtain excellent agreement for $\ell = 1, 2, 3$ for tensor and circular polarization (\cite{Mingarelli:2013dsa, Kato:2015bye}), confirming the reliability of our integrator.

We notice the high sensitivity of the network to vector polarization mode for all the multipole considered. 

\begin{figure}[t!]
    \centering
    \includegraphics[scale=0.45]{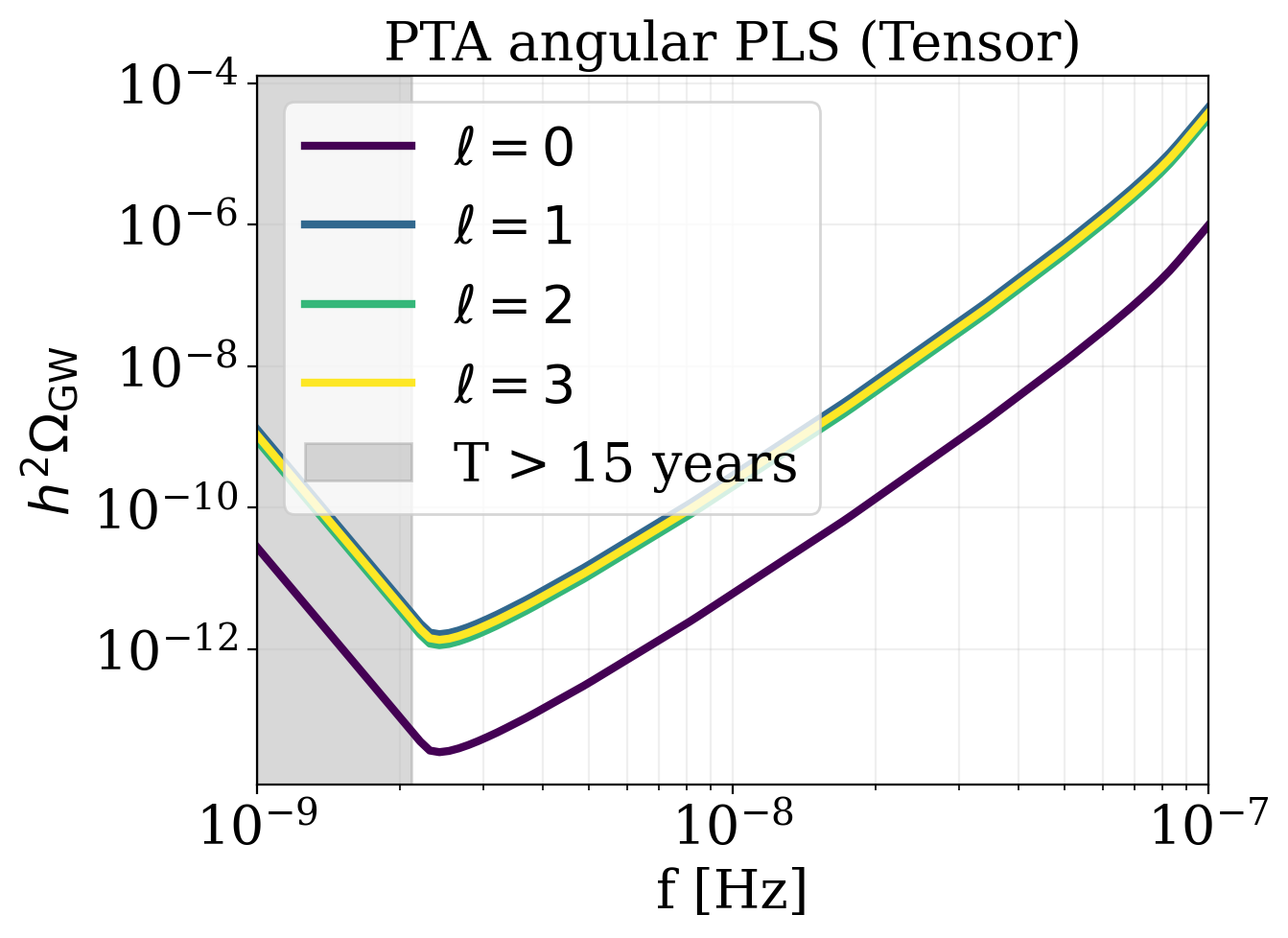}
    \includegraphics[scale=0.45]{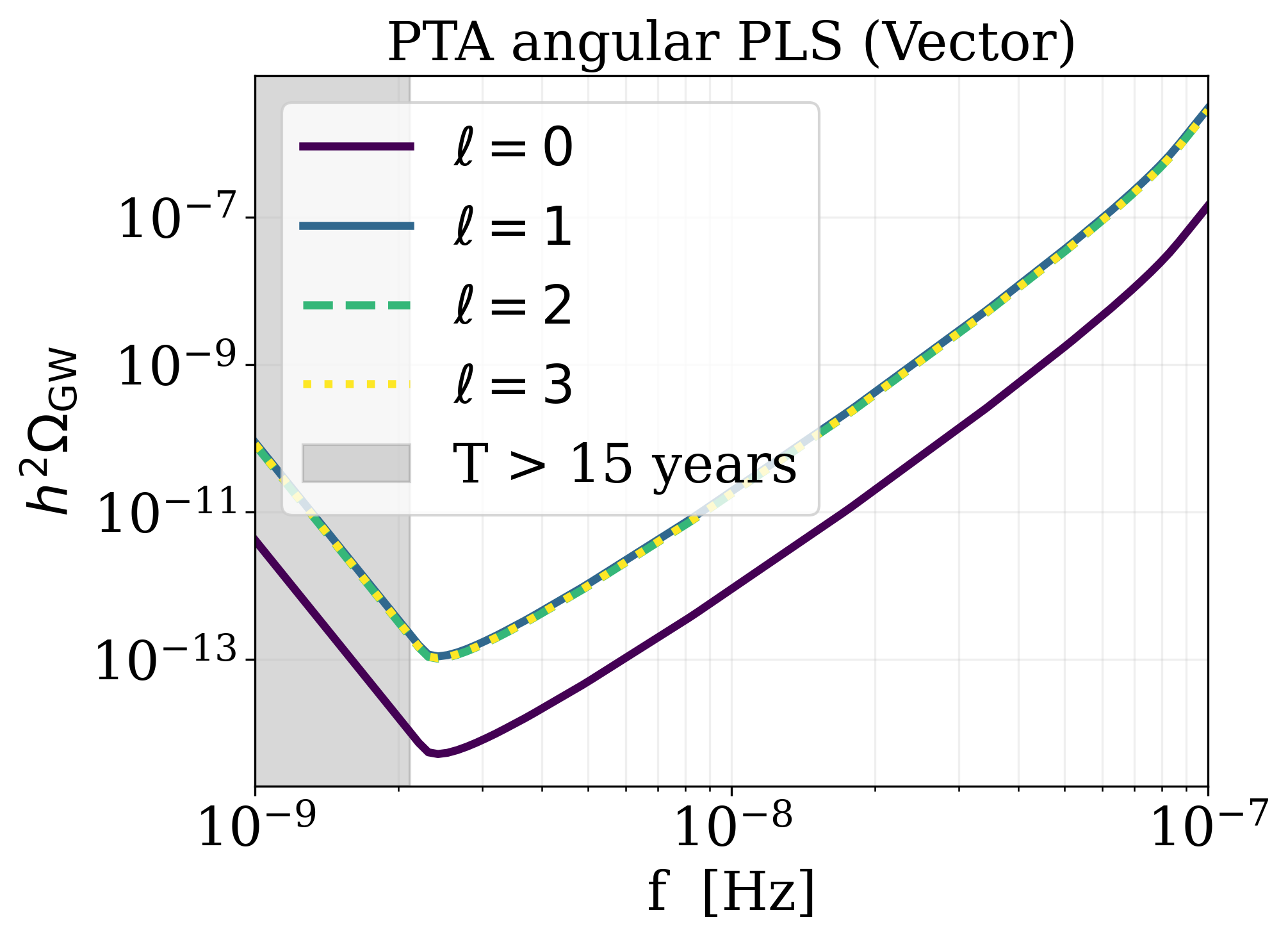}
    \includegraphics[scale=0.45]{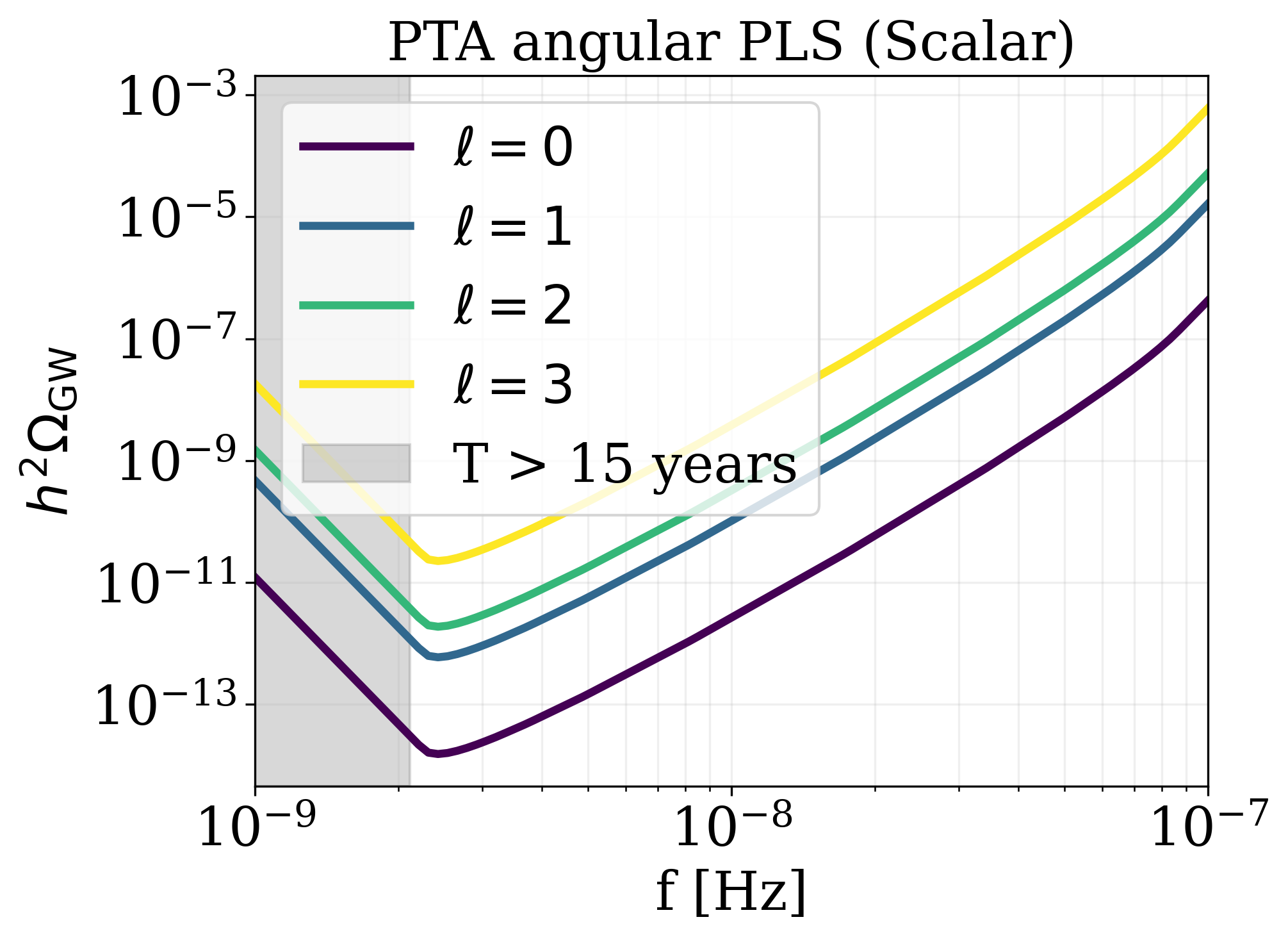}
    \includegraphics[scale=0.45]{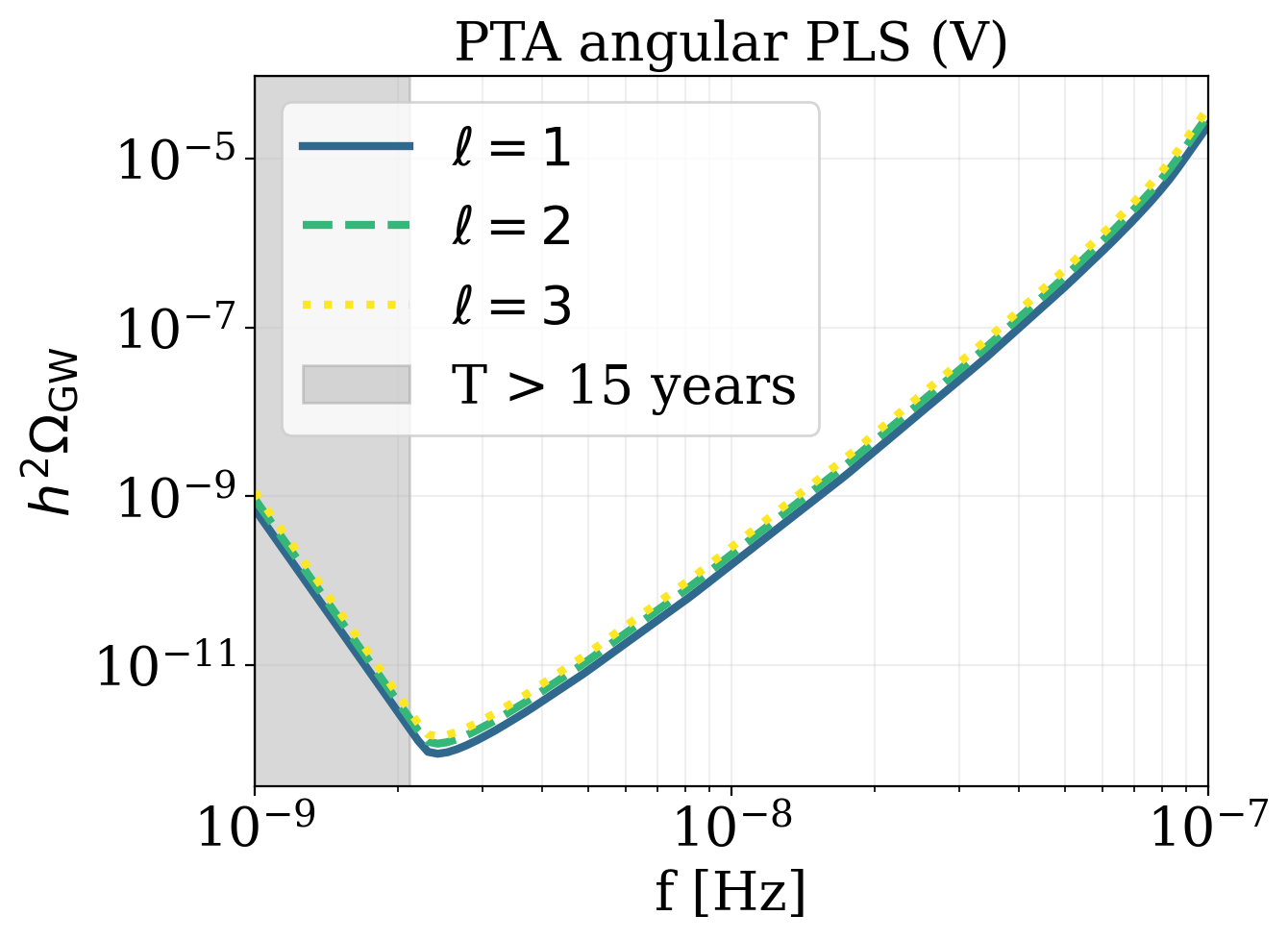}
    \caption{Angular power law integrated sensitivity curves for PTA catalog for $\ell = 0, 1, 2, 3$. We set $\langle\rm SNR \rangle_{\ell, th}=1$, $T= 15 {\rm yrs}$, $C_{\ell = 1, 2, 3}^{\rm GW}=10^{-3}$.}
    \label{fig:PTA_Nell}
\end{figure}

Finally, we show in Figure~\ref{fig:PTA_Nell} the corresponding APLS for the array of pulsars considered. 
Vector modes exhibit the highest sensitivity across all multipoles considered (Table~\ref{tab:aplsminPTA}), in agreement with the predictions of~\cite{daSilvaAlves:2011fp}.

\begin{table}[t!]
    \centering
    \begin{tabular}{ccccc}
    \multicolumn{5}{c}{PTA} \\
    \hline
    \hline
    \multicolumn{5}{c}{$h^2\Omega_{\rm GW}$} \\
    \hline
                 & Tensor                & Vector               & Scalar            & V \\
    \hline
        $\ell=1$ & $1.53 \times 10^{-12}$   & $1.05 \times 10^{-13}$   & $5.61\times 10^{-13}$ & $8.93\times 10^{-13}$  \\
        & {\scriptsize$@ \, 2.11 \times 10^{-9} \, \rm Hz$} &  {\scriptsize$ @ \, 2.11 \times 10^{-9} \, \rm Hz$} &  {\scriptsize$ @ \, 2.11 \times 10^{-9} \, \rm Hz$} & {\scriptsize$ @ \, 2.11 \times 10^{-9} \, \rm Hz$} \\
        $\ell=2$ & $1.08 \times 10^{-12}$  & $9.71\times 10^{-14}$    & $1.73\times 10^{-12}$  & $1.18\times 10^{-12}$ \\
        & {\scriptsize$@ \, 2.11 \times 10^{-9} \, \rm Hz$} &  {\scriptsize$ @ \, 2.11 \times 10^{-9} \, \rm Hz$} &  {\scriptsize$ @ \, 2.11 \times 10^{-9} \, \rm Hz$} & {\scriptsize$ @ \, 2.11 \times 10^{-9} \, \rm Hz$} \\
        $\ell=3$ & $1.27 \times 10^{-12}$  & $9.73\times 10^{-14}$    & $6.67\times 10^{-12}$  & $1.43\times 10^{-12}$\\
        & {\scriptsize$@ \, 2.11 \times 10^{-9} \, \rm Hz$} &  {\scriptsize$ @ \, 2.11 \times 10^{-9} \, \rm Hz$} &  {\scriptsize$ @ \, 2.11 \times 10^{-9} \, \rm Hz$} & {\scriptsize$ @ \, 2.11 \times 10^{-9} \, \rm Hz$} \\
    \hline
    \end{tabular}
    \caption{Minimum energy density spectrum amplitude required ($\langle\rm SNR \rangle_{\ell, th}=1$, $T= 15 {\rm yrs}$ and  $C_{\ell=1,2,3}=10^{-3}$) for a SGWB made of tensor, vector or scalar polarization modes to spot for anisotropies.}
    \label{tab:aplsminPTA}
\end{table}

\section{Conclusions}
\label{Conclusion}

In this paper, we have presented \texttt{GWBird}, a self-comprehensive and modular Python-based code designed for the characterization of the SGWB across a broad spectrum of detectors, including ground-based, space-based, and PTAs detectors. The code incorporates all possible GW polarizations: tensor, vector, and scalar, in addition to circular polarization modes, thus enabling studies not only within GR but also in the context of alternative theories of gravity.

Our implementation includes tools for computing ORFs, angular response functions, isotropic and anisotropic Power Law Sensitivities Curves (PLS and APLS), SNR, and sky-maps for detector responses. These functionalities are fully customizable, offering flexibility in choosing detector geometries, orientations, noise curves, and network configurations, particularly suited to explore the performance landscape for third-generation ground-based detectors whose final layouts are still under discussion.

We have applied \texttt{GWBird} to several representative detector configurations, including the current LIGO-Virgo-KAGRA network, ET in different layouts contemplated in literature, LISA, and PTA under the assumption of considering a specific NANOGrav catalog.  The results recover known benchmarks and extend naturally to scenarios involving non-standard polarizations and anisotropic backgrounds.
We show how angular sensitivity varies across detector types and configurations, highlighting the potential of next-generation detectors to map anisotropies and distinguish between astrophysical and cosmological sources.

A particularly novel feature of \texttt{GWBird} is its capacity to compute angular sensitivity to different GW polarization states for both interferometers and PTAs, including circular polarization, a crucial observable for probing parity violation in the early universe and astrophysical discrete sources. 
Crucially, \texttt{GWBird} offers a coherent and efficient environment to evaluate the angular and polarization sensitivity of future networks such as the Einstein Telescope, providing a valuable perspective on their discovery potential under different design scenarios.

In summary, \texttt{GWBird} delivers a fast, unified, and versatile platform for the characterization of detector networks sensitivity to the stochastic gravitational-wave background, supporting both current and next-generation detectors. The code is publicly available, fully documented, and accompanied by a tutorial notebook that allows for reproducibility of all the results presented in this work, making it an accessible and powerful resource for the gravitational-wave community.

\acknowledgments

We thank Gabriele Astorino, Andrea Begnoni, Patrick M. Meyers, Gabriele Perna, Mauro Pieroni and Lorenzo Valbusa Dall'Armi for valuable discussions and feedback on the draft.

We acknowledge usage of the Python programming
language~\cite{10.5555/1593511}, as well as the following Python packages: \texttt{numpy}~\cite{harris2020array}, \texttt{healpy}~\cite{Zonca2019, 2005ApJ...622..759G}, \texttt{matplotlib}~\cite{Hunter:2007}, \texttt{astropy}~\cite{astropy:2013, astropy:2018, astropy:2022}, \texttt{mpmath}~\cite{mpmath}, \texttt{scipy}~\cite{2020SciPy-NMeth}, \texttt{PINT}~\cite{2021ApJ...911...45L, 2024ApJ...971..150S}.

\appendix

\section{Ground based detectors coordinates}

\begin{table}[H]
\centering
\renewcommand{\arraystretch}{1.3}
\setlength{\tabcolsep}{8pt}
\begin{tabular}{|p{4cm}|l|l|}
\hline
\textbf{Detector} & \textbf{Component} & \textbf{Coordinates} \\
\hline
\hline
\multirow{3}{*}{LIGO Hanford} 
  & Vertex       & [ -0.338, -0.600, 0.725 ] \\
  & First arm    & [ -0.224, 0.800, 0.557 ] \\
  & Second arm   & [ -0.914, 0.0261, -0.405 ] \\
\hline
\multirow{3}{*}{LIGO Livingston} 
  & Vertex       & [ -0.017, -0.861, 0.508 ] \\
  & First arm    & [ -0.955, -0.142, -0.262 ] \\
  & Second arm   & [ 0.298, -0.488, -0.821 ] \\
\hline
\multirow{3}{*}{Virgo} 
  & Vertex       & [ 0.712, 0.132, 0.690 ] \\
  & First arm    & [ -0.701, 0.201, 0.684 ] \\
  & Second arm   & [ -0.049, -0.971, 0.236 ] \\
\hline
\multirow{3}{*}{KAGRA} 
  & Vertex       & [ -0.591, 0.546, 0.594 ] \\
  & First arm    & [ -0.390, -0.838, 0.382 ] \\
  & Second arm   & [ 0.706, -0.006, 0.709 ] \\
\hline
\multirow{3}{*}{ET Sardinia $\Delta$} 
  & Center       & [ 0.750, 0.124, 0.650 ] \\
  & First vertex & [ 0.462, -0.801, -0.381 ] \\
  & Second vertex& [ -0.640, -0.106, 0.762 ] \\
  & Third vertex & [ 0.178, 0.908, -0.381 ] \\
\hline
\multirow{3}{*}{ET Sardinia 2L} 
  & Vertex       & [ 0.750, 0.124, 0.650 ] \\
  & First arm    & [ -0.640, -0.106, 0.762 ] \\
  & Second arm   & [ 0.165, -0.987, 0.000753 ] \\
\hline
\multirow{3}{*}{ET Netherlands 2L} 
  & Vertex       & [ 0.628, 0.0625, 0.776 ] \\
  & First arm    & [ -0.443, -0.788, 0.429 ] \\
  & Second arm   & [ 0.639, -0.613, -0.466 ] \\
\hline
CE & \multicolumn{2}{c|}{Same as LIGO Hanford} \\
\hline
\end{tabular}
\caption{Ground-based detector positions.}
\label{tab:detector_positions}
\end{table}

\newpage
\bibliographystyle{JHEP}
\bibliography{bibliography.bib}


\end{document}